\documentclass[12pt]{article}

\RequirePackage{amsthm,amsmath,amsfonts,amssymb}
\RequirePackage[round]{natbib}
\RequirePackage[colorlinks,citecolor=blue,urlcolor=blue]{hyperref}
\RequirePackage{graphicx}
\usepackage{authblk}
\usepackage[english]{babel}
\usepackage[normalem]{ulem}
\usepackage{csquotes}
\usepackage{multirow}
\usepackage{latexsym}
\usepackage{tabularx}
\usepackage{array}
\usepackage{lmodern}
\usepackage{url}
\usepackage{multirow}
\usepackage{hyperref}
\usepackage{booktabs}
\usepackage{nicefrac}
\usepackage{microtype}
\usepackage{mathtools}
\usepackage{algorithm}
\usepackage{algpseudocode}
\usepackage{appendix}
\usepackage{subcaption}
\usepackage{soul}

\providecommand{\bysame}{\leavevmode\hbox to 3em{\hrulefill}\,}

\begin{document}

\title{Modelling spatial heterogeneity in the effects of area-level covariates on income distributions using Bayesian nonparametric methods}


\author[1]{Ziyou Wang}
\author[2]{Jim Griffin}
\author[1]{Maria Kalli}

\affil[1]{Department of Mathematics, King's College London}

\affil[2]{Department of Statistical Science, University College London}

\date{}

\maketitle

\begin{abstract}
Understanding the how the distribution of an economic outcome, such as income, changes with respect to space and covariates is a key concern for policy makers. 
To address this, we develop a Bayesian nonparametric model, the Normalised Latent Measure Factor Model with Covariates (NLMFM-C), which expresses a 
large collection of related densities as mixtures of latent factor densities and
allows for spatial and covariate effects.
We propose an adaptive Gibbs sampler to automatically infer the number of latent factor distributions, and a rotation method to make posterior inference on different data sets comparable.
We apply the NLMFM-C model to Public Use Microdata Sample (PUMS) data, focusing on 
income distributions for sub-areas of four U.S. states over to different years,  2016 and 2020. We show that the latent factor distributions can be interpreted by income level
({\it e.g.}, low, medium, and high) and investigate 
 the spatially- and time-changing impact of three covariates: gender, race and educational attainment.


\end{abstract}


\section{Introduction}\label{sec:introduction}

Economists have long been interested in understanding socioeconomic disparities and spatial patterns by analyzing how economic covariates differ across geographical regions. Spatial variations in these covariates provide insights into the effectiveness of public policies and the allocation of resources \citep{cartone2022regional, moucque2000survey, eva2022spatial}. Income distributions in different areas, which we call \textit{areal income distributions},
play a central role in characterizing inequality and guiding policy evaluation. 
Traditional approaches that rely on summary statistics or pointwise comparisons \citep{thas2010comparing, engmann2011comparing} are limited in scalability and interpretability due to their focus on simplified measures rather than capturing the full shape of distributions, particularly when comparing hundreds of groups.
We adopt a Bayesian nonparametric (BNP) approach which allows spatially-varying demographic covariate effects without making parametric assumptions.

We analyse personal income data from 
the American Community Survey (ACS), an annual nationwide survey conducted by the United States (US) Census Bureau, collecting detailed economic, demographic, and social data at fine geographic scales. Its Public Use Microdata Sample (PUMS) offers anonymized individual-level records 
and has become a key data source for studying income inequality and spatial heterogeneity in socioeconomic outcomes.
Several recent studies have leveraged PUMS data to analyze income variation.  \citet{simpson2023interpolating} use a model-based framework to interpolate tract-level income distributions and examine the relationship between income inequality and segregation. However, their approach focuses on estimating marginal income densities and does not consider demographic covariates or cross-group comparisons. \citet{hu2023bayesian} proposed a spatial clustering method for state-level Lorenz curves to uncover similarities, but this approach omits distributional shape differences and prevents direct comparison of full income distributions across groups. Both aforementioned studies contribute to understanding areal income variation, yet neither is designed to capture full distributional heterogeneity or to assess the effects of demographic covariates within an integrated framework. A related but distinct line of research is exemplified by \citet{Klein_2015}, who analyzed German socio-economic panel data using a structured additive distributional regression framework. While this approach enables interpretation of covariate effects on distributional features, its expressiveness is constrained by the underlying parametric assumptions, which  may not be sufficiently flexible to model skewed, heavy-tailed or multimodal distributions. 

Existing methods are not designed to jointly address both full distributional heterogeneity and covariate effects within a single, integrated framework, and we address this gap.
While frequentist approaches such as kernel density estimation are popular, they often struggle with bandwidth selection and capturing complex multimodal structures, especially in high-dimensional settings. BNP methods overcome these limitations by assuming an infinite number of parameters and
allowing the number of components of mixture modules to grow adaptively with the data. \citet{lo1984class} introduced an infinite mixture model for density estimation where the mixing measure is the Dirichlet process (DP) of \citet{ferguson1973bayesian}. This 
Dirichlet Process Mixture Model (DPMM) has been extended to allow dependence on covariates and group-specific structures, enabling richer modelling of heterogeneity; see \cite{quintana2022dependent} for a review.

Efficiently summarising distributional features shared across many distributions in an interpretable way
is fundamental to our approach and 
 motivates the use of the Normalised Latent Measure Factor Model (NLMFM) \citep{beraha2023normalised}.
 This model expresses a collection of related distributions as a weighted sum of common underlying factor distributions leading to a factor-analytic structure which enables  comparison of hundreds of distributions revealing global latent ``traits''  across groups.
This is a task that standard nested or spatially smoothed processes struggle to perform efficiently.
 By comparing the weights of the factor distributions across groups, we can 
 estimate covariate effects and their spatial variation.
The model allows us to address three  objectives. First, motivated by extensive evidence linking demographic covariates, such as gender, race, and educational attainment, to income outcomes \citep{tinbergen1972impact, blau1994rising, shaikh2014race}, 
we extend the NLMFM by including such covariates. Second, by comparing income distributions across groups and time periods, 
we identify evolving spatial patterns. Finally, we consider multiple US states, and describe a cross-state analysis.


The remainder of the paper is structured as follows. In Section \ref{sec:data}, we introduce the PUMS data from the ACS. Section 3 describes the NLMFM with covariates (NLMFM-C) together with a sampler that adaptively chooses the number of latent factors. Sections 4 and 5 provide a simulated example and the application of the NLMFM-C to PUMS data, respectively. Section 6 summarises our findings. 


\section{Public Use Microdata Sample}\label{sec:data}
The ACS PUMS data contain detailed demographic information such as age and education attainment, alongside continuous measures like income.  To protect confidentiality, PUMS aggregates individuals into PUMAs, non-overlapping geographic regions, each covering at least 100,000 residents. We focus on comparing income distributions (rather than summaries) across PUMAs, US states and years while accounting for three demographic covariates: gender, race, and educational attainment, which are
fundamental drivers of income inequality, see
 \citep{xie2022influence, liu2019income,  mckinney2022us}. Table \ref{tab:cov description1} summarises these covariates. 
\begin{table}[htbp]
\centering
\caption{Individual-level categorical covariates of PUMS data}
\begin{tabular}{lll} 
\hline
\textbf{Variable} & \textbf{Description} & \textbf{Category Codes} \\
\hline
SEX   & Sex                     & \textbf{1} = Male  \\
               &                          & \textbf{2} = Female \\
\hline
RAC1P & Recoded Race Code        & \textbf{1} = White  \\
               &                          & \textbf{2+} = Other Races \\
\hline
SCHL  & Educational Attainment   & \textbf{1-20} = Below Bachelor's Degree \\
               &                          & \textbf{21-24} = Bachelor's Degree and Above \\
\hline
\end{tabular}
\label{tab:cov description1}
\end{table}

We consider the \href{https://www.census.gov/programs-surveys/acs/microdata/access.2016.html}{2016} and \href{https:
//www.census.gov/programs-surveys/acs/data/experimental-data/2020-1-year-pums.html}{2020} ACS PUMS datasets, and analyse the income variable (PINCP) for individuals in California (265 PUMAs), Florida (151 PUMAs), New York (145 PUMAs), and Washington (56 PUMAs). We randomly sample 200 individuals from each PUMA for our analysis. The four states selected illustrate the political and economic diversity within the US. California is known for its strong liberal political orientation, particularly in urban centers such as San Francisco and Los Angeles. The state's economy is driven by the high-tech and agricultural sectors, leading to pronounced income inequality and wealth concentration \citep{ppicCaliEconomy}. Washington state also tends to be liberal with substantial redistributive policies \citep{acsB19083WA2022,waMinWage}, especially in metropolitan areas such as Seattle, although some rural areas tend to be more conservative. New York state is politically liberal overall, with economic activity concentrated in financial, legal and hospitality services, and notable urban–rural divides in both income and political orientation, \citep{nycComptrollerEconomy}. Florida has a diversified economy, including tourism, agriculture and real estate \citep{visitFloridaResearch,floridaRealtorsStats}. Predominately conservative, it is widely regarded as a political swing state due to its close elections. Such labour force variation and political orientation provides a natural setting to investigate how these factors interact with demographic composition to influence income inequality.

\section{Model}\label{sec:model}

We assume that there are $g$ areas and that, in the $j$-th area, a sample 
$\boldsymbol{y}_j = (y_{j,1},\dots, y_{j, n_j})$ is observed that is modelled by $y_{j,i}\stackrel{i.i.d.}{\sim} p_j$ for $i=1,\dots, n_j$ where $p_1, \dots, p_g$ are densities. 

\subsection{Normalised latent measure factor model}
The Normalised Latent Measure Factor Model (NLMFM) \citep{beraha2023normalised}  parsimoniously models 
$p_j$ as 
\begin{equation}\label{eq:pj_simple}
p_j = \sum_{h=1}^H s_{j,h}\, p_h^\star,
\end{equation}
where $p_1^\star, \dots, p_H^\star$ are densities,
$s_{j,h}\ge0$ and $\sum_{h=1}^H s_{j,h}=1$.
The densities $p_1^\star, \dots, p_H^\star$ are interpreted as $H$ shared latent factor densities and 
$s_{j,h}$ is the weight of  $p_h^{\star}$ in the $j$-th area. This construction borrows strength while still allowing flexibility in distributional shape. 
 A factor structure can be built by assuming that 
\[
s_{j,h} = \frac{\lambda_{j,h}\sum_{k=1}^\infty m_{h,k}J_k}
              {\sum_{l=1}^H \lambda_{j,l}\sum_{k=1}^\infty m_{l,k}J_k}, \qquad j = 1, \dots, g, \quad h = 1, \dots, H,
\]
and
\[
p_h^\star(\cdot) = \frac{\sum_{k=1}^\infty m_{h,k}\, J_k  \, f(\cdot \mid \theta_k^\star)}
                 {\sum_{k=1}^\infty m_{h,k}\, J_k}, \qquad h=1,\dots,H.
\]
where $J_k > 0$, $\lambda_{j,h}>0$ and $m_{h, k} > 0$ for $j = 1, \dots, g$, $k = 1,\dots,\infty$ and $h = 1,\dots, H$, $\sum_{k=1}^{\infty} J_k < \infty$, and 
 $f(\cdot\mid\theta)$ denotes a density with parameters $\theta$. We refer to $\boldsymbol\lambda_h = (\lambda_{1, h}, \dots, \lambda_{g, h})$ as the $h$-th factor loading.
 The shared jumps $J_1, J_2, \dots$ are given a gamma process prior, while $m_{h, k}$ is given a gamma prior distribution. 
The structure of the weights and latent factor densities define a factor-type structure for $\lambda_{j, h}$ and $m_{h, k}$ and  lead to the simplified form 
\[
p_j(\cdot) = \frac{\sum_{l=1}^H \lambda_{j,l}\sum_{k=1}^\infty m_{l,k}J_k f(\cdot \mid \theta_k^\star)}
           {\sum_{l=1}^H \lambda_{j,l}\sum_{k=1}^\infty m_{l,k}J_k}, \qquad j = 1,\dots, g.
\]

\subsection{Normalised latent measure factor model with covariates}\label{sec:model cov}
The NLMFM does not account for demographic and socioeconomic covariates and spatial location. We assume that a vector of $p$ covariates $\boldsymbol{c}_j = (c_{j, 1}, \dots, c_{j, p})$ are observed in the $j$-th area.
 The NLMFM can be seen as reducing the heterogeneity in the densities $p_1\dots, p_g$ to heterogeneity in the $H$-dimensional vectors of unnormalized weights $\boldsymbol{s}_1, \dots, \boldsymbol{s}_j$ where $\boldsymbol{s}_j = (\boldsymbol{s}_{j, 1}, \dots, \boldsymbol{s}_{j, H})$, which
determine how strongly each factor density contributes to $p_j$. This suggests modelling 
 the effects of covariates on area-specific  densities through the unnormalized weights. Specifically, we model the $h$-th factor loading by a  log-linear model with 
log–Gaussian Markov random field (MRF) errors to give
 \begin{equation}
 \log \boldsymbol{\lambda}_{h} = \psi+ \boldsymbol{C}\boldsymbol{\zeta}_h + \sum^p_{m=1}\boldsymbol{C}_{\cdot,m}\boldsymbol{\gamma}_{h,m} + \boldsymbol{\epsilon}_{h},
 \qquad h = 1,\dots, H
 \label{equ:lambda prior cov}
 \end{equation}
 where $\boldsymbol{\epsilon}_{h}\sim\mathcal{N}\left(0, \tau^{-1}\, (\boldsymbol{F}-\rho \boldsymbol{W})^{-1}\right)$ for a $(g\times g)$-dimensional adjacency matrix $\boldsymbol{W}$
 and $\boldsymbol{F}$ is a diagonal matrix with  $F_{i,i}=\sum_{j=1}^g W_{i,j}$, and 
$\boldsymbol{C}$ is  the $g\times p$-dimensional design matrix of covariates.
The model includes a baseline log-intensity $\psi$, main covariate effects, $\boldsymbol{\zeta}_h\in \mathbb{R}^{p}$,
a spatial interaction for the $m$-th covariate $\boldsymbol{\gamma}_{h,m}\in \mathbb{R}^{g}$, 
and an error $\boldsymbol{\epsilon}_h$. We define $\boldsymbol\eta_{h, m} = \boldsymbol\zeta_h + \boldsymbol\gamma_{h, m}$ to be the area-specific covariate effect.
The baseline log-intensity $\psi$ sets a reference level of $\log \boldsymbol{\lambda}_h$ in the absence of covariate effects.  
In our simulated and real data examples, the log-Gaussian MRF is specified by 
 an 
adjacency matrix $\boldsymbol{W}$ with $W_{j,l}=1$ if the $j$-th and $l$-th areas share a boundary and 0 otherwise, 
a precision parameter $\tau$, which controls the smoothness of $\boldsymbol{\lambda}_h$, 
and $\rho\in (0,1)$ which regulates the strength of correlation between neighboring areas. The model suffers from two forms of non-identifiability.
Firstly, the model is invariant to re-labelling of the $H$ factor densities. This can be addressed by standard post-processing algorithms for mixture models.
Secondly, the weights $s_{j, h}$ are invariant to additive changes to $\psi$, and to a shift of all regressions effects for the $j$-th covariate $\zeta_{1, j}, \dots, \zeta_{h, j}$. We address this by choosing a reference factor $h$ and 
 setting $\boldsymbol{\zeta}_h = \mathbf{0}$ in a post-processing step.
Both methods are described in the posterior inference section.

\subsection{Residual factor density} \label{sec:residual factor density}
If the related densities $p_1, \dots, p_g$ are fairly similar, the estimated factor densities $p_1^{\star}, \dots, p_H^{\star}$ can also be similar, making it difficult to see differences. In these cases, to easily interpret the factors, we follow \citet{beraha2023normalised} and work with residual factor densities. Specifically, we decompose the density of the $j$-th area, $p_j,$ into two parts: 1) an overall average factor density across areas, $\bar{p}(\cdot)$, and 2) deviations explained by latent factor densities
\begin{equation}\label{eq:residual-density}
r_h(\cdot)\;=\;p_h^\star(\cdot) - \bar{p}(\cdot),\qquad h = 1,\dots, H.
\end{equation}
 Since each of the probability densities $p_h^\star$ and $\bar p$ integrates to one, $r_h$ integrates to zero
and is a signed measure.
A positive value of $r_h(x)$ implies that $p_h^{\star}$ places more mass than the average factor density at $x$ (and similarly, a negative value implies that $p_h^{\star}$ places less mass than the average factor density at $x$). 
\begin{figure}[htbp]
    \centering
    \includegraphics[width=0.8\textwidth]{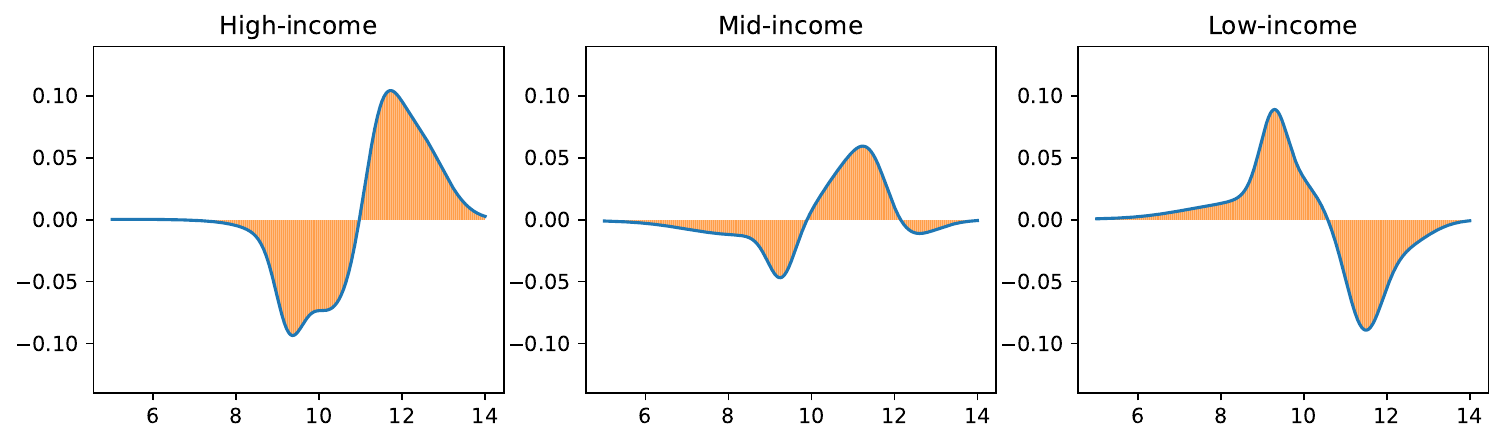}
    \vspace{-0.2cm}
        \caption{Plots of the logarithm of income against the
        posterior mean residual factor densities for California estimated using 
         the NLMFM-C 
        with $H=3$. }
        \label{fig:residual_latent_factor}
\end{figure}
Figure \ref{fig:residual_latent_factor} shows the posterior mean residual densities from the NLMFM-C model applied to log personal income in PUMAs of California.
The first latent factor is labeled high income factor since $r_1(x)$ is positive for higher incomes (with 
a peak  near $x=12$ with $r_1(x)\approx\ 0.10$) and 
negative for lower incomes (with a peak near $x=9$ with $r_1(x) \approx -0.09$).
 Similarly, the second latent factor is labeled mid-income factor since $r_h(x)$ is positive for middle incomes, and the third factor is labeled low-income factor since $r_h(x)$ is positive for lower incomes.

\subsection{Posterior inference}

Posterior inference is made by extending the Gibbs sampler for the NLMFM model 
\citep{beraha2023normalised}. 
We improve computational efficiency and interpretability of their  WAIC-based model selection procedure \citep{watanabe2013widely} 
with an adaptive Gibbs sampler that estimates the number of latent factors $H$ and prunes redundant components in the MCMC run. 

\subsubsection{Adaptive Gibbs sampler}\label{sec:adaptive gibbs}
\cite{beraha2023normalised} infer the number of latent factors in the NLMFM by choosing the value that minimizes the WAIC.
This requires refitting the model multiple times with different numbers of latent factors, which is computationally expensive, and may yield closely related factors, leading to redundancy. To eliminate redundant latent factors, we adopt the adaptive Gibbs sampler for infinite mixture models \citep{murphy2020infinite}. Our adaptive sampler both
adds and removes factor densities. A factor density is removed if two factor residual densities become too similar. 
We define $\boldsymbol{r}_h = (r_{h}(x_1), \dots, r_{h}(x_n))$ to be the $h$-th residual factor density evaluated on a
fixed grid $x_1, x_2 \dots, x_n$, which are 
  normalized to integrate to one under the Riemann approximation $\sum_i r_{h,i}\,(x_i - x_{i-1}) \approx 1$. We quantify the similarity of factor densities $p^{\star}_h(\cdot)$ and $p^{\star}_{\ell}(\cdot)$
using cosine similarity defined as
\begin{equation}\label{eq:cosine}
\cos(u,v)=\frac{\langle \boldsymbol{u},\boldsymbol{v}\rangle}{\|\boldsymbol{u}\|_2\,\|\boldsymbol{v}\|_2},\,\mbox{ for }\,\boldsymbol{u},\boldsymbol{v}\in\mathbb{R}^n
\end{equation}
where $\langle \boldsymbol{u},\boldsymbol{v}\rangle$ denotes the Euclidean inner product and $\|\cdot\|_2$ the
Euclidean norm. Applied to the residual factor densities $(\boldsymbol{r}_h,\boldsymbol{r}_\ell)$ it yields $\cos(\boldsymbol{r}_h,\boldsymbol{r}_\ell)$, with an
associated angular separation measure,
\begin{equation}
\beta_{h, \ell} =\arccos\left(\cos(\boldsymbol{r}_h,\boldsymbol{r}_\ell)\right).
\end{equation}
The $\beta_{h, \ell} $ captures the directional similarity between $p^{\star}_h(\cdot)$ and $p_{\ell}^{\star}(\cdot)$ with smaller values indicating greater similarity. If $\boldsymbol{r}_h$ and $\boldsymbol{r}_\ell$ have small $\beta_{h, \ell}$, then one factor is redundant, since its residual density captures nearly identical deviations from $\bar p(\cdot)$. To remove overlapping (redundant) factors, we  set a threshold $\beta_0$. If $\beta_{h, \ell}$ is less than $\beta_0$ then one of either $h$-th or $\ell$-th factor densities will be removed at random. We found that setting 
the threshold angle $\beta_0$ to 0.1 radians provides a good balance between removing redundant factors and retaining sufficient flexibility in the model. Algorithm \ref{algo} presents our adaptive Gibbs sampling scheme.
\begin{algorithm}
\caption{Gibbs sampler for Adaptive Normalised Latent Factor Model}
\label{algo}
\begin{algorithmic}[1]
\State \textbf{Input:} Number of iterations \(N\), burn-in iterations $N_b$, initial number of latent factors $H$, number of clusters $K$, threshold angle \(\beta_0\), prior distribution parameters, and initial latent factor values.
\State \textbf{Initialize:} Set initial latent factors, clusters, and associated parameters. Define the state of the Markov chain as $\gamma$.
\For {$i = 1, \dots, N$}  \Comment{Main Gibbs sampling loop}
    \If {$i \leq N_b$} $\rightarrow$ Burn-in phase
            \State Calculate the adaptation probability $p_{adapt}=\mbox{exp}(-b_0-b_1\mbox{iter})$
            \If {$\mbox{Uniform}(0,1)<p_{adapt}$}
                \State Compute residual densities and angles $\beta_{h, \ell}$ between each pair of latent factors.
                \If {angle \(\beta_{h, \ell}] > \beta_0\)}
                    \State Add a new latent factor and initialize its corresponding atoms and loadings from their priors.
                \ElsIf {angle \(\beta_{h, \ell} < \beta_0\)}
                    \State Remove the corresponding latent factor, keeping only one from the pair.
            \EndIf
        \EndIf
    \EndIf
    \State Update \(\{m_{hk}\}\), \(\{J_k\}\), and other relevant parameters (e.g., atoms) using the conditional posterior distributions.
\EndFor
\State \textbf{Output:} Updated latent factors, clusters, and associated parameters after the final iteration.
\end{algorithmic}
\end{algorithm}

\subsubsection{Post-processing}\label{sec:post-processing}

We address the invariance of the factors to re-labelling using an equivalent class representation algorithm \citep{PapIll10, RodWal2014}.
To address the invariance of $s_{j, h}$  to a multiplicative change in the unnormalized factor weights $\lambda_{j, h}$,
we choose a baseline reference factor $h^{\star}$ and set $\boldsymbol{\zeta}^{\star}_{h^{\star}} = \mathbf{0}$. 
All MCMC samples are then adjusted as 
$\boldsymbol{\zeta}^{\star}_{h} = 
\boldsymbol{\zeta}_{h} -
\boldsymbol{\zeta}_{h^{\star}}
$. The adjusted main effect $\boldsymbol\zeta^{\star}$ is interpreted as a contrast to the baseline. Similarly, we define post-processed spatial interactions
 $\boldsymbol{\gamma}^{\star}$ 
 where $\boldsymbol{\gamma}^{\star}_{h} = 
\boldsymbol{\gamma}_{h} -
\boldsymbol{\gamma}_{h^{\star}}$.


\section{Simulated Example}\label{sec:simulation}
We use a simulated example to show how the NLMFM-C model  can recover  the factor densities $p_h^{\star}$, 
the main factor effects $\boldsymbol\zeta_h$, and  
the area-specific covariate effects $\boldsymbol\eta_{h, m}$. 
We simulated $n_j=100$ observations for each of $g=100$ areas with two area-specific covariates $C_{j, 1}\sim \text{Beta}(2, 5)$ and $C_{j, 2}\sim \text{Beta}(2, 5)$. This choice 
  reflects the distribution of the covariates in the PUMA data. Observations are generated 
using the following three component mixture model
$$
y_{j,i}\sim w_{j,1}\,\text{N}(-2,1.5^2)+w_{j,2}\,\text{N}(0,1.5^2)+w_{j,3}\,\text{N}(2,1.5^2), \qquad i = 1, \dots ,n_j
$$
which has distinct but slightly overlapping components. 
The mixture weights $w_{j,h}$ are given by 
\[
w_{j,h} = \frac{
\exp\!\left(\boldsymbol{C}_j \,\boldsymbol{\zeta}^\star_h + \sum_{m=1}^p C_{j,m}\,\gamma_{j,h,m}^\star\right)
}{
\sum_{l=1}^H \exp\!\left(\boldsymbol{C}_j \,\boldsymbol{\zeta}^\star_l + \sum_{m=1}^p C_{j,m}\,\gamma^\star_{j,l,m}\right)
}, \quad j=1,\dots,g, \quad h=1, 2, 3.
\]
where the main factor effects are   
$\boldsymbol{\zeta}^\star_1 = (1, -2)$, $\boldsymbol{\zeta}^\star_2 = (0, 0)$, and $\boldsymbol{\zeta}^\star_3 = (-2, 1)$ so that the second component acts as a reference.
These settings lead to the covariates having very different effects on the different mixture weights.
 The spatial covariate interactions $\boldsymbol{\gamma}^{\star}_{j, h, m}$ are drawn from a standard normal distribution. 

We run the adaptive Gibbs sampler for 15,000 iterations with a burn-in of 5,000. The sampler finds three factors which we re-label so that factor 1 corresponds to mixture component 1, and so on. 
\begin{figure}[htbp]
    \centering
    \includegraphics[width=0.8\linewidth]{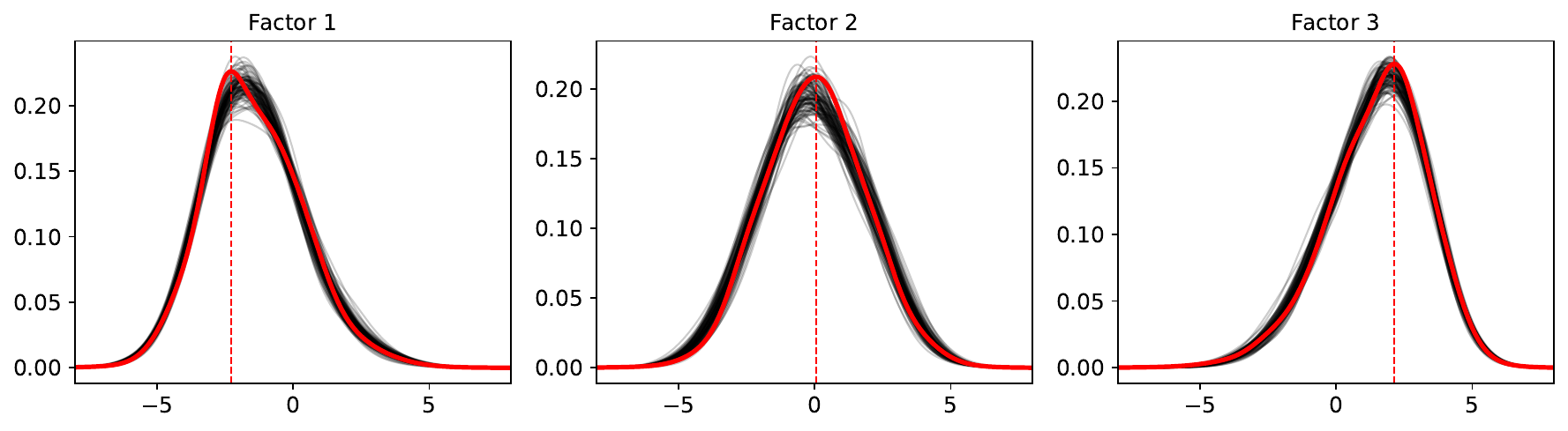}
    \caption{Simulated example posterior draws of factor densities (black) and true mixture components (red line).}
    \label{fig:latent_factors}
\end{figure}
Figure~\ref{fig:latent_factors} shows the posterior draws for each factor density which 
concentrate tightly around the corresponding mixture component demonstrating that the model successfully recovers the true factor densities.
Table~\ref{table:zeta_estimates} contains the posterior means and 95\% credible intervals of $\boldsymbol{\zeta}_{h}$ (using factor 2 as a baseline in the post-processing).
\begin{table}[htbp]
\centering
\caption{Simulated Example: 
Posterior mean and 95\% CI of the main factor effects
$\boldsymbol{\zeta}$.}
\begin{tabular}{|c|c|c|}
\hline
    &Factor 1 &Factor 3 \\ \hline 
$\boldsymbol{\zeta}_1$ & 1.15 [0.40, 1.95]  & -1.91 [-2.70, -1.14]  \\ 
$\boldsymbol{\zeta}_3$ & -1.95 [-2.77, -1.16] & 1.20 [0.39, 2.02]      \\ \hline
\end{tabular}
\label{table:zeta_estimates}
\end{table}
The model is able to recover the true values, which are  close to the posterior means and 
covered by the 95\% credible intervals for all main factor effects.



%



Next, we assess how well the NLMFM-C recovers differences in the area-specific covariates effects for the factors by comparing the posterior of 
$\eta_{j, h, m} - \eta_{j, l, m}$ using  the empirical coverage ({\it i.e.} the proportion of difference for which the true value is contained in  the corresponding 95\% credible interval), the average bias and the average mean absolute error (MAE). 
 \begin{table}[htbp]
\begin{center}
\caption{Simulated Example: Coverage probabilities (95\% CI), bias and MAE of the posterior mean main and area covariate effects.}
\label{tab:coverage_bias}
\begin{tabular}{|c|ccc|ccc|}\hline
Covariate & \multicolumn{3}{c|}{Main Effects} & \multicolumn{3}{c|}{Area Effects}\\
\hline
 & Coverage (\%) & Bias & MAE & Coverage (\%) & Bias & MAE\\
\hline
 1 & 99 & 0.07 &0.51 & 88 & 0 &1.12\\
 2 & 99 & 0.02 &0.45 & 87 & 0 &1.39\\
\hline
\end{tabular}
\end{center}
\end{table}
 Table~\ref{tab:coverage_bias} shows that credible intervals are slightly conservative with a coverage of 99\% for both covariates, with very small bias.
To assess the model’s ability to capture spatial heterogeneity, we also examine the differences between area-specific covariates for effects for the same factor across different areas, $\eta_{j,h,m} -\eta_{d,h,m}$.
Although there is zero bias the between-area contrasts are less well-estimated with coverage slightly less than the nominal level.
Finally, we assess the effectiveness of our 
adaptive sampler to prune redundant factors during the MCMC run. We compare to an extension of the sampler of \cite{beraha2023normalised} to the NLMFM-C  with the number of latent factors $H$ fixed at three (baseline sampler). The adaptive sampler effectively recovers three factors without requiring manual tuning. To compare the output from the two samplers we consider the following two metrics: 1) Kullback–Leibler (KL) divergence \citep{joyce2011kullback}, which measures how closely the model-based predictive densities approximate the true data-generating densities, and 2) WAIC, which estimates expected out-of-sample predictive accuracy while penalizing model complexity. For KL divergence lower values indicate better performance; for WAIC (reported on the log predictive density scale), values closer to zero indicate better performance.
\begin{table}[htbp]
\centering
\caption{Simulated Example: KL divergence and WAIC for the baseline Beraha-Griffin Gibbs sampler and adaptive Gibbs sampler}
\begin{tabular}{|c|c|c|}
\hline
    &Baseline Gibbs sampler&Adaptive Gibbs sampler \\ \hline 
KL Divergence& 3.29  & 3.27      \\    
WAIC & -21002.96&-20999.18\\ \hline
\end{tabular}
\label{table:kl_waic}
\end{table}
The results in Table~\ref{table:kl_waic} show that the adaptive Gibbs sampler performs better on both metrics, indicating that the adaptive sampler can automatically find a closer representation of the true densities.
 MCMC trace plots of the main factor effects $\boldsymbol\zeta_{h}$ and spatial interaction effects $\boldsymbol\gamma_{j,h}$ are provided in 
Appendix \ref{sec:appendix simulation}, which
confirm that our adaptive Gibbs sampler mixes well. 
\section{Application of NLMFM-C to PUMS data}\label{sec:posterior inference pums}
 
We separately fitted an extended NLMFM-C for pairwise comparisons between years (we chose 2016 and 2020 ACS) to the logarithm of personal income in four states: California, Florida, Washington and New York. The extended model allows us to 
consider changes in demographic covariate effects on the income distribution from 2016 to 2020 and we use a factor rotation method to compare 
covariate effects across states.

Let $\boldsymbol{C}^t$ be the $(g\times p)$-dimensional covariate matrix for year $t$.
In the extended NLMFM-C model, the distribution in the $j$-th PUMA for year $t$ is 
\[
p_j^{t} = \sum_{h=1}^H s^t_{j,h}\, p_h^\star,\quad j = 1,\dots,g,
\]
where the year-dependent weight is
\[
s^t_{j,h} = \frac{\lambda^t_{j,h}\sum_{k=1}^\infty m_{h,k}\,J_k}
              {\sum_{l=1}^H \lambda^t_{j,l}\sum_{k=1}^\infty m_{l,k}\,J_k}, \qquad j = 1, \dots, g, \quad h = 1, \dots, H,
\]
 the factor loadings for the $h$-th factor in the $t$-th year are
\[
\log \boldsymbol\lambda_h^t = \left(
\log\lambda^{t}_{1,h},
\dots, \log\lambda^{t}_{g,h}
\right)^T
=
\psi \mathbf1_g+\mathbf C^{t}\,\boldsymbol{\zeta}^{t}_h + \mathbf C^{t}\,\boldsymbol{\gamma}^t_h + \mathbf\epsilon_h^t,
   \]
and $\mathbf\epsilon_h^t \stackrel{i.i.d.}{\sim} \mathcal{N}\left(0, \tau\left( F-\rho W)^{-1}\right)\right)$. The main effects and
the spatial interactions in  year $t$ for the $h$-th factor are $\boldsymbol\zeta_h^t$ and   $\boldsymbol{\gamma}^{t}_h\in \mathbb{R}$ respectively. We define 
$\Delta s_{j,h} = s_{j,h}^{2020}-s_{j,h}^{2016}$ 
to measure the change in the factor weights between 2016 and 2020.

Our chosen covariates, gender, race and education summarise the demographics in each PUMA. Gender is the average female income minus average male income standardized by overall average income, race is the proportion of white residents, and education is the proportion of residents with a bachelor's degree or higher.

\subsection{Income distributions across states and years}

\begin{figure}[htbp]
    \begin{center}
    \includegraphics[width=0.8\linewidth]{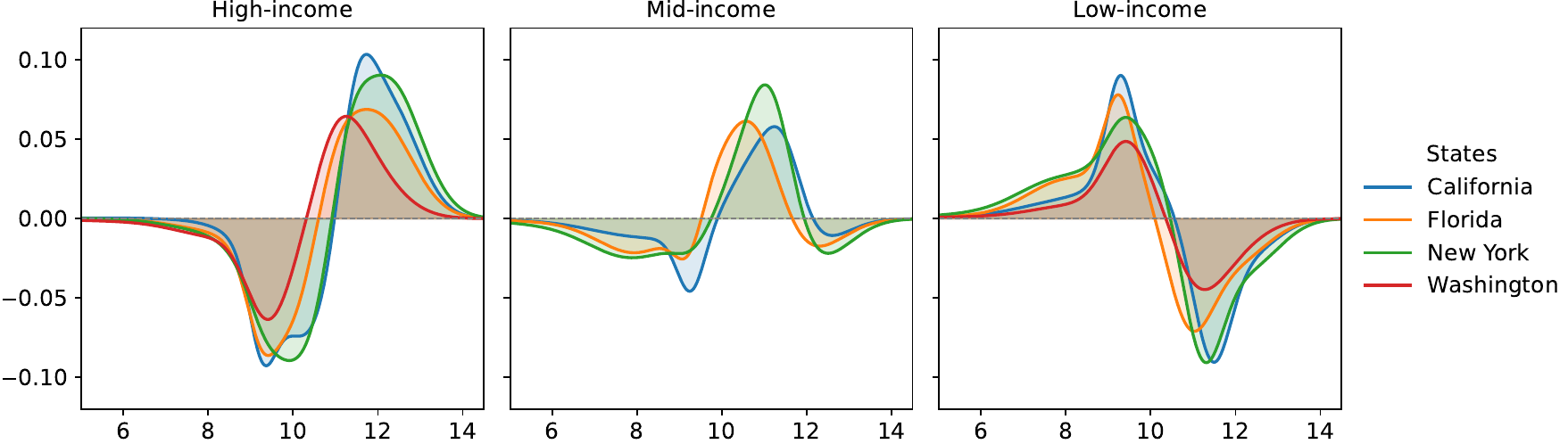}
    \end{center}
    \vspace{-0.2cm}
    \caption{Posterior mean residual factor densities $r_h$ in California, Florida, New York, and Washington (the mid-income factor is only shown for states where this factor is present).}
    \label{fig:residual_latent_factor_combined}
\end{figure}

The adaptive Gibbs sampler identifies three factors for California, Florida and New York, and two factors for Washington.
Figure~\ref{fig:residual_latent_factor_combined} displays the posterior mean residual factor densities. All residual densities have similar shapes which leads to labeling  the factors as high, mid and low income respectively, see Section 3.3. 
The size of peaks and troughs in the low-income and high-income factors differ across states with California and New York showing larger peaks than Florida. Washington has the smallest peaks indicating more homogeneous income distributions compared to the other states, which is consistent with their socioeconomic profiles (see Section~\ref{sec:data}) and levels of income inequality. We choose California as a running example and present analyses for Florida, New York, and Washington  in Appendix~\ref{sec:app cross state}.

\subsection{Analysis for California}\label{sec:cali}

\subsubsection{Factor weights}

\begin{figure}[htbp]
    \centering
    \includegraphics[width=\textwidth]{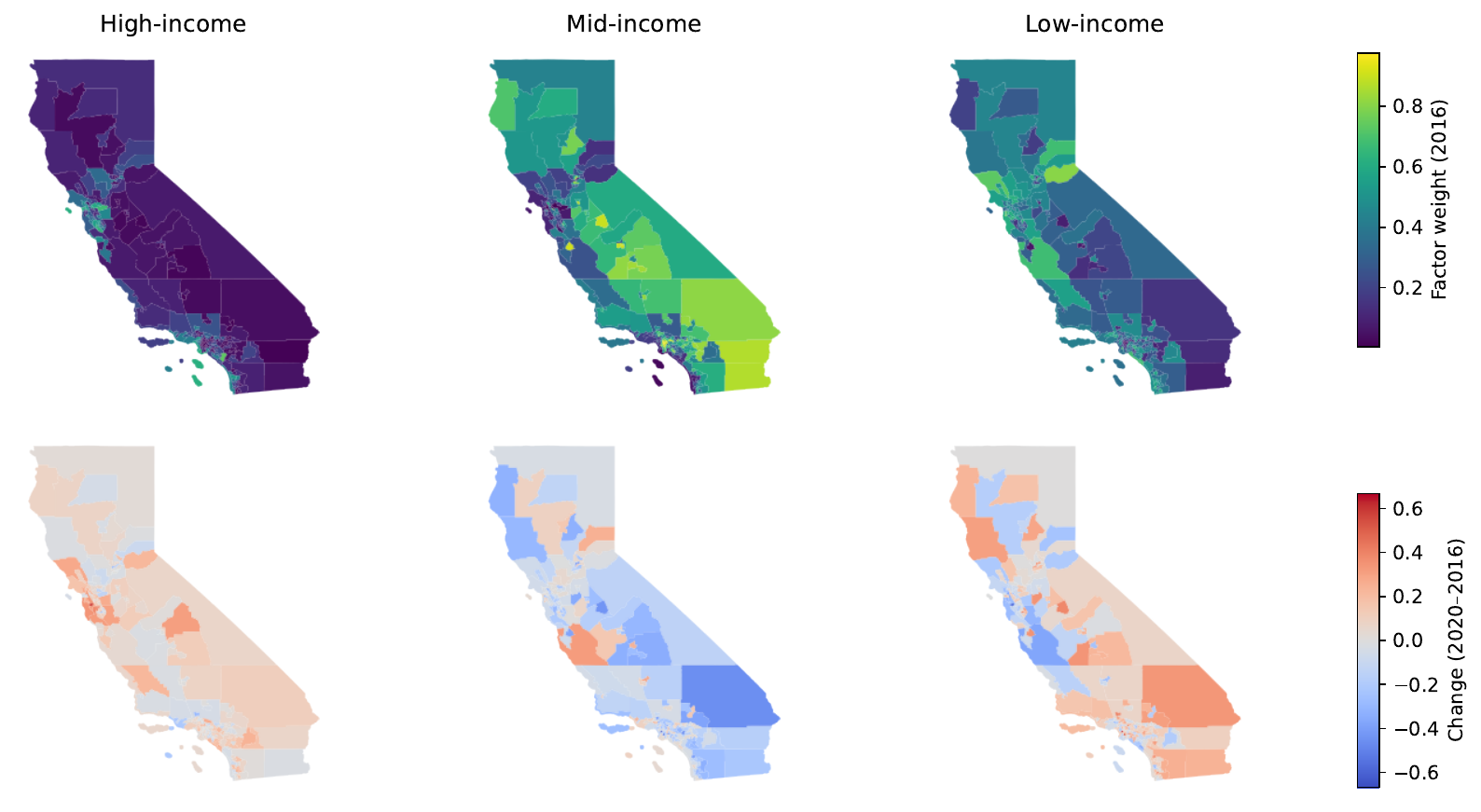}
    \vspace{-0.3cm}
    \caption{Posterior mean factor weights for California. First row displays the 2016 $s_{j,h}$'s. Second row displays $\Delta s_{j,h}$, the change in weights  between 2020 and 2016.}
    \label{fig:california_comparison}
\end{figure}

Figure \ref{fig:california_comparison} displays
 graphs of the spatial distribution of the posterior mean factor weights, $s_{j,h}$ for 2016 (top row), and the posterior mean factor weight change, $\Delta s_{j,h}$, between 2020 and 2016 (bottom row). 
For 2016 we observe large weights for the high-income factor in the major coastal metros of San Francisco Bay Area and Los Angeles–Orange County, whereas Nevada and Sierra counties and El Dorado Hills have large weights on the low income factor. This is in line with the labor market composition of these counties. In the high income pair the majority of jobs are in entertainment and digital media, heavy industry and engineering, and technology where the median wage is \$ 89000 contrasted with the low income pair where the majority of jobs are in education, health and social assistance, restaurant and food services and construction where the median wage is \$ 23000. San Bernardino and Imperial counties have large weights for the mid-income factor, but what is more interesting is their change in weights, $\Delta s_{j,h}$, for the mid and low income factors. To better illustrate the effect of weight changes between 2016 and 2020, we zoom-in on San Bernardino and Monterey South East(SE) PUMAs, see Figure~\ref{fig:pdf_sanbernardino_years} and discuss Imperial county in Appendix B. It is clear there is an inverse weight change between the mid-income and low factors for the PUMAS in these counties. We have a negative change in low income factor and a positive change in the mid-income factor for Monterey and vise-versa for San Bernardino.
This is probably due to the difference in the rate of change in minimum wage between the two counties. San Bernardino had a lower rate of increase in minimum wage (following federal guiedlines), whereas Monterey had a higher rate of increase. The estimated densities of these two counties are left-skewed with that of Monterey SE having a shift in mode towards higher incomes. Regarding the change in factor weight for the high income factor, the bottom row of Figure \ref{fig:california_comparison} reveals an increase in weight in the coastal areas and metro which inline with the increase in executive wage and skilled labor wage of their core industries such as tech, digital media and entertainment, and engineering.



\begin{figure}[htbp]
    \centering
    \includegraphics[width=0.49\linewidth]{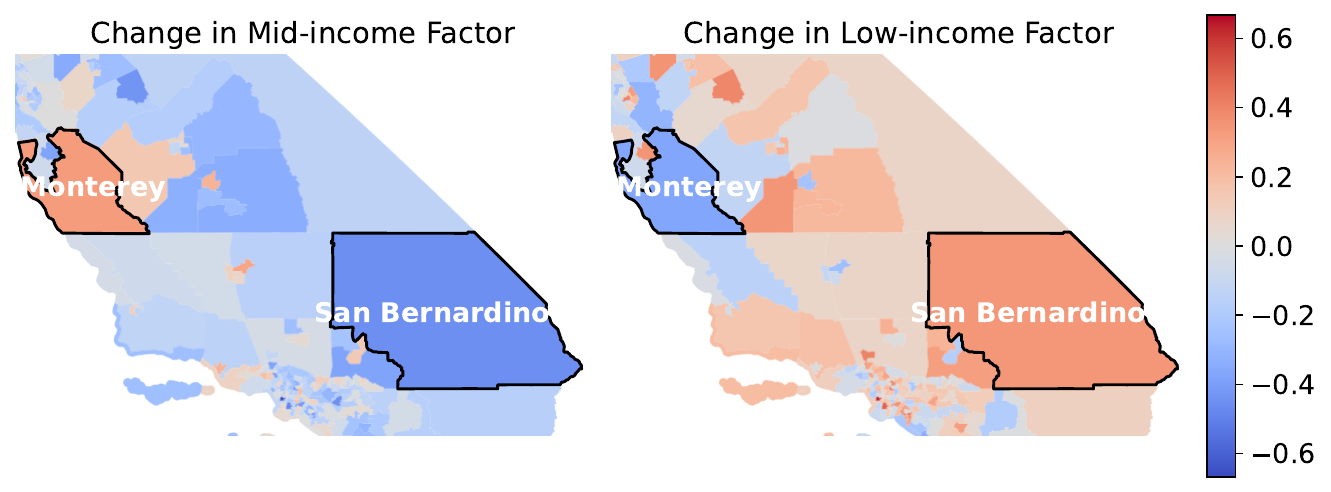}
    \includegraphics[width=0.49\linewidth]{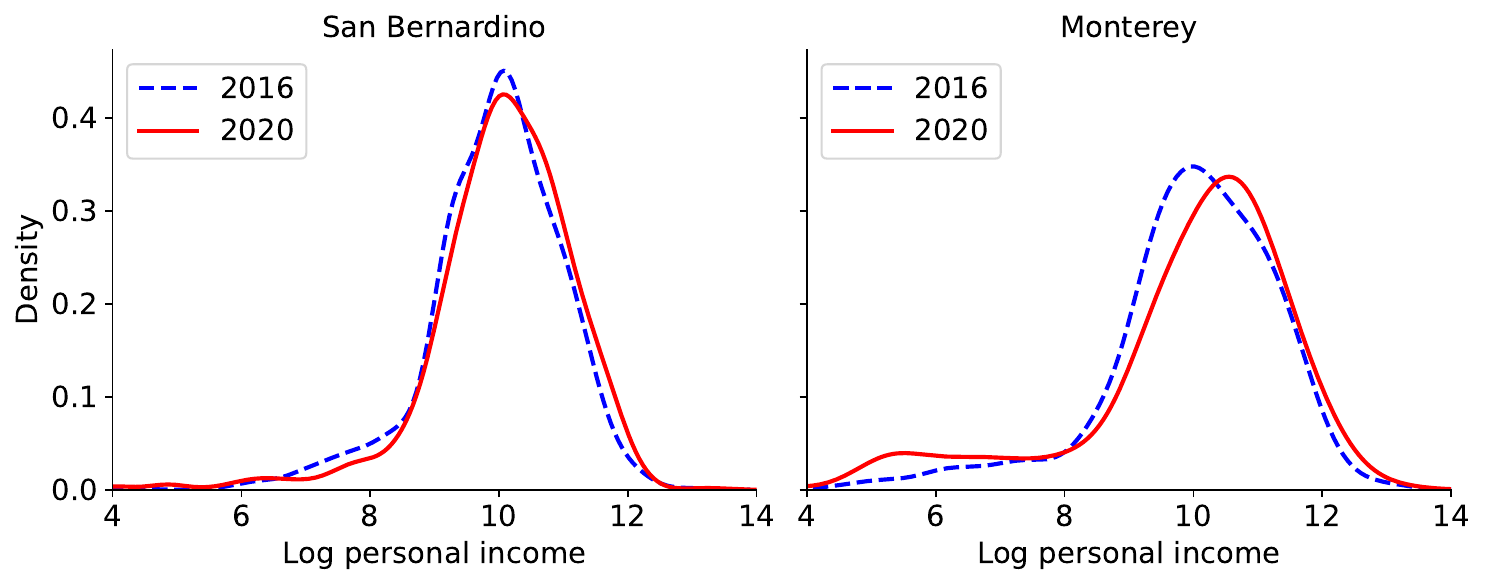}
    \caption{California (zoom-in on San Bernardino and Monterey): Posterior mean of
    $\Delta s_{j,h}$ and posterior mean
    income distributions in 2016 and 2020. }
    \label{fig:pdf_sanbernardino_years}
\end{figure}
\vspace{-0.15in}

\subsubsection{Effects of demographic covariates on personal income distribution}\label{sec:cali cov}

Table~\ref{table:globaleffects} reports the posterior means and 95\% credible intervals (CIs) of the main factor effects $\boldsymbol{\zeta}$ for each covariate. Setting the high income factor as baseline, positive contrasts indicate that large covariate effects are associated with larger weights on the mid and low income factors, whereas negative contrasts indicate a larger weight on the high income factor. We interpret an effect as having strong posterior evidence when its 95\% CI excludes zero.
\begin{table}[htbp]
\centering
\caption{California: Posterior means and 95\% CIs of main effects $\boldsymbol{\zeta}_h$ for each latent factor (high-income factor as baseline) for 2016 and 2020.}
\begin{tabular}{|l|l|l|l|l|}
\hline
  Covariate  & Year & Education & Race & Gender \\ \hline
  Mid-income factor & 2016 & -2.5 [-3.1, -1.8] & 0.7 [0.1, 1.3]  & 0.8 [0.3, 1.3]\\
 & 2020 & -2.0 [-2.6, -1.3]& -1.7 [-2.3, -1.1] & 0.7 [0.1, 1.2] \\ \hline
Low-income factor & 2016 & -5.8 [-6.4, -5.2]& 0.1 [-0.4, 0.6] & 0.9 [0.4, 1.3]    \\ 
 & 2020 &  -5.0 [-5.6, -4.3] & -2.0 [-2.5, -1.5] & 0.6 [0.2, 1.1]  \\ \hline
\end{tabular}
\label{table:globaleffects}
\end{table}
 The contrasts for education are high in magnitude and negative for both mid and low income factors with their 95\% CI's excluding zero for both 2016 and 2020. This suggests that higher education attainment leads to larger factor weight for the high income factor rather than the mid and low income factors, which is in line with labor market composition of the PUMAs with high weights on the high income factor (San Francisco Bay Area and LA and Orange counties). The contrasts for race are lower in magnitude compared to those of education, and differ between mid and low income factors and across years for the mid income factor. For the low income factor we have negative contracts for both years with their 95\% CI's excluding zero. This suggests that race has less of an impact on factor weight. The picture for the mid income factor differs between the two years, with only 2020 having a substantial negative contract with its 95\% CI's excluding zero. This suggests a stronger association between a higher white share and a redistribution of mass away from the mid/low factors towards the high income factor. Gender contrasts are small but consistently positive, with 95\% CIs entirely above zero in both 2016 and 2020.
 
\begin{figure}[htbp]
    \centering
    \includegraphics[width=0.6\linewidth]{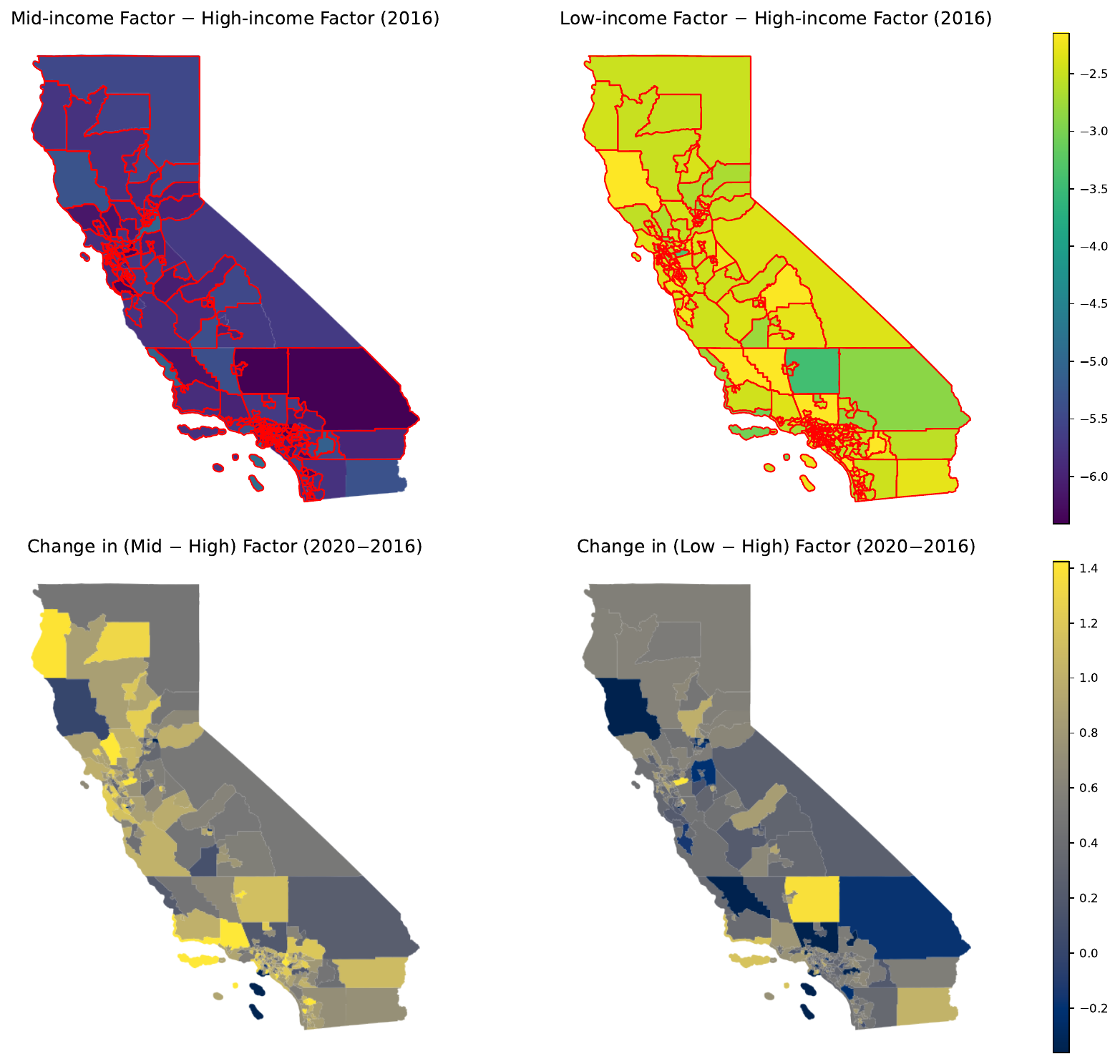}
    \caption{Posterior mean PUMA-specific effects of Education for California in 2016 (top row) and changes from 2016 to 2020 (bottom row). High income factor is the baseline. Red boundaries indicate that the 95\% CIs  exclude zero. }
    \label{fig:schl_2016_2020}
\end{figure}

Since gender contrasts for main factor effects are weak, we focus our attention to education and race for our analysis of the spatial variation in covariate effects. To assess how covariate effects vary spatially, we 
consider the PUMA-specific covariate effects $\boldsymbol\eta_{h, m}$ in 2016 and the change from 2016 to 2020. Figure~\ref{fig:schl_2016_2020} displays the education effects where the red boundaries in the top row indicate that the 95\% CIs exclude zero. These effects are sizable in 2016 for the majority of PUMAs for both mid and low income contrasts. The bottom row has no red boundaries indicating that none of the changes in PUMA specific education effects between 2016 and 2020 pass the 95\% CI threshold.  

\begin{figure}[htbp]
    \begin{center}
  \begin{tabular}{cc}
  Mendocino & Kern East \\
    \includegraphics[width=0.35\linewidth]{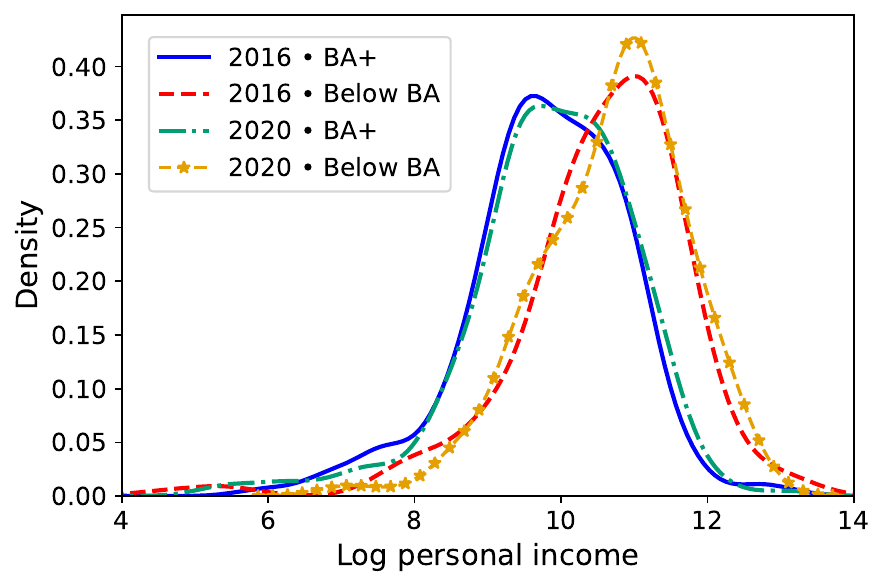}
    &
    \includegraphics[width=0.35\linewidth]{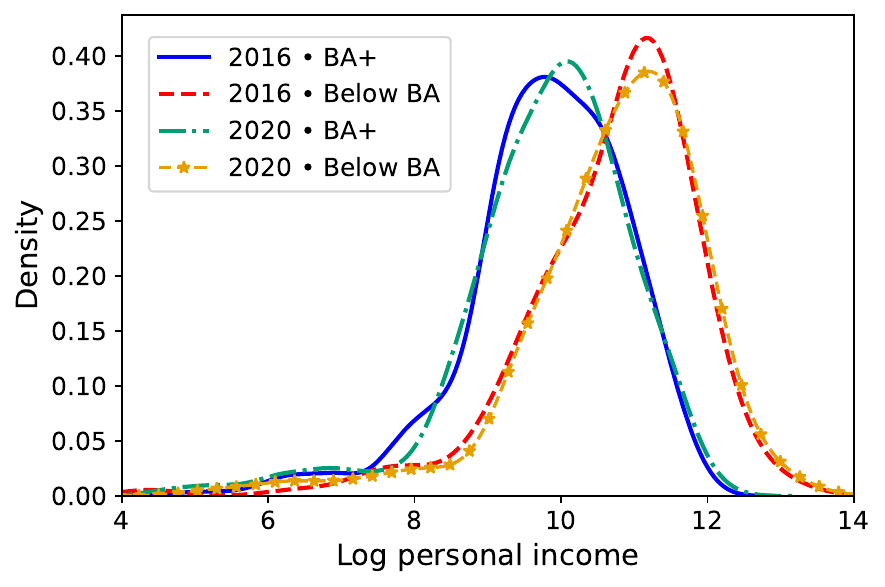}
       \end{tabular}
   \end{center}
  \caption{Mendocino and Kern East, California: Posterior mean densities of log personal income by educational attainment (BA+: bachelor’s degree or higher; Below BA: less than a bachelor’s degree) in 2016 and 2020.}
    \label{fig:schl_pdf_mendocino_kern_east}
        
\end{figure}

Figure~\ref{fig:schl_pdf_mendocino_kern_east} confirms that there is little change in the income distribution for both educational attainment groups in the Mendocino and Kern East.

\begin{figure}[htbp]
    \centering
        \includegraphics[width=0.6\linewidth]{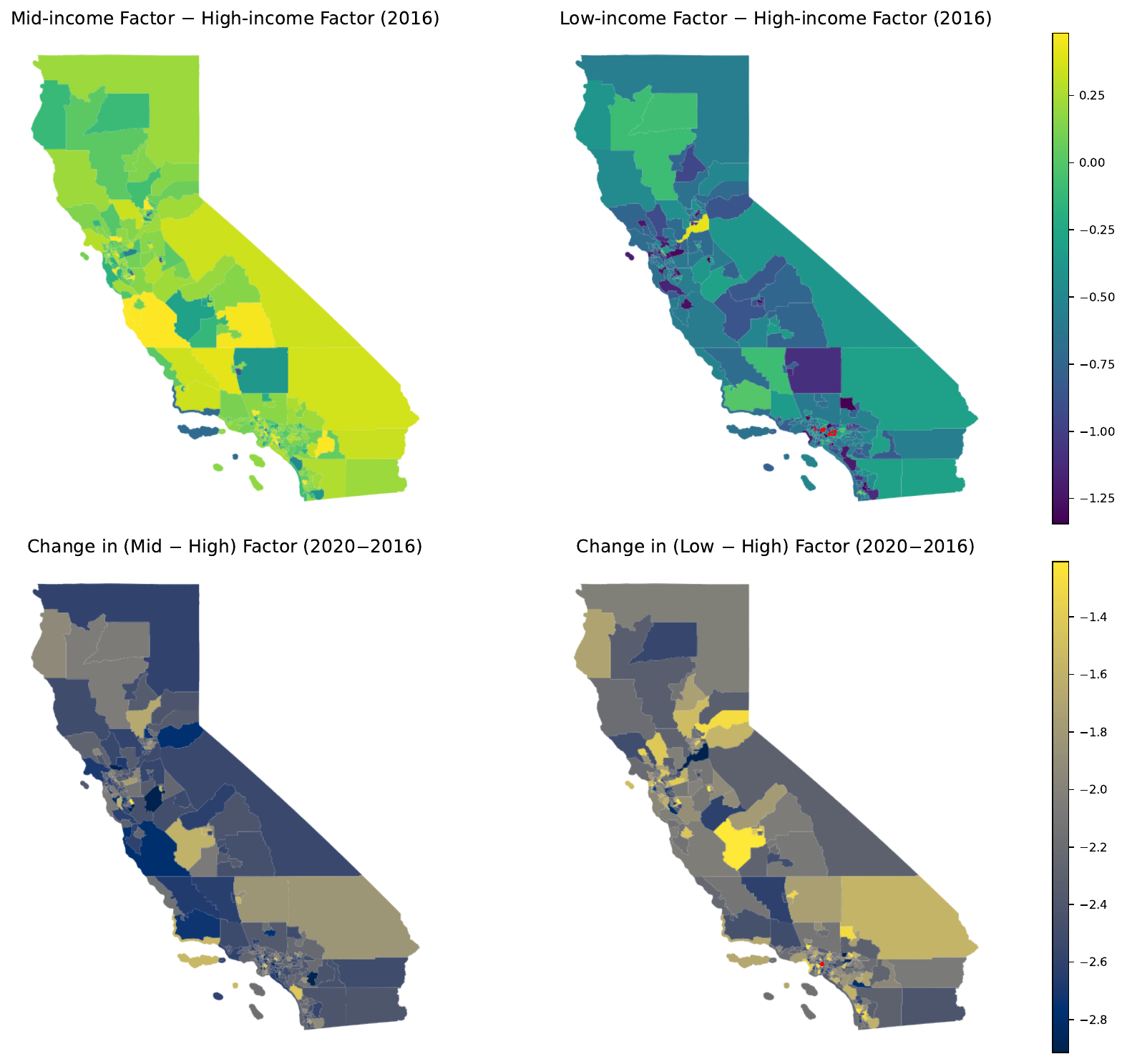}
        \caption{California: Posterior mean PUMA-specific effects of Race (high-income factor as baseline)
       in 2016 (top row) and changes from 2016 to 2020 (bottom row). Strong posterior evidence in a PUMA is indicated by a red boundary.}
        \label{fig:race_2016_2020}
\end{figure}

 Figure~\ref{fig:race_2016_2020} shows the race covariate has strong posterior evidence in 
 only a few PUMAs (e.g. in the Los Angeles basin) for the 
``Low--High'' factor contrast.
From 2016 to 2020, the maps display weak and spatially patchy changes, with most estimates showing high posterior uncertainty. Nevertheless, the posterior mean differences over this period are modestly negative in many regions, typically ranging from –1 to –2, suggesting a downward shift in race-related effects. 
\begin{figure}[htbp]
  \centering
  \includegraphics[width=0.55\textwidth]{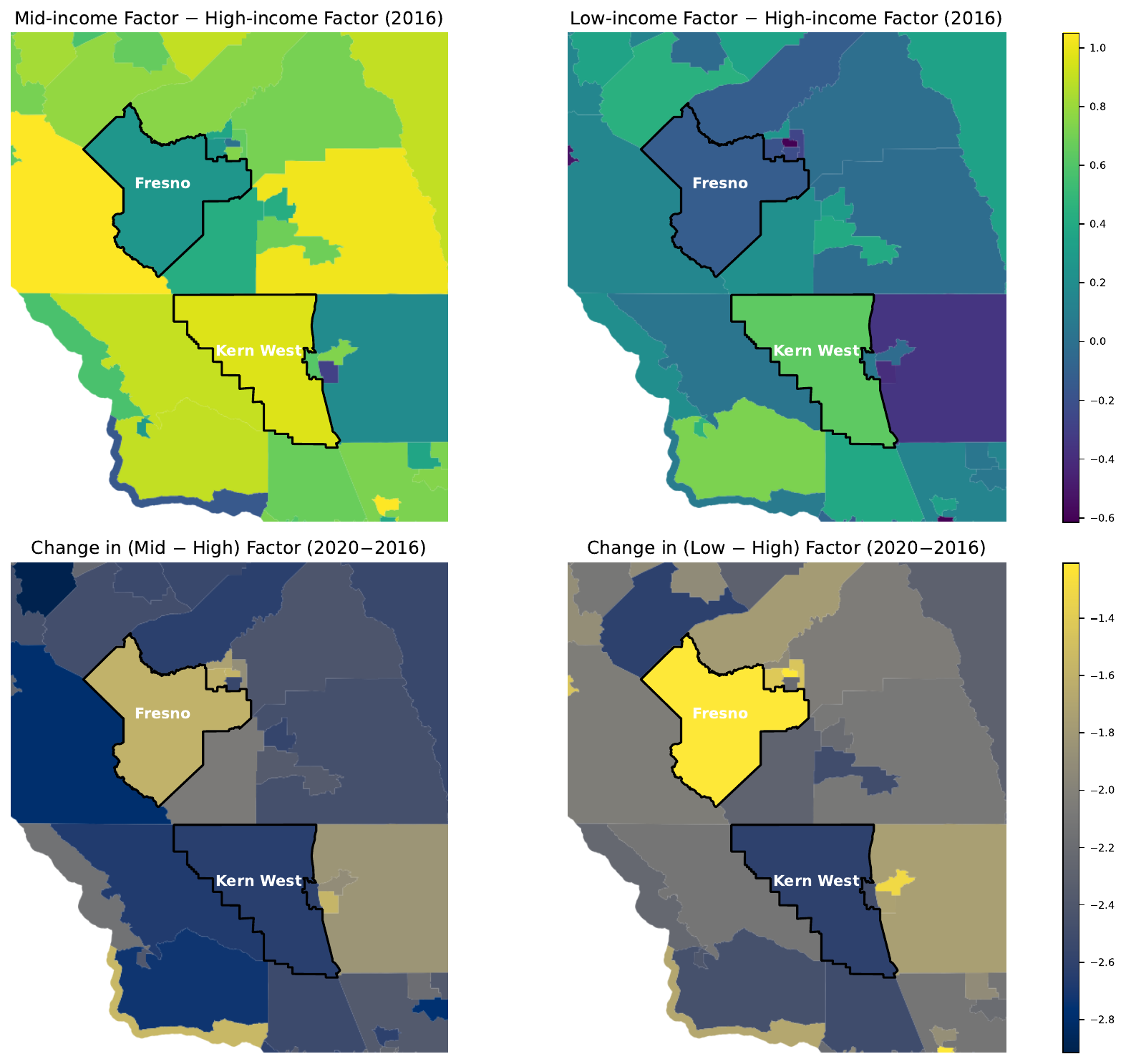} 
  \caption{California (zoom-in on Fresno and Kern West): Posterior mean PUMA-specific effects of Education (high-income factor as baseline) in 2016 (top row) and changes from 2016 to 2020 (bottom row). Strong posterior evidence in a PUMA is indicated by a red boundary.}
  \label{fig:race_2016_diff_zoomin_fresno}
\end{figure}
Figure \ref{fig:race_2016_diff_zoomin_fresno} shows the spatial race effect on the central part of California and highlights two PUMAs in Fresno and Kern West for interpretation. In Fresno, the 2016 ``Low--High'' panel shows pronounced negative values, meaning that a higher White share is associated with a greater weight on the high-income factor (relative to the low-income factor). By contrast, Fresno exhibits only a mild negative trend in the ``Mid--High'' change map (2020–2016), suggesting a limited shift toward the high-income factor relative to the mid-income factor. Kern shows the opposite effect in 2016: the 2016 ``Mid--High'' panel is bright yellow, indicating that a higher White share is associated with a greater weight on the mid-income factor relative to the high-income factor compared to other PUMAs. From 2016 to 2020, the race effect in Kern West turns negative in the change of the ``Mid-High'' Factor, indicating a shift in association toward the high-income factor relative to the mid-income factor (i.e., the high-income factor association stronger across years). Figure~\ref{fig:race_pdf_two_PUMAs} 
corroborate these patterns. In both Fresno and Kern West, the distribution for the White group shifts to the right, whereas the Non-White group hardly changes.


\begin{figure}[htbp]
  \begin{center}
  \begin{tabular}{cc}
  Fresno & Kern West \\
  \includegraphics[width=0.35\textwidth]{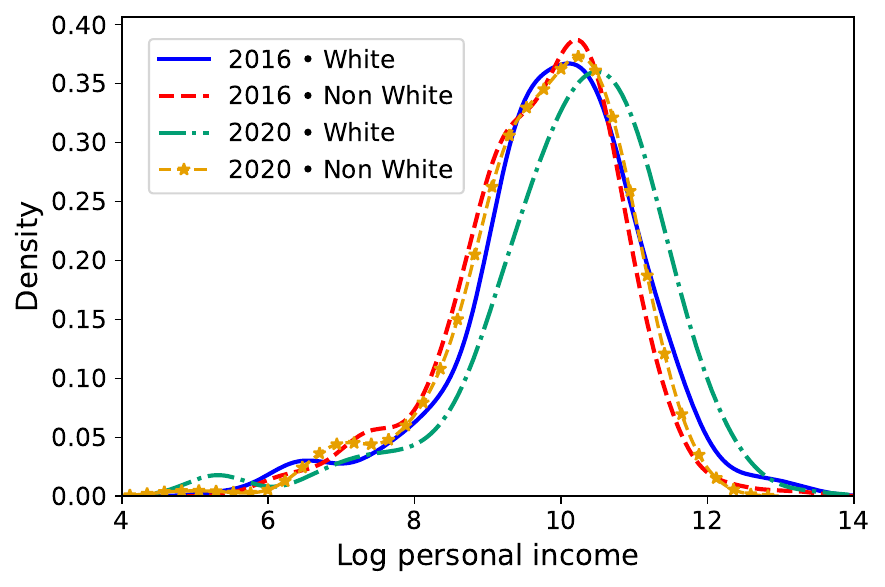}
  &
   \includegraphics[width=0.35\textwidth]{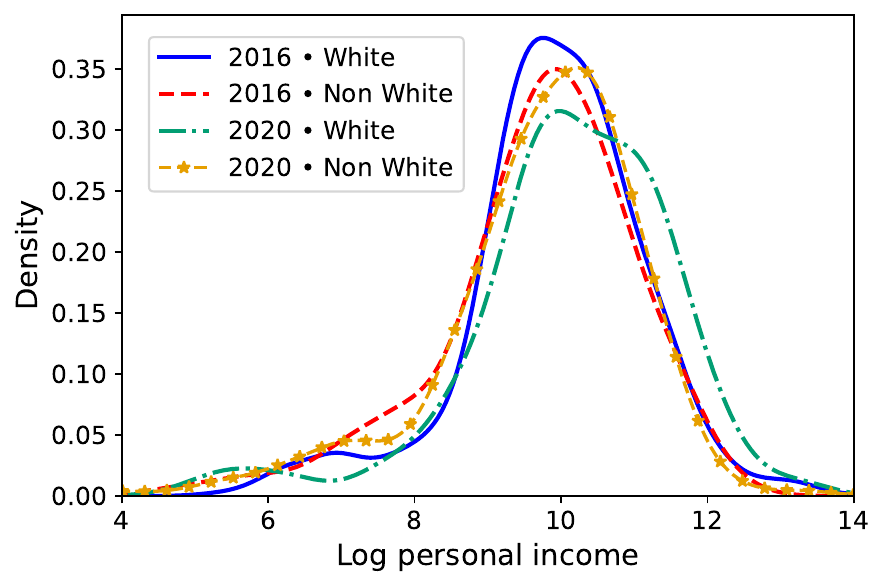}
   \end{tabular}
   \end{center}
  \caption{Fresno and Kern West, California: Posterior mean densities of log personal income for White and Non-White individuals in 2016 and 2020.}
  \label{fig:race_pdf_two_PUMAs}
\end{figure}

\begin{figure}[htbp]
    \centering
    \includegraphics[width=0.6\linewidth]{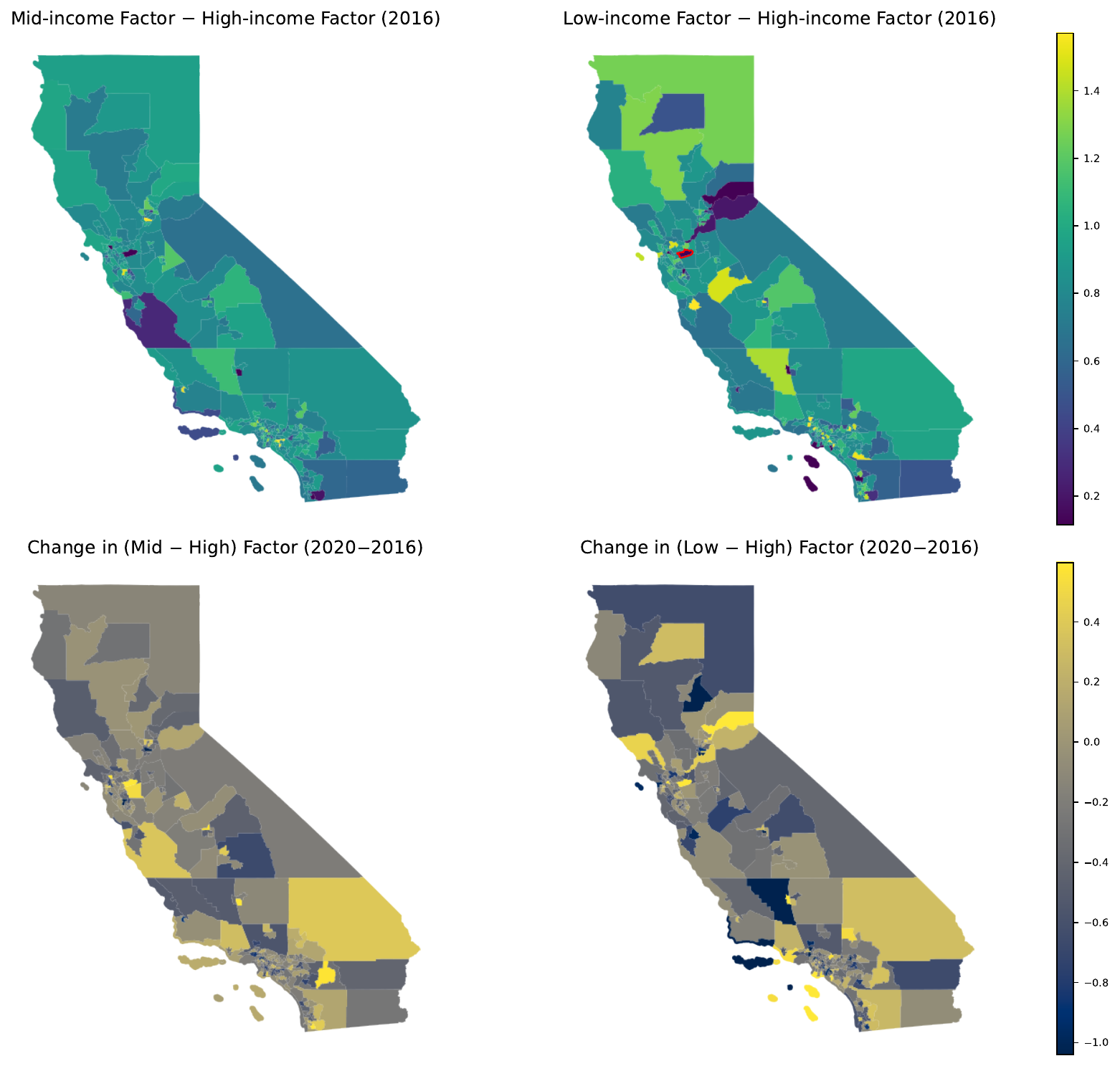}
    \caption{
    California: Posterior mean PUMA-specific effects of Gender (high-income factor as baseline)
       in 2016 (top row) and changes from 2016 to 2020 (bottom row). Strong posterior evidence in a PUMA is indicated by a red boundary.
    }
    \label{fig:gender_2016_2020}
\end{figure}

\begin{figure}[htbp]
  \centering
  \includegraphics[width=0.6\textwidth]{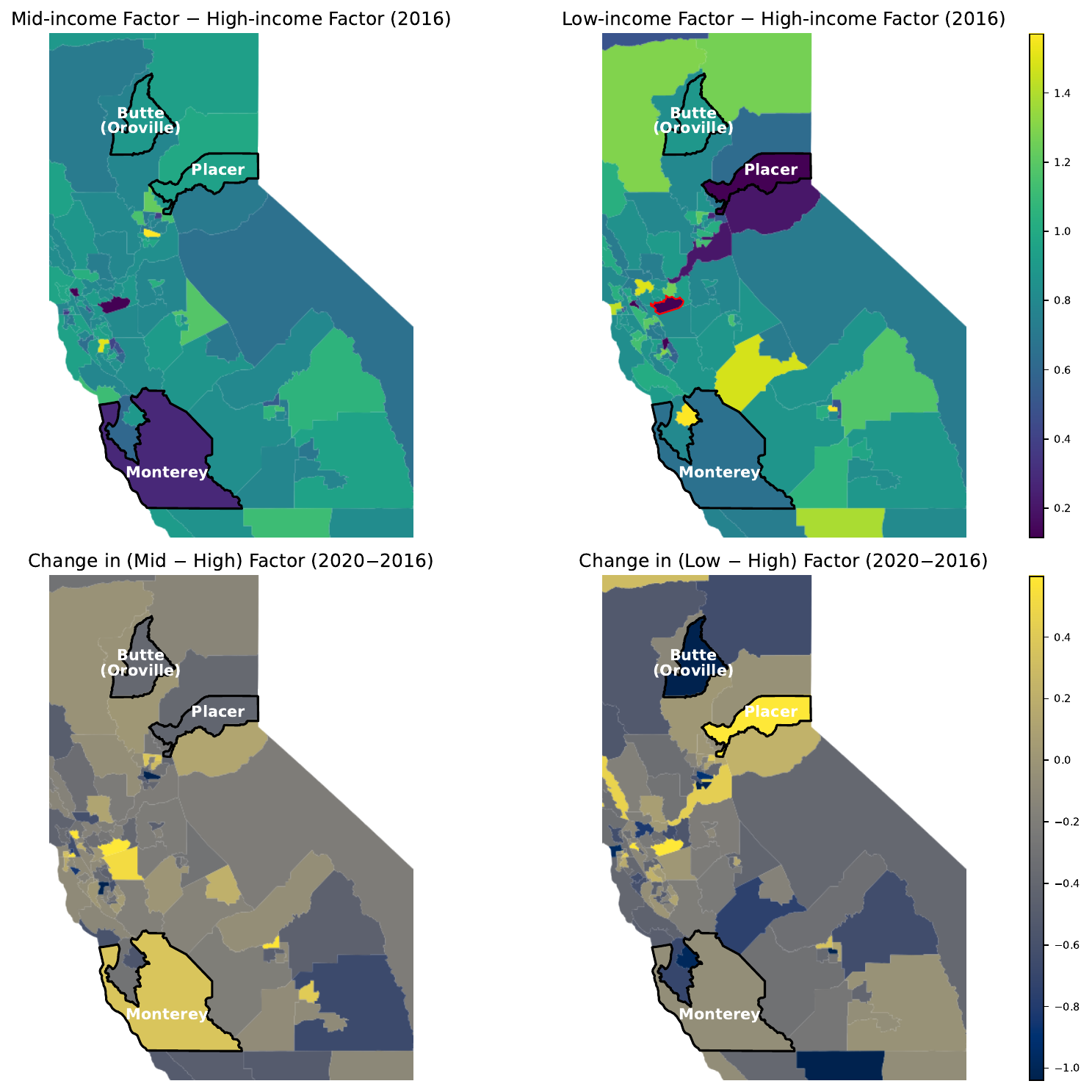}
  \caption{California (zoom-in on Butte, Placer and Monterey): Posterior mean PUMA-specific effects of Gender (high-income factor as baseline)
       in 2016 (top row) and changes from 2016 to 2020 (bottom row). Strong posterior evidence in a PUMA is indicated by a red boundary.}
  \label{fig:gender_2016_diff_zoomin}
\end{figure}

\begin{figure}[htbp]
 \centering
 \begin{tabular}{ccc}
 Butte (Oroville) & Monterey & Placer \\
  \includegraphics[width=0.31\textwidth]{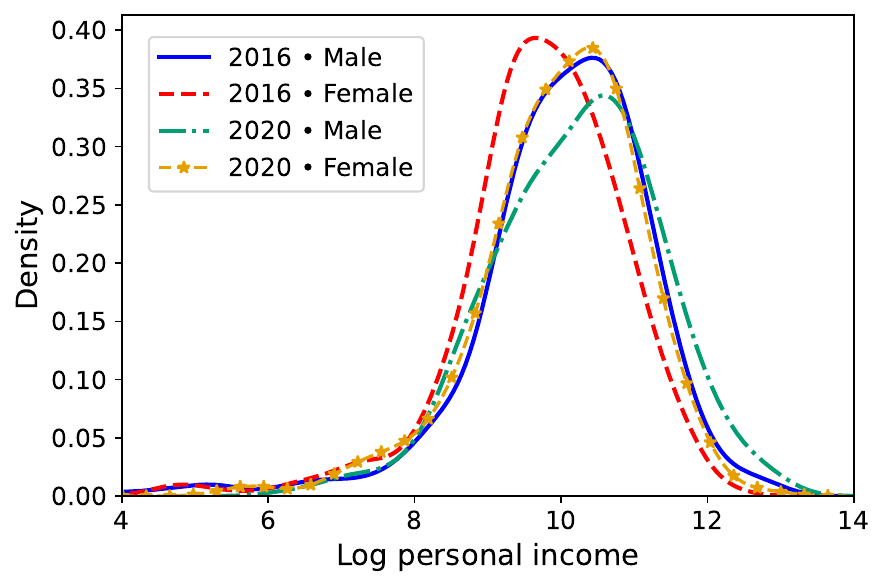} &
 \includegraphics[width=0.31\textwidth]{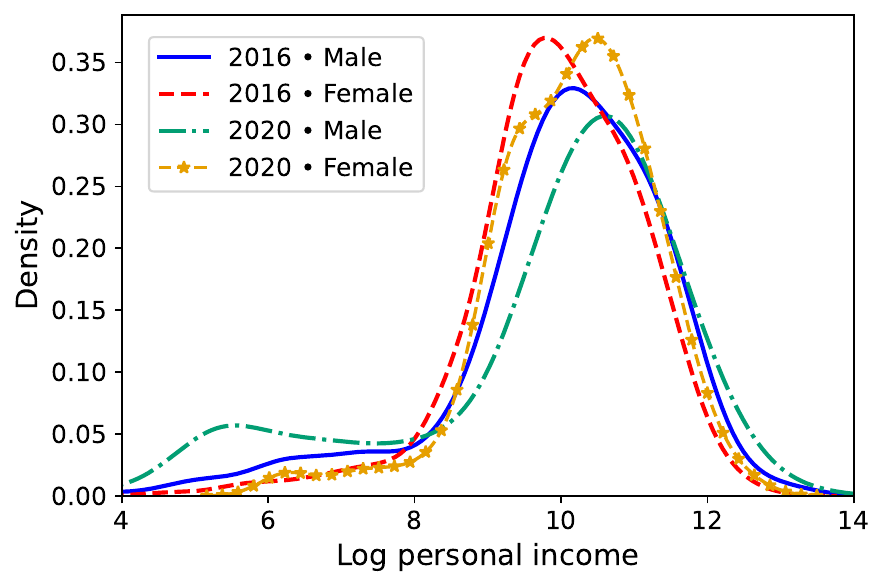} &
 \includegraphics[width=0.31\textwidth]{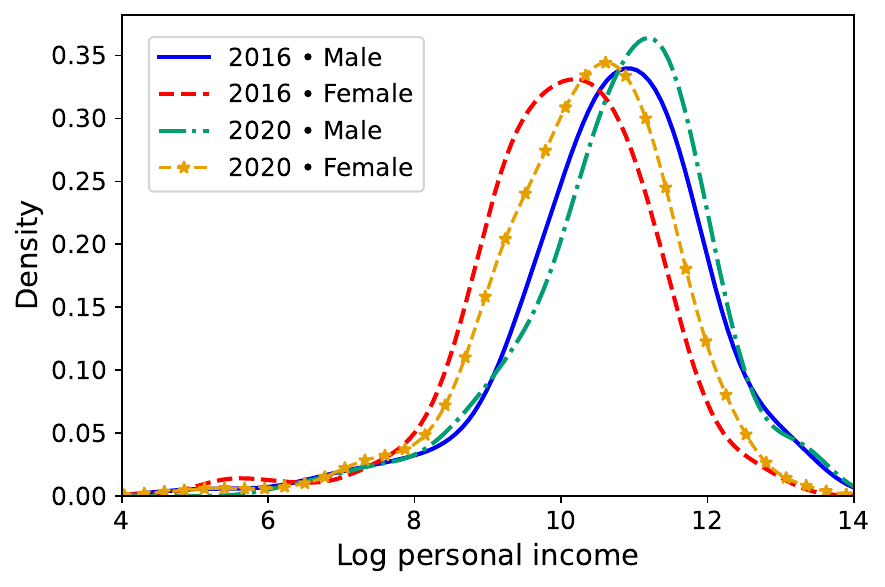}
 \end{tabular}
 \caption{Butte (Oroville), Placer and Monterey, California: Posterior mean densities of log personal income for females and males in  2016 and 2020.}
 \label{fig:gender_2016_diff_zoomin_two_PUMAs}
\end{figure}

 Figure~\ref{fig:gender_2016_2020} suggests that the gender pay difference has a largely weak and spatially uniform pattern for both contrasts. The changes map shows that these effects are largely unchanged from 2016 to 2020. Zooming in on central California (Figure~\ref{fig:gender_2016_diff_zoomin}) reveals more localized heterogeneity, with Butte (Oroville), Monterey and Placer PUMAs explicitly outlined in the map. In Monterey, the 2016 ``Mid–High'' contrast is markedly negative, implying that a larger female–male income difference is associated with greater weight on the high-income factor relative to the mid-income factor. However, Monterey is strongly positive in the ``Mid–-High'' change map (2020–2016), indicating that the (female–male) income gap in 2020 is more strongly associated with the mid-income factor relative to the high-income factor than in 2016. Similarly, in Placer the 2016 ``Low--High'' panel is deep blue, indicating that a larger (female–male) income difference is associated with a greater weight on the high-income factor relative to the low-income factor. From 2016 to 2020 Placer shows positive values in the change of ``Low--High'', indicating a shift in association toward the low-income relative to high-income (i.e., the high-income factor association weakens across years). By contrast, Butte (Oroville) is mildly positive in both the ``Mid--High'' and ``Low--High'' contrasts, suggesting that a larger female--male income gap is associated with relatively less weight on the high-income factor than in Monterey or Placer. The ``Low--High'' change map is strongly negative, indicating that in 2020 the association becomes more strongly concentrated on the high-income factor relative to the low-income factor. Figure~\ref{fig:gender_2016_diff_zoomin_two_PUMAs} presents the estimated male and female income densities for Butte (Oroville), Monterey and Placer in 2016 and 2020, providing distributional evidence consistent with the patterns discussed above.

\subsection{Comparison of latent factors across states via rotation}\label{sec:across states} We are interested in comparing latent factors and their main effects $\zeta_{h}$ across states to reveal regional commonalities and differences. We find that latent factors are similar across states; see Figure 22 in Appendix C, which motivates a simpler model with latent factors shared across states. Fitting the NLMFM-C across multiple states and years is computationally demanding and so we consider aligning results from separately fitted NLMFM-C models to a common reference state using a rotation-based matching and rescaling approach.

\subsubsection{Rotation-based alignment via Hungarian matching and diagonal Procrustes}
With California as the reference state we fit the NLMFM-C model to the other three states using the number of latent factor densities inferred by the California NLMFM-C fit. We then use a two-step alignment method to match and rescale each states' factor densities to those of California's keeping latent factors and their weights $s_{j,h}$ non-negative.  
Recall, that $\mathbf{r}_h = \left(r_h(x_1), \dots, r_h(x_n)\right)$ is the vector holding values of the $h$-th residual factor density evaluated on a grid $x_1, \dots, x_n$ (see Section~\ref{sec:adaptive gibbs}). Let $\hat{\mathbf{r}}^S_h$ represent the posterior mean of $\mathbf{r}_h$ for state $S$ then, 
 $\hat{\mathbf R}^{S}=\big(\hat{\mathbf r}_1^{S},\dots,\hat{\mathbf r}_H^{S}\big)$ is the stacked vector of factor residual densities for state $S$. 

We first match the factors of each state $S$ to the reference state $A$ (California in our application) using the Hungarian algorithm \citep{kuhn1955hungarian}. We define a cost matrix $C$ as $$C_{i,j} = 1-\cos\big(\hat{\mathbf r}^{A}_i,\hat{\mathbf r}^{S}_j\big)$$ where $\cos\big(\hat{\mathbf r}^{A}_i,\hat{\mathbf r}^{S}_j\big)$ is the cosine similarity is defined in Eq.~\eqref{eq:cosine}. This similarity is invariant to overall scale, and small values indicate similar residual density shapes between pairs of factors. The Hungarian algorithm utilizes the cost matrix $C_{i,j}$ to yield a permutation matrix $P$ that minimizes the total cost and therefore maximizes the similarity.

We then use a simplified version of Procrustes alignment \citep{schonemann1966generalized}, restricted to non-negative diagonal rescaling, to resolve scale indeterminacy. 
Let $(P\hat{\mathbf{R}}^S)_h$ refer to the $h$-th column of $\hat{\mathbf{R}}^S$ after permutation by $P$ and $d_h$ refer to the non-negative scaling coefficient estimated as\[
d_h = \frac{\langle \hat{\mathbf r}^A_h, (P \hat{\mathbf R}^S)_h \rangle}{\|(P \hat{\mathbf R}^S)_h\|_2^2}, 
\]
such that $d_h$ is the rescaling that best matches the factors of 
state $S$ and reference state $A$. We collect all the $d_h$'s into a diagonal matrix $D = \mathrm{diag}(d_1,\dots,d_H),$ and define the full alignment operator $O = DP$. We collect the factor loadings of state $S$ into a $g\times H$ matrix $\Lambda^{S}=(\boldsymbol\lambda^{S}_{1}, \dots, \boldsymbol\lambda^S_H)$. The aligned factors and weights for state $S$ are then, 
\[
\hat{R}^{S}_{\text{rotated}} = O\,\hat{\boldsymbol{R}}^{S}, 
\qquad 
\Lambda^{S}_{\text{rotated}} = \Lambda^{S} O^{-1}.
\]
We obtain a posterior sample of the main factor effects $\boldsymbol{\zeta}_h$ and the spatial interaction $\boldsymbol{\gamma}_{h,m}$ by sampling the rotated factor loadings are from their full conditional distributions conditioning on the posterior means of the rotated $\lambda_{j,h}$, rotated $m_{h,k}$, $J_k$, and $\theta_k$. 

\subsubsection{Comparing California, New York, Florida and Washington}
\begin{figure}[htbp]
    \centering
    \includegraphics[width=\textwidth]{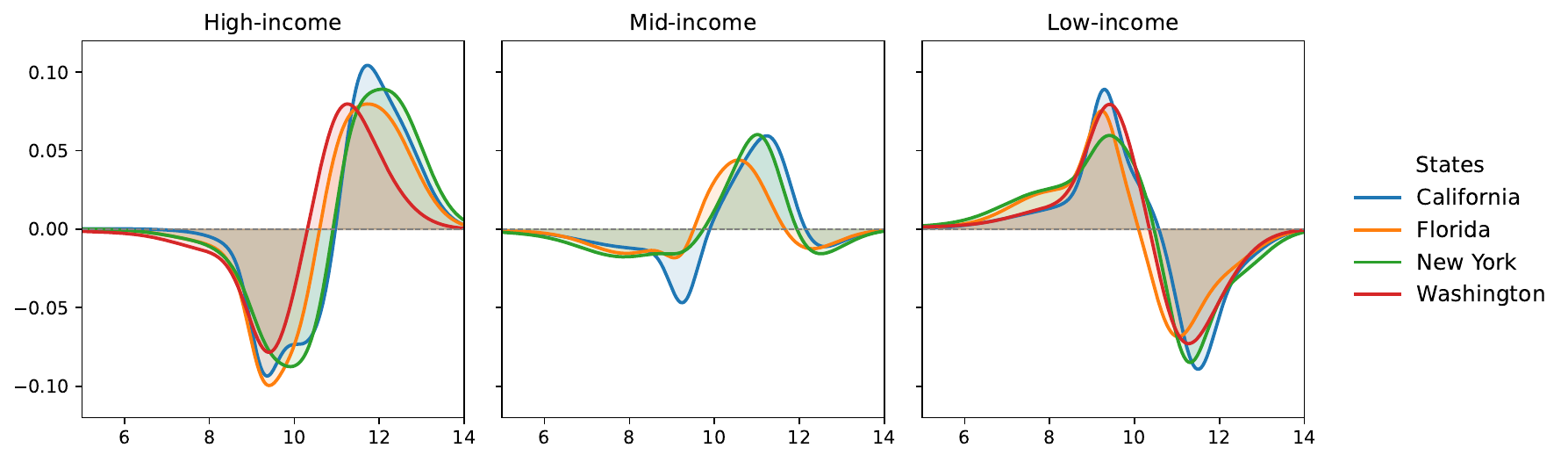}
    \vspace{-0.3cm}
    \caption{Residual latent factor densities (left to right: high, mid and low income factor) of California, and after rotation for Florida, New York and Washington (to align with California).}
    \label{fig:residual_factor_all_rotated}
\end{figure}

With California as the reference state we use our rotation method to compare the main factor effects $\boldsymbol{\zeta}_h$, and the area specific covariate effects $\boldsymbol{\eta}_{h,m}$ of Florida, New York, and Washington. The NLMFM-C identifies three latent factors (high, mid and low income) for California, Florida, and New York whereas for Washington it identifies two factors (high and low income). We align the three factors of Florida, New York, and the two factors of Washington to those of California. The residual factor densities after rotation are displayed in Figure~\ref{fig:residual_factor_all_rotated}. The shapes are similar, an indication that rotation improves factor alignment across states. This is confirmed by the Root Mean Square Error between the residual factor densities before and after rotation displayed in Table~\ref{tab:rmse_states_rotation} in Appendix \ref{sec:app cross state}. 

Table~\ref{table:globaleffects_states} reports the posterior means and 95\% CIs of the contrasts (after rotation with the high income factor as baseline) of the main factor effects $\boldsymbol{\zeta}_h$ for each covariate in each of the four states for 2016 and 2020. Positive values indicate a larger weight for the mid and low income factors whereas negative values indicate a larger weight for the positive factor. 
\begin{table}[htbp] 
\centering 
\caption{Posterior means and 95\% CIs of contrasts of main factor effects $\boldsymbol{\zeta_h}$ (after rotation) across the four states for 2016 and 2020. High-income factor is the baseline.} 
\begin{tabular}{|l|l|l|c|c|c|} 
\hline
\textbf{state} & \textbf{Factor} & \textbf{Year} & \textbf{Education} & \textbf{Race} & \textbf{Gender} \\ 
\hline
\multirow{4}{*}{California} 
& \multirow{2}{*}{Low} & 2016  & -5.8 [-6.4, -5.2]  & 0.1 [-0.4, 0.6] & 0.9 [0.4, 1.3] \\ 
& & 2020 & -5.0 [-5.6, -4.3]   & -2.0 [-2.5, -1.5] & 0.6 [0.2, 1.1]  \\ 
& \multirow{2}{*}{Mid} & 2016  & -2.5 [-3.1, -1.8] & 0.7 [0.1, 1.3] & 0.8 [0.3, 1.3]  \\ 
& & 2020  & -2.0 [-2.6, -1.3]  & -1.7 [-2.3, -1.1] & 0.7 [0.1, 1.2]  \\  \hline 
\multirow{4}{*}{\shortstack[l]{Florida\\(Re-estimated)}} 
& \multirow{2}{*}{Low} & 2016  & -4.0 [-4.3, -3.6]  & -0.3 [-0.9, 0.3] & 0.4 [0.0, 0.8]  \\ 
& & 2020 & -3.8 [-4.1, -3.4] & -1.6 [-2.1, -1.1] & 0.4 [0.0, 0.8]  \\ 
& \multirow{2}{*}{Mid} & 2016  & -2.5 [-2.9, -2.1]  & 0.1 [-0.4, 0.7] & 0.4 [-0.0, 0.9] \\ 
& & 2020  & -2.3 [-2.7, -1.9]  & -0.7 [-1.2, -0.1]  & 0.9 [0.4, 1.2] \\  \hline

\multirow{4}{*}{\shortstack[l]{New York\\(Re-estimated)}}
& \multirow{2}{*}{Low} & 2016  & -6.6 [-7.0, -6.2] &  -2.2 [-2.6, -1.7] & 0.7 [0.2, 1.0] \\ 
& & 2020  & -6.2 [-6.6, -5.7]  & -2.4 [-2.9, -2.0] & 0.7 [0.3, 1.2]  \\ 
& \multirow{2}{*}{Mid} & 2016  & -3.6 [-4.0, -3.3] &  0.5 [-0.1, 1.0] & 0.0 [-0.4, 0.4]\\ 
& & 2020 & -3.3 [-3.7, -2.8] &  0.0 [-0.5, 0.5] & 0.7 [0.3, 1.2] \\ 
\hline 

\multirow{2}{*}{\shortstack[l]{Washington\\(Re-estimated)}} 
& \multirow{2}{*}{Low} & 2016 & -0.3 [-1.0, 0.4]  & -0.2 [-0.8, 0.4] & 0.2 [-0.4, 0.7]  \\ 
& & 2020 &  -0.4 [-1.0, 0.3] & -0.6 [-1.2, 0.1] & 0.2 [-0.3, 0.7]  \\ \hline 
\end{tabular} 
\label{table:globaleffects_states} 
\end{table}
  Education-related effects are negative, with 95\% CI's excluding zero, for both low and mid income contrasts across the four states for both years. This pattern suggests that higher levels of educational attainment are closely aligned with the high income factor than with the mid or low income factors across states. From 2016 to 2020, these negative contrasts become less pronounced, suggesting a weakening relationship between the level of educational attainment and the high income factor across years. This is consistent with the flattening of the college wage premium since the mid-2010s, 
and the supply of college-educated labor outpacing demand,
\citep{frbsf_collegepremium_2023,clevfed_demand_supply_2025}. New York exhibits the largest separation in the contrasts, followed by California, Florida, and Washington, which reflect New York’s high concentration of credential-intensive, top-paying sectors (finance and professional services) strengthening the link between educational attainment level and the upper tail of the income distribution \citep{nyba_finance_impact_2023}. For 2016 race-related effects are not as important as education effects for California and Florida since the 95\% CIs for both the low and the mid income contrasts include zero. There is a marked change for these states in 2020 with high negative contracts, suggesting more weight is placed on the high income factor, with their 95\% CIs excluding zero. This is consistent with \citet{chetty_covid_inequality_2020} and \citet{mongey_pilossoph_weinberg_2021}, who found that the COVID-19 shock disproportionately affected racial minority workers, who are overrepresented in lower-wage frontline and hospitality jobs, contributing to larger income losses. The expansion of remote work during that period insulated higher-income workers who are disproportionally white. The effects of race are not important and small in magnitude for both years for Washington, whereas for New York we have similar in magnitude negative contrasts for both years for the low income factor. This suggests that race places more weight on the high income factor which is consistent with the disproportionally white make up of finance, fintech, and other professional service jobs driving the third largest state economy in the U.S. 
Gender related effect are small in magnitude across the four states and the two years. For Florida, New York and Washington there is no change between years for the low income factor. For the mid income factor, for Florida and New York, the changes from 2016 to 2020 are important (95\% CIs exclude zero) but relatively small magnitude. This implies that by 2020 the gender pay gap became associated with middle income jobs, such us education, health care support and other middle management and administration roles. This is consistent with \citep{albanesi_kim_2021,bls_covid_gender_2022} who suggest that the labor market adjustment during the covid-pandemic and shortly after led to more women retaining and gaining employment in middle income jobs whereas top paying jobs in financial, engineering and high tech sectors remained male dominated.

\section{Discussion}\label{sec:discussion}
In this paper we introduce the NLMFM-C model, a flexible and interpretable Bayesian nonparametric framework for analyzing related distributions allowing for spatially-varying and longitudinal-change covariate effects. The NLMFM-C expresses the distributions as weighted sums of latent factor densities which can be interpreted as distributional sources of variation. We propose an adaptive Gibbs sampling scheme which ensures parsimony in the number of latent factors, avoiding redundancy and overfitting. To facilitated analyses across different data sets we
suggest a factor alignment post-processing approach based on the Hungarian algorithm and Procrustes alignment.
This approach provides a practical alternative to full joint estimation, which would be computationally infeasible on large-scale datasets.

We apply the NLMFM-C to the ACS PUMS data across four states: California, Florida, New York and Washington. The model identifies variation in factor composition and covariate influence, both within and across states, for personal income distributions. 
 In particular, we find that educational attainment has a strong influence on the income distribution across all states and years, but the magnitude of weight placed on the high-income factor is not the same across states. This highlights a fundamental aspect of regional economics where high educational attainment does not guarantee uniform economic prosperity across different geographical areas. Additionally, race-related effects are more pronounced in California for 2020 than 2016 and like educational attainment more weight is placed on the high income factor. This could be attributed to the disproportional shock of the Covid-19 pandemic. Racial minority workers faced bigger income losses because they make up a larger proportion of lower-wage frontline and hospitality jobs, whereas the expansion of remote work during this period insulated high income earners who are disproportionally white. Gender effects, while subtler, show temporal shifts across the states between 2016 and 2020. These findings highlight the importance of modeling spatial heterogeneity and covariate interactions when studying income inequality.

Nevertheless, limitations remain. The NLMFM-C is difficult to apply to a lot (or all) of US states and PUMAS at the same time. This is due to the way we model the factor loadings via a log-Gaussian Markov random field. Even though we have considered the changes in main factor effects between two years, we do not have a spatiotemporal model for the factor loadings. We took a simpler approach and fitted the NLMFM-C to each of our chosen states and relied on an alignment approach that assumes communalities in factor densities, which works only when the densities are well aligned.


Future extensions of the NLMFM-C framework could consider spatiotemporal models for the factor loadings and  more efficient computational methods
such as stochastic variational inference or other approximation methods could make the model more tractable at larger spatial and temporal scales, especially when incorporating additional covariates or years. This will allow the model to be applied across several years and states without the need for post-hoc alignment. 


\bibliography{bibliography.bib} 

\newpage
\appendix
\section{Simulation example and California area specific covariate effects}\label{sec:appendix simulation}
\subsection{Convergence check for simulated data}
\begin{figure}[htbp]
    \centering
    \begin{subfigure}[b]{0.7\linewidth}
        \centering
        \includegraphics[width=\linewidth]{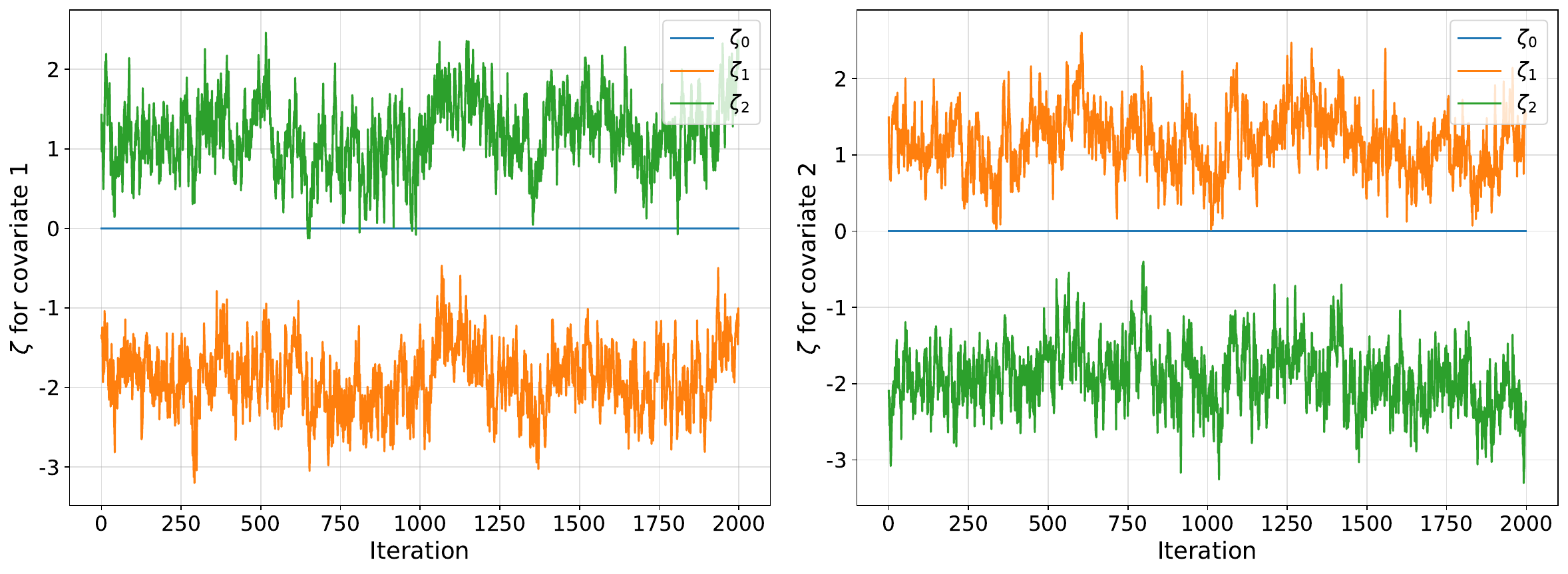}
        \label{fig:simulation_gamma_trace_plot}
    \end{subfigure}
    \hfill
    \begin{subfigure}[b]{0.7\linewidth}
        \centering
        \includegraphics[width=\linewidth]{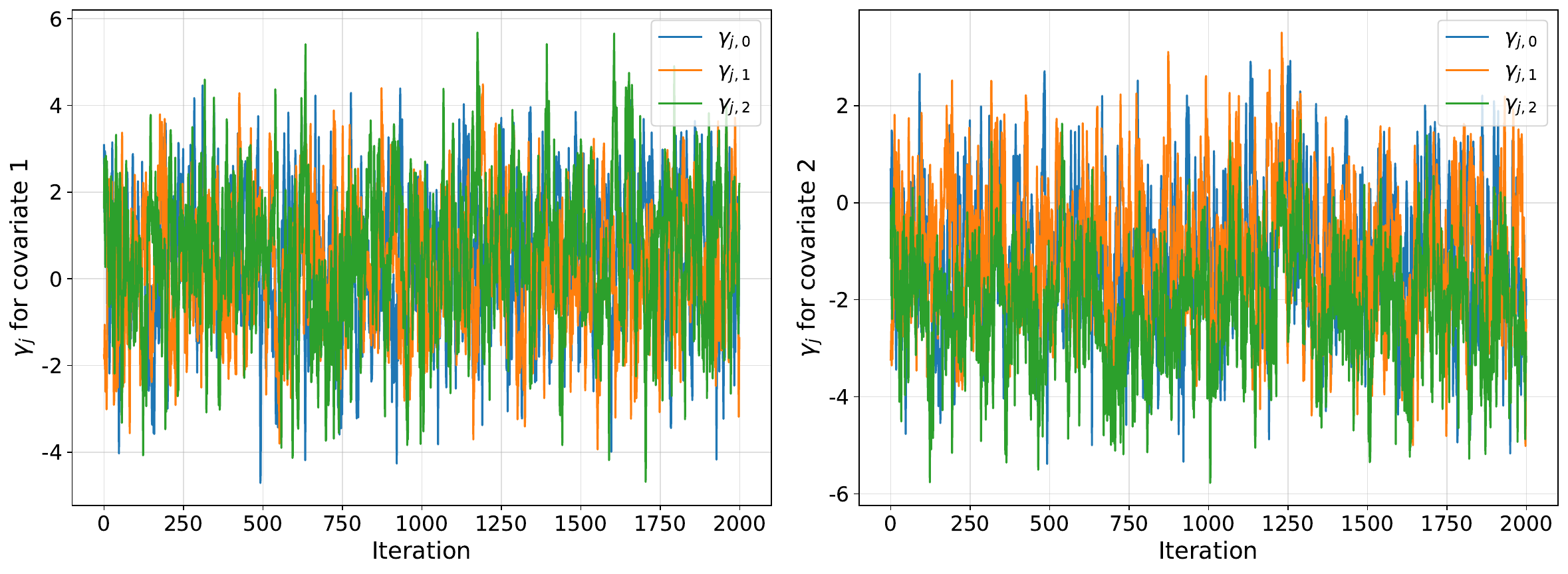}
        \label{fig:simulation_gammaj_trace_plot}
    \end{subfigure}
    \vspace{-0.12in}
    
    \caption{Trace plots for the main factor effects $\zeta_{h}$ and spatial interaction effects $\gamma_{h,m}$.}
    \label{fig:simulation_trace_plots}
\end{figure}

To evaluate the convergence of our adaptive Gibbs sampler and the recovery of the main factor effects $\zeta_h$ and the spatial interaction of the covariates $\gamma_{h,m}$, we produce trace plots. A stable, noisy horizontal band without visible trends or drifts suggests good mixing and convergence. Figure~\ref{fig:simulation_trace_plots} shows the trace plots of the main factor effects $\zeta_{h}$ and the spatial interaction effects $\gamma_{h,m}$ for the eighth group (randomly selected). Both figures display good mixing.
 \vspace{-0.12in}



\subsection{Educational attainment effect for California metro areas}
\begin{figure}
    \centering
    \includegraphics[width=\linewidth]{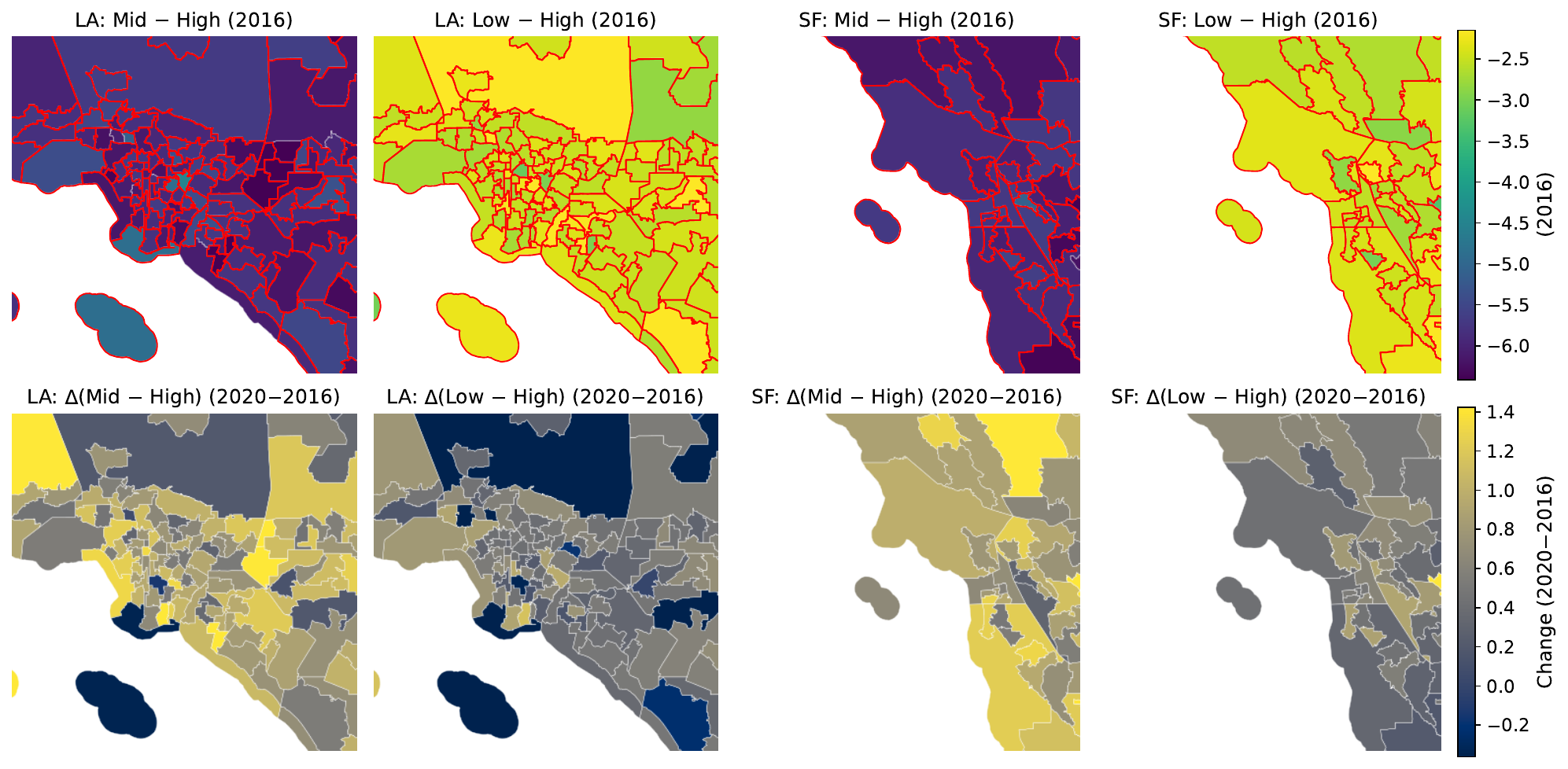}
    \caption{Posterior mean PUMA-specific effects of Education for Los Angeles and San Francisco in 2016 (top row) and their changes between 2016 and 2020 (bottom row). High income factor is the baseline. Red boundaries indicate the 95\% CIs exclude zero.}
    \label{fig:schl_2016_2020_lasf}
\end{figure}

Figure \ref{fig:schl_2016_2020_lasf} presents zoom-in views for PUMAs in Los Angeles and San Francisco metros, where pronounced year-to-year variation was observed. In both counties, the 2016 maps show that all PUMAs exhibit negative contrasts for the mid and low income factors where the high income factor is the baseline. This indicates that high educational attainment leads to higher weight on the high-income factor. Based on \citep{USCensusS0501_2016,USCERI_SOILA} this pattern suggests that, despite of migrant and low-income populations, educational attainment is a consistent gateway to higher-paying opportunities, particularly in professional and knowledge-intensive sectors that dominate both the Los Angeles and San Francisco labor markets. Between 2016 and 2020, the changes are modest and geographically scattered, with no systematic reversal, consistent with evidence that the returns to education in LA have remained stable over the period. 
From 2016 to 2020, the negative contrasts become slightly less pronounced in some southern and peripheral areas, suggesting a modest weakening in the association between educational attainment and access to high-paying opportunities in these areas.

\newpage
\section{Cross-state comparison}\label{sec:app cross state}
This section examines changes of factor weights and the effect of covariates on area specific latent densities for Florida, New York and Washington between 2016 and 2020. These results are based on the rotation and alignment method described in Section \ref{sec:across states}. Table~\ref{tab:rmse_states_rotation} displays the root mean squared errors (RMSE) of the factors of Florida, New York and Washington before and after rotation with California used as the reference state. Rotation improves cross-state factor alignment for all factors and states. The major improvements (above 80\%) can be seen in New York's mid and low income factors and Washington's low income factor. The relatively small improvements for Florida and for New York's high income factor, suggest that these are already close to those of our reference state, California.

\begin{table}[htbp]\centering
\caption{RMSE before and after rotation for Florida, New York and Washington factors}
\begin{tabular}{ccccc}
\hline
\textbf{State} & \textbf{Factor} & \textbf{Before rotation} & \textbf{After rotation} & \textbf{Improvement (\%)} \\
\hline
Florida & High-income & 0.0187 & 0.0175 & 6.23\% \\
        & Mid-income  & 0.0180 & 0.0167 & 7.52\% \\
        & Low-income  & 0.0177 & 0.0177 & 0.22\% \\
\hline
New York & High-income & 0.0090 & 0.0089 & 0.47\% \\
         & Mid-income  & 0.0604 & 0.0101 & 83.33\% \\
         & Low-income  & 0.0597 & 0.0109 & 81.69\% \\
\hline
Washington & High-income & 0.0701 & 0.0330 & 52.92\% \\
         & Low-income  & 0.0690 & 0.0097 & 86.00\% \\
\hline
\end{tabular}
\label{tab:rmse_states_rotation}
\end{table}

\subsection{Factor weight changes, $\Delta s_{j,h}$, in Florida, New York, and Washington}\label{sec: app sjh changes across states}
\begin{figure}[htbp]
    \centering
    \includegraphics[width=\textwidth]{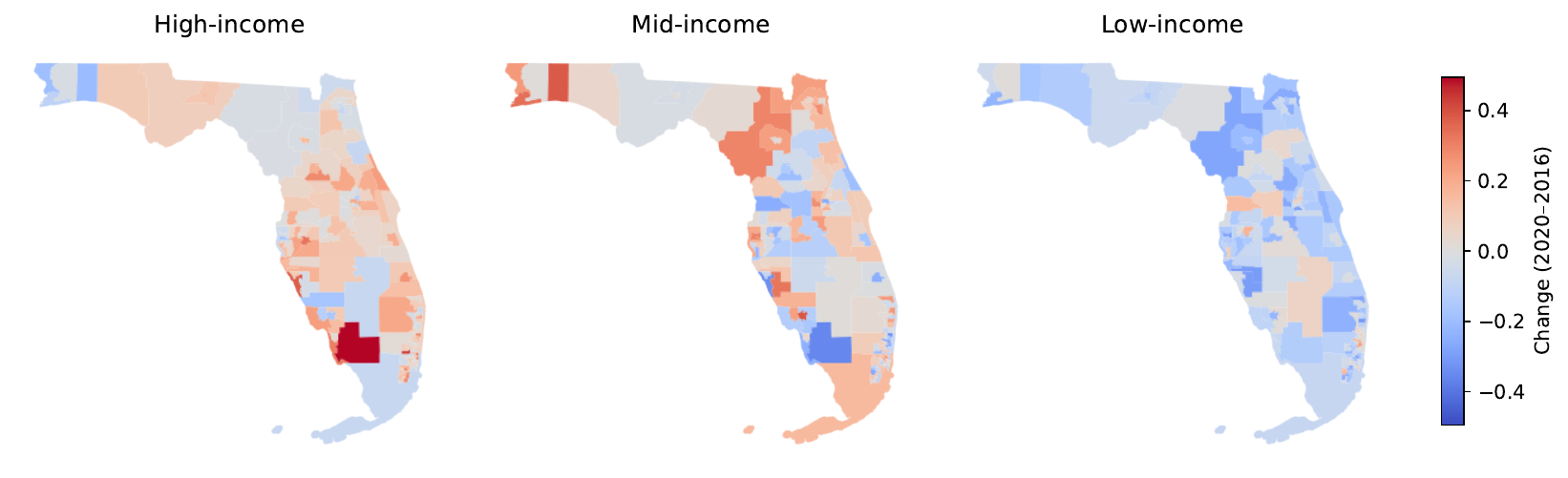}
    \vspace{-0.3cm}
    \caption{Florida: Heat maps of the changes in the posterior mean factor weights, $\Delta s_{j,h}$, between 2016 and 2020.} 
    \label{fig:florida_lam_comparison}
\end{figure}

Figures \ref{fig:florida_lam_comparison}, \ref{fig:ny_lam_comparison} and \ref{fig:washington_lam_comparison} display the spatial distributions of the changes of factor weights $s_{j,h}$ between 2016 and 2020 for Florida, New York, and Washington. We provide zoom-ins of PUMAs with high positive (red color) and negative (blue color) factor weight changes, $\Delta s_{j,h},$ for an in depth analysis for each state.

Figure \ref{fig:florida_lam_comparison} displays the changes in factor weights between 2016 and 2020 for Florida. From the heat map we can see that the highest changes occur in the high and mid income factor for Collier East and Saratosa East PUMAs. 
\vspace{-0.2in}

\begin{figure}[htbp]
    \centering
    \begin{tabular}{ccc}
        \multicolumn{3}{c}{Collier (East)} \\
        \multicolumn{2}{c}{\includegraphics[width=0.50\textwidth]{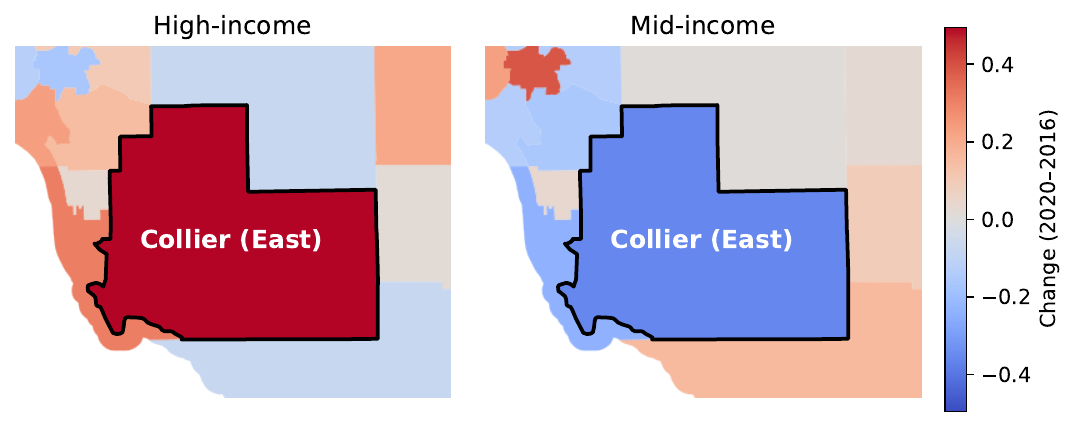}} &
        \includegraphics[width=0.3\textwidth]{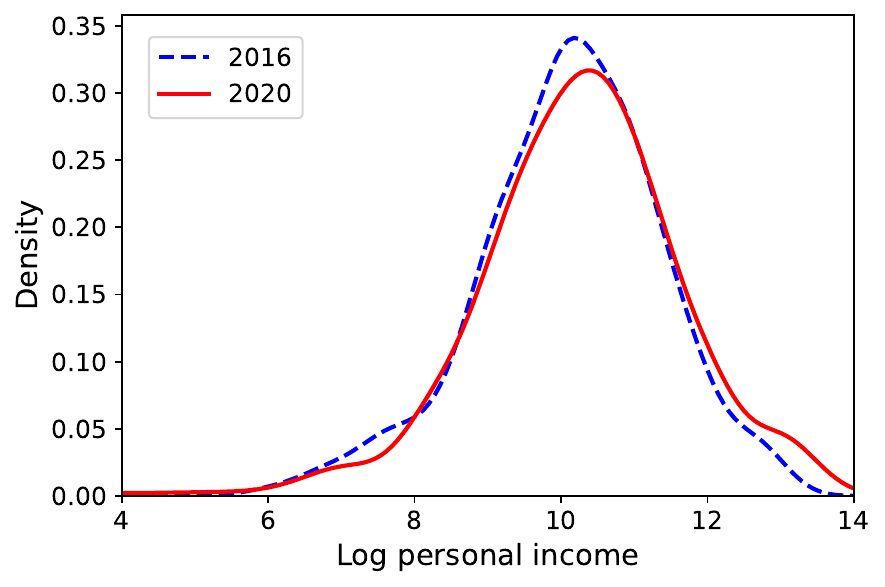} \\
        \multicolumn{3}{c}{Sarasota (East)}\\
        \multicolumn{2}{c}{\includegraphics[width=0.50\textwidth]{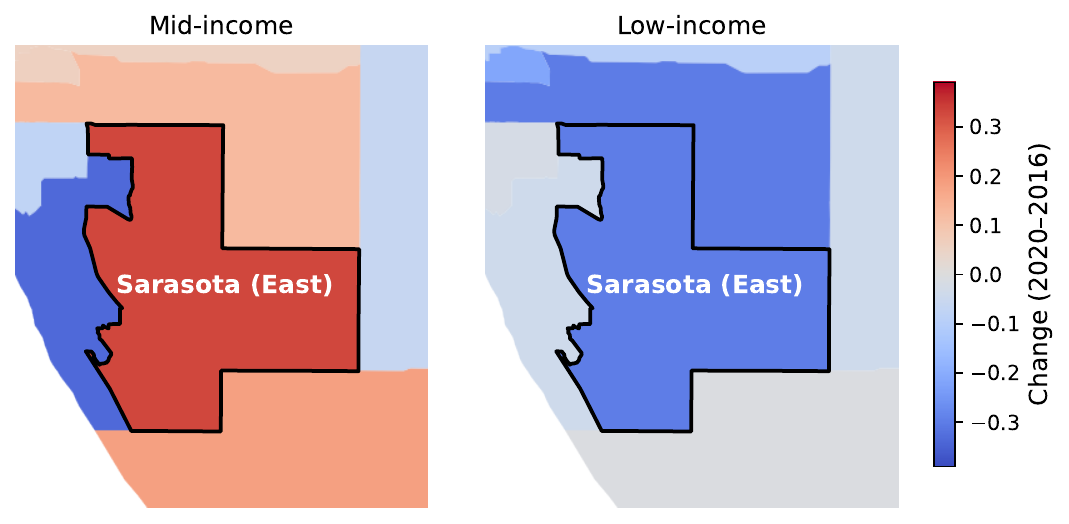}} &
        \includegraphics[width=0.3\textwidth]{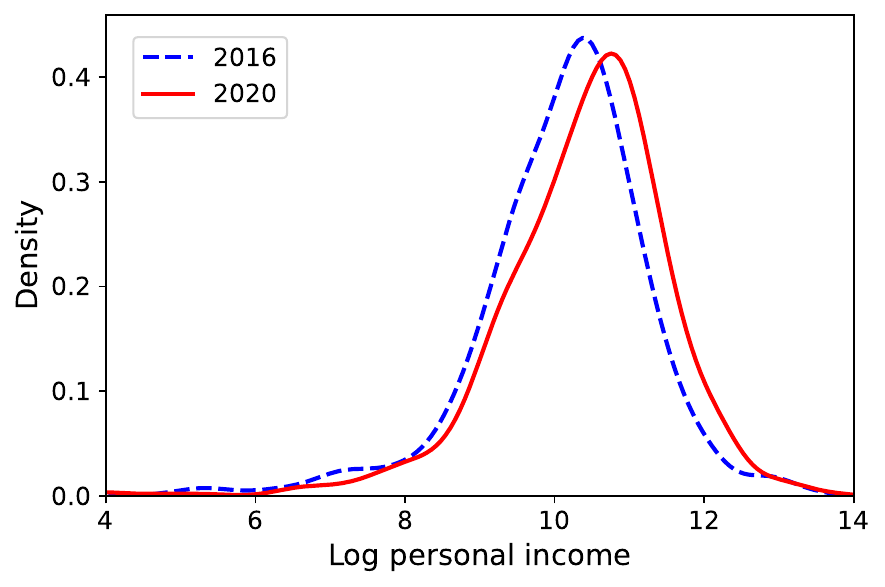}
    \end{tabular}
    \caption{Florida: Results for PUMAs with the biggest change, $\Delta s_{j,h},$ between 2016 and 2020 (Collier (East) and Sarasota (East)). Left-hand graphs: heat map zoom ins. Right-hand graph: posterior mean density of log income.}
    \label{fig:florida_collier_sarasota_combined}
\end{figure}
\vspace{-0.2in}

Figure \ref{fig:florida_collier_sarasota_combined} provides the zoom-in to highlight these changes in factor weights for the Collier East (top row) and Sarasota East PUMAs (bottom row) together with the corresponding posterior mean densities of log personal income in 2016 and 2020. 
For Collier East, we have an increase in weight for the high income factor and a decrease in weight for the mid income factor. This suggests a shift in weight toward the upper end of the income distribution over time. For Sarasota East we have an increase in weight for the mid income factor and a decrease in weight for the low income factor. This indicates a movement away from the lower-income range toward the middle of the distribution. The density plots on the right are consistent with these spatial patterns: in Collier East the 2020 density is slightly shifted toward higher income values relative to 2016, whereas in Sarasota East the 2020 density shows reduced mass in the lower-income range and greater concentration around middle-income values. These results are in line with the sharp increase of the income inequality ratio \footnote{the ratio of the mean income for the highest quintile (top 20\%) of earners divided by the mean income of the lowest quintile (bottom 20\%) of earners} in both Collier East and Saratosa East from 2018 to 2020 produced by \cite{collier}. It is worth mentioning that Collier and Saratosa counties are among the wealthiest in Florida with pronounced income polarisation and a Gini coefficient close to 0.55. \footnote{measures the inequality among the values of a frequency distribution, such as income levels. Values near 0 reflect perfect equality, whereas values near 1 reflect maximal inequality}



Figure \ref{fig:ny_lam_comparison} displays the changes in factor weights between 2016 and 2020 for New York state. From the heat maps we can see that the highest changes occur in the mid and low income factors for Columbia and Green and New York City-Bronx PUMAs. 

\begin{figure}[htbp]
    \centering
    \includegraphics[width=\textwidth]{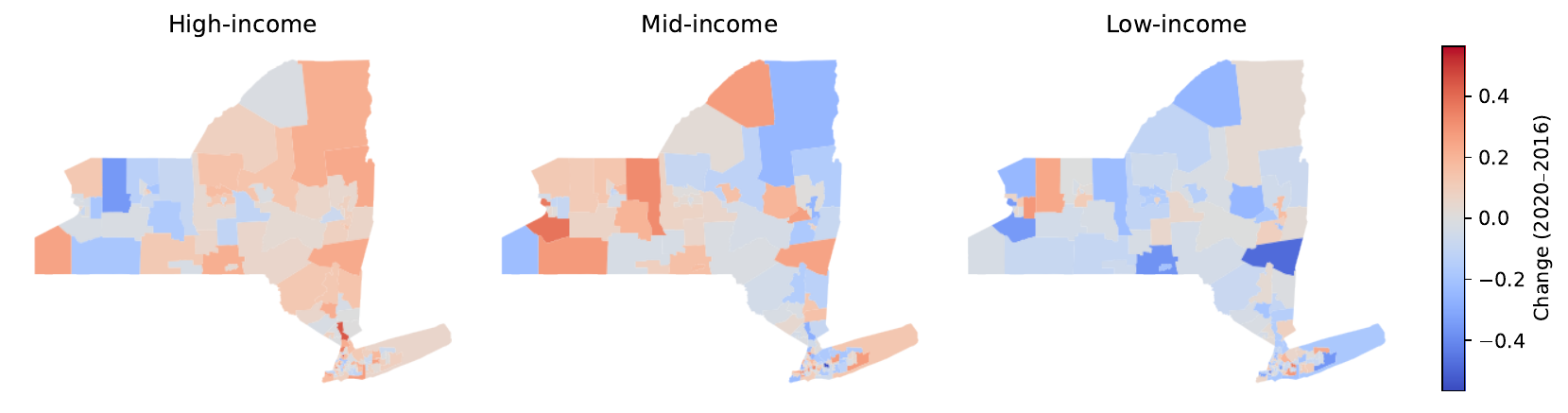}
    \vspace{-0.3cm}
    \caption{New York: Heat maps of the changes in posterior mean factor weights, $\Delta s_{j,h}$, between 2016 and 2020.} 
    \label{fig:ny_lam_comparison}
\end{figure}
\vspace{-0.2in}

\begin{figure}[htbp]
    \centering
    \begin{tabular}{ccc}
        \multicolumn{3}{c}{Columbia \& Greene} \\
        \multicolumn{2}{c}{\includegraphics[width=0.50\textwidth]{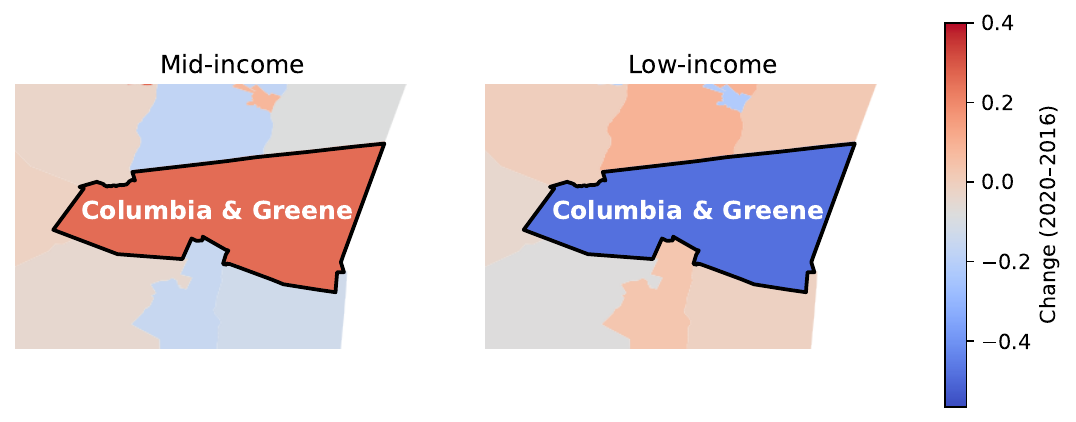}} &
        \includegraphics[width=0.3\textwidth]{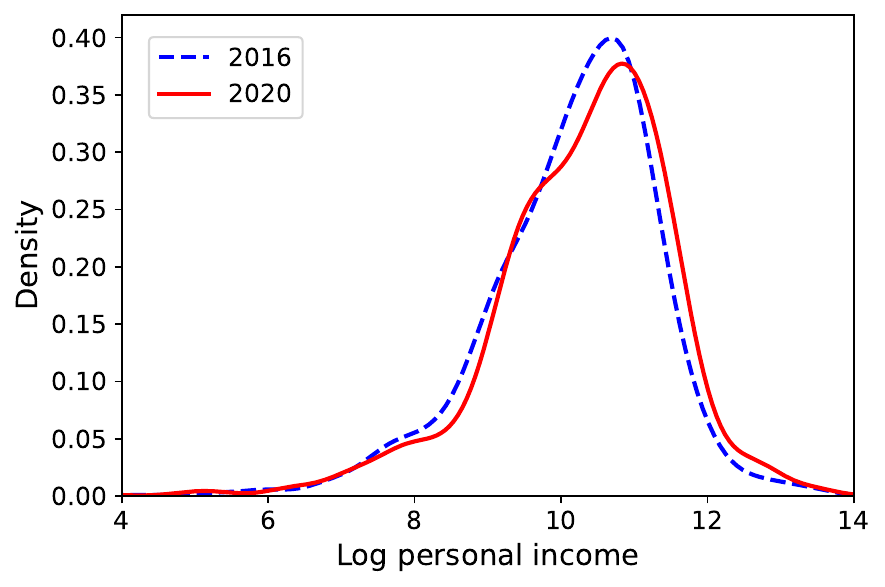} \\
        \multicolumn{3}{c}{New York City - Bronx}\\
        \multicolumn{2}{c}{\includegraphics[width=0.50\textwidth]{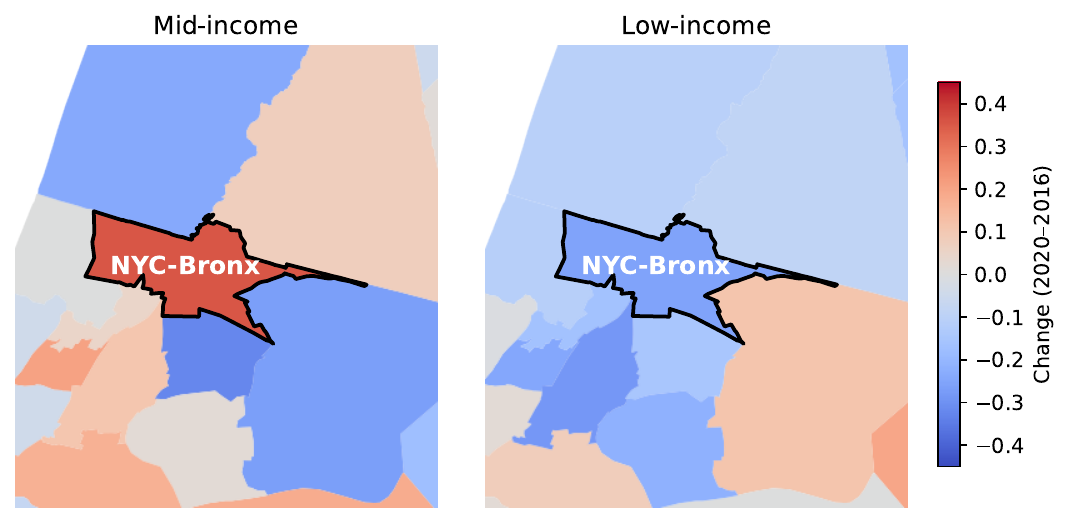}} &
        \includegraphics[width=0.3\textwidth]{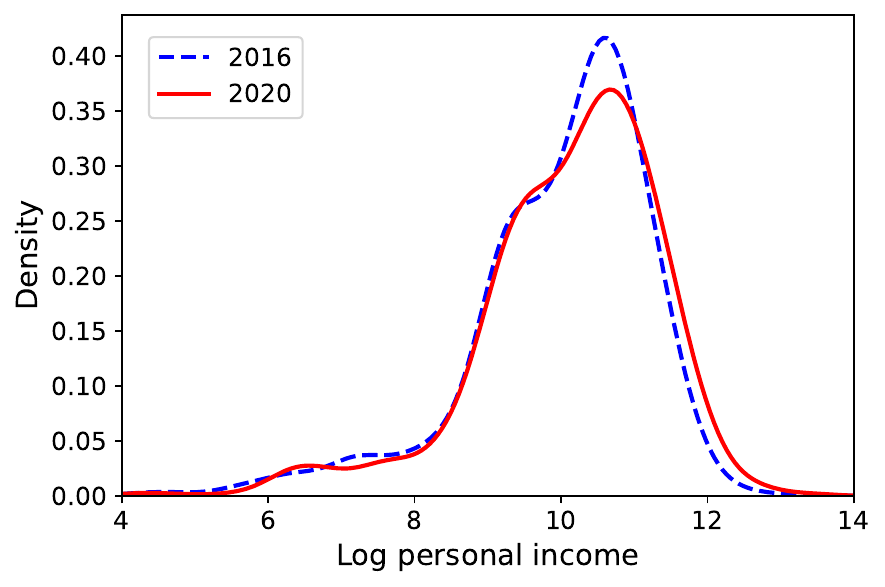}
    \end{tabular}
    \caption{New York: Results for PUMAs with the biggest change, $\Delta s_{j,h}$, between 2016 and 2020 (Columbia \& Greene and 
    New York City-Bronx). Left-hand graphs: heat map zoom ins. Right-hand graph:  posterior mean density of log income.}
    
    \label{fig:ny_columbia_nyc}
\end{figure}

Figure \ref{fig:ny_columbia_nyc} provides the zoom-in to highlight these changes in factor weights for Columbia and Green (top row) and New York City-Bronx (bottom row) together with the corresponding posterior mean densities of log personal income in 2016 and 2020.
For Columbia \& Greene we have an increase in weight for the mid-income factor and a decrease in weight for the low-income factor. This suggests a shift in probability mass away from the lower-income range and toward the middle of the income distribution over time. A similar pattern is observed for New York City - Bronx, where the mid income factor weight increases and the low income factor weight decreases, again indicating a transition from lower to middle income levels. The density plots in the third column are consistent with these spatial patterns, showing that in both areas for 2020 we have more mass around mid-income values compared to 2016 and less mass around the low income values. Columbia and Green's economy is driven by agriculture and tourism, sectors that can contribute to income disparities. The Gini coefficient in has been steady around 0.47 over the years leading to 2020, with a jump to 0.53 attributed to the Covid-19 pandemic.
\vspace{0.2in}


\vspace{0.15in}

\begin{figure}[htbp]
    \centering
    \includegraphics[width=\textwidth]{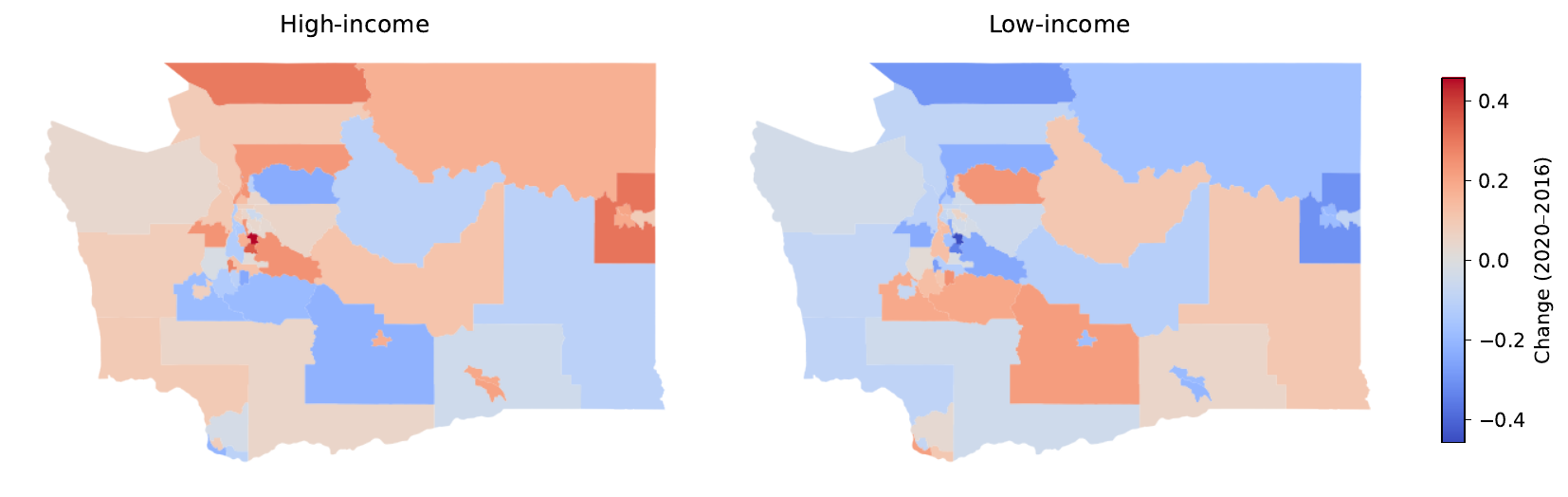}
    \vspace{-0.3cm}
    \caption{Washington: Heat maps of the changes in posterior mean factor weights, $\Delta s_{j,h}$, between 2016 and 2020.} 
    \label{fig:washington_lam_comparison}
\end{figure}
\vspace{0.15in}

\begin{figure}[htbp]
    \centering
    \begin{tabular}{ccc}
        \multicolumn{3}{c}{King (Central)} \\
        \multicolumn{2}{c}{\includegraphics[width=0.50\textwidth]{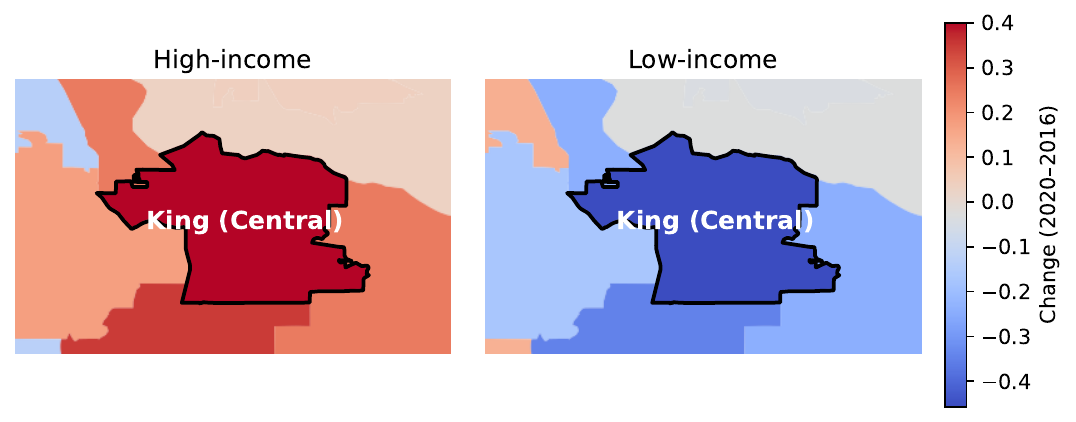}} &
        \includegraphics[width=0.3\textwidth]{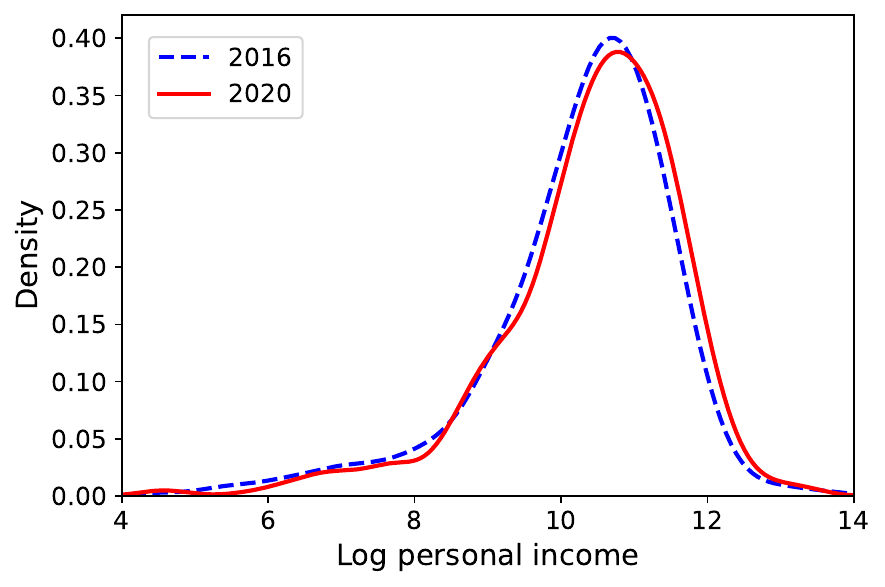} \\
        \multicolumn{3}{c}{Spokane (Outer)}\\
        \multicolumn{2}{c}{\includegraphics[width=0.50\textwidth]{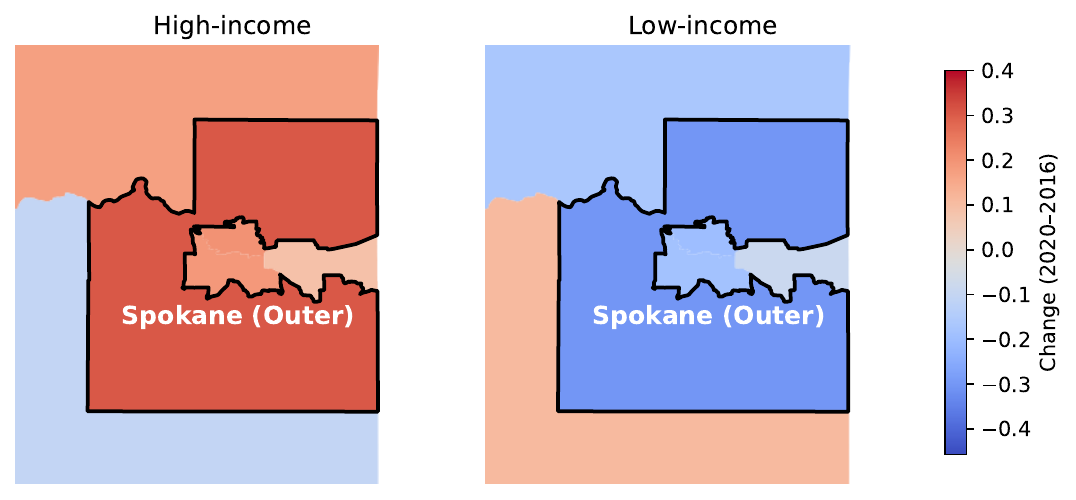}} &
        \includegraphics[width=0.3\textwidth]{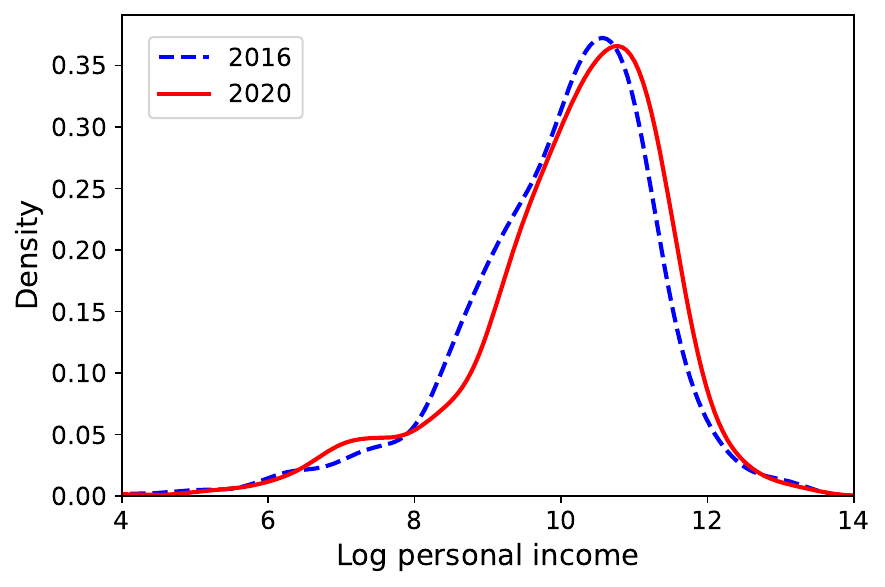}
    \end{tabular}
    \caption{Washington: Results for PUMAs with the biggest changes, $\Delta s_{j,h}$, between 2016 and 2020 (King (Central) and Spokane (Outer)). Left-hand graphs: heat map zoom ins. Right-hand graph:  posterior mean density of log income.}
    
    \label{fig:wa_king_spokane}
\end{figure}

Figure \ref{fig:washington_lam_comparison} presents the changes in factor weights across Washington from 2016 to 2020. Figure \ref{fig:wa_king_spokane} zooms in on King Central and Spokane Outer PUMAs, where the biggest weight changes are observed. For both PUMAs, we see a positive change in the high-income factor and a negative change in the low income factor from 2016 to 2020. Taken together, these patterns indicate an upward shift in the income distribution, with probability mass moving away from lower-income levels toward higher-income levels over time. The density estimates shown in Figure \ref{fig:wa_king_spokane} support this interpretation by showing a modest rightward shift in log personal income in 2020 relative to 2016 for both PUMAs.

\newpage
\subsection{Area specific covariate effects ($\eta_{h,m}$) on personal income distribution for Florida, Washington and New York} \label{sec: app total effect across states}
To examine how our chosen covariates affect the income distributions in Florida, New York, and Washington, we produce Figures \ref{fig:covariates_2016_2020_florida}–\ref{fig:wa_race_combined}. Figures \ref{fig:covariates_2016_2020_florida}, \ref{fig:covariates_2016_2020_ny} and \ref{fig:covariates_2016_2020_washington} display the estimated changes in PUMA specific covariate effects, $\Delta\, \eta_{h,m}$ between 2016 and 2020 for Florida, New York and Washington respectively. As in the case of California we use the high income factor as the baseline and so the plots provide the contrasts of the mid and low income factors, see Section \ref{sec:cali}. For a more granular study of the biggest changes in PUMA specific covariate effects we provide zoom in maps and the corresponding personal income densities .

\subsubsection{Florida} 
\begin{figure}[htbp]
    \centering
    \begin{subfigure}[b]{0.7\linewidth}
        \includegraphics[width=\linewidth]{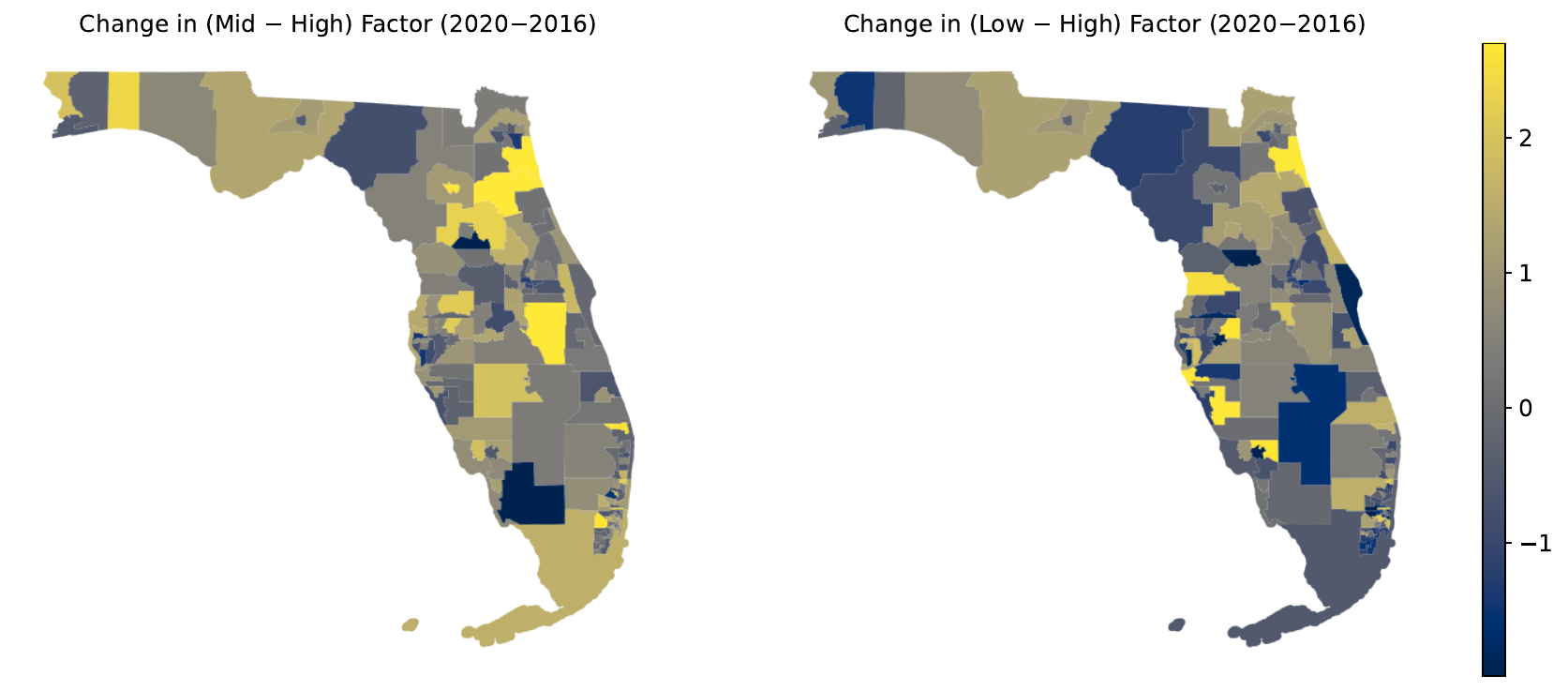}
        \caption{Education-related effects}
        \label{fig:schl_2016_2020_fl}
    \end{subfigure}
    \hfill
    \begin{subfigure}[b]{0.7\linewidth}
        \includegraphics[width=\linewidth]{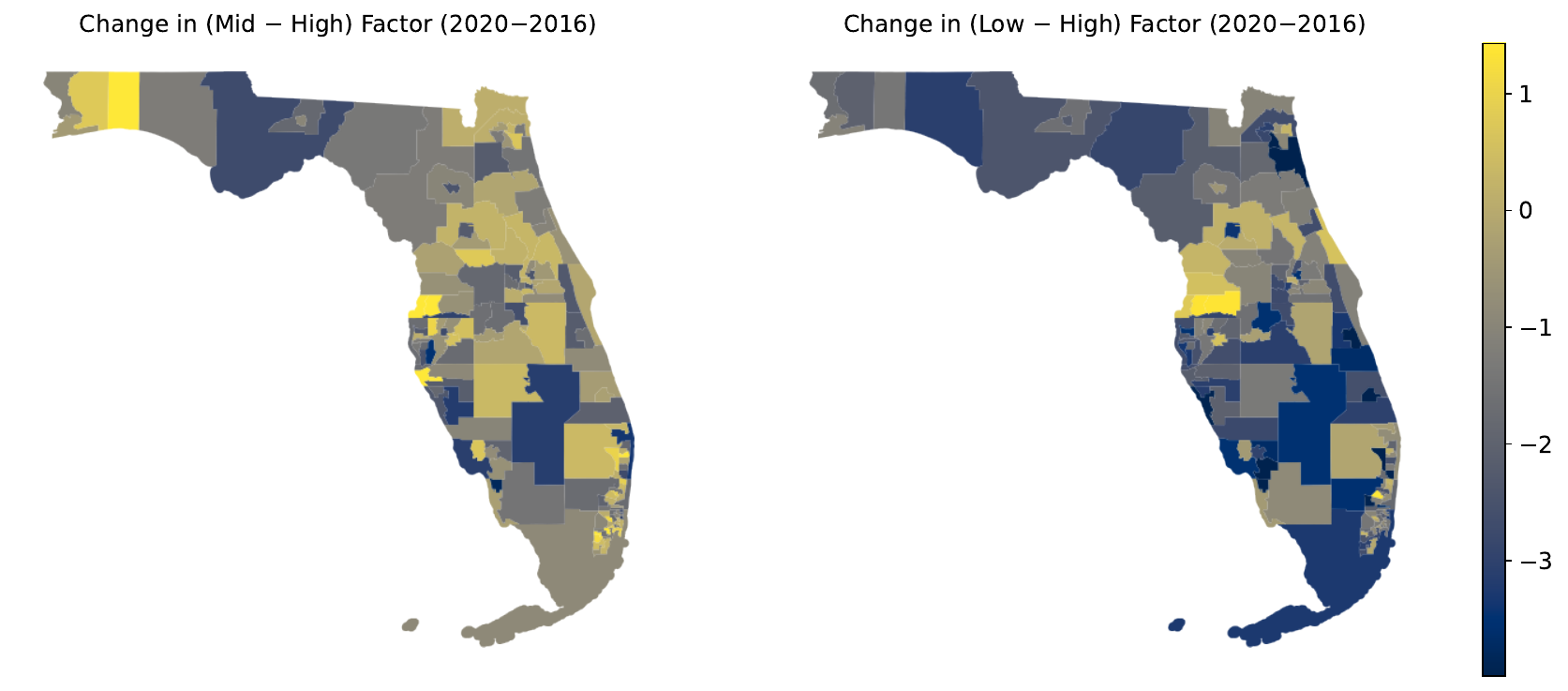}
        \caption{Race-related effects}
        \label{fig:race_2016_2020_fl}
    \end{subfigure}
    \hfill
    \begin{subfigure}[b]{0.7\linewidth}
        \includegraphics[width=\linewidth]{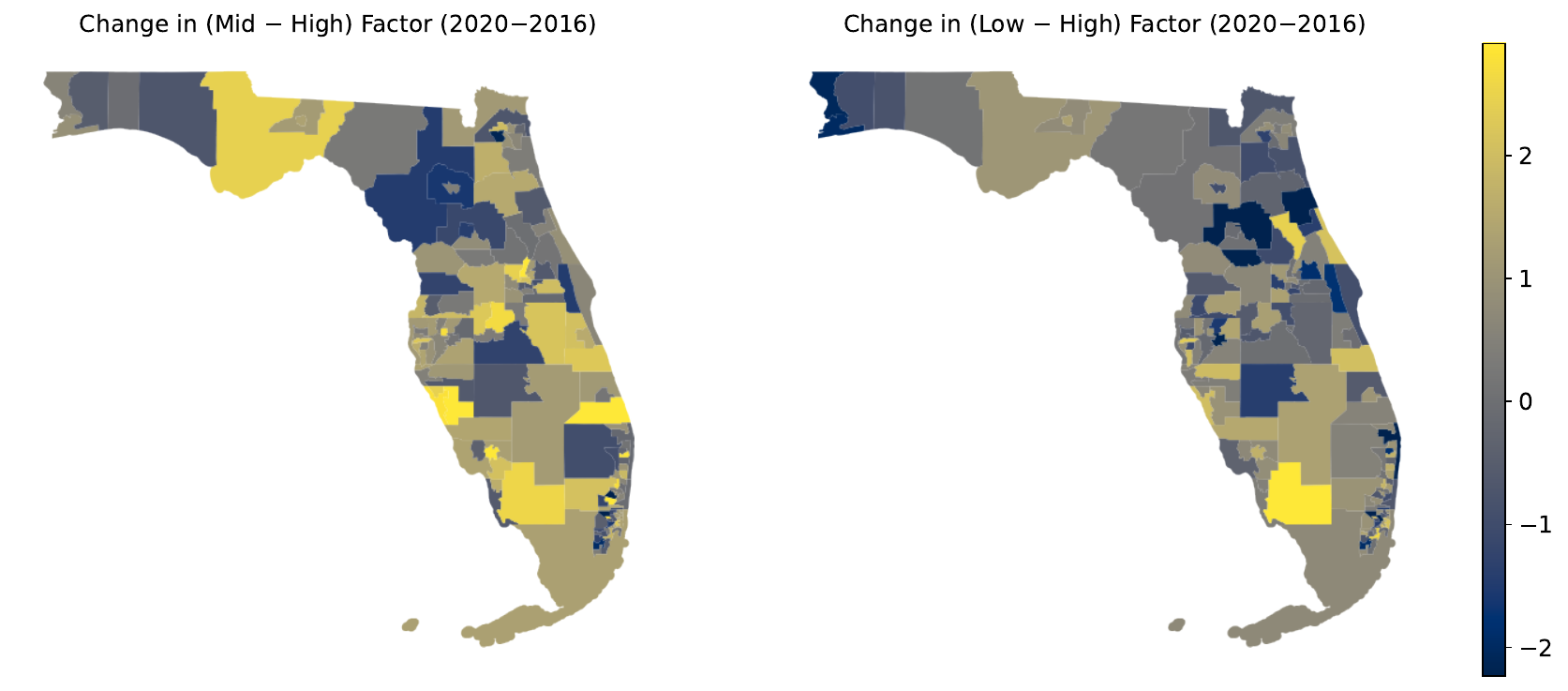}
        \caption{Gender-related effects}
        \label{fig:gender_2016_2020_fl}
    \end{subfigure}
    \caption{Florida: Heat maps of contrast changes of PUMA specific covariate effects from 2016 to 2020 for mid and low income factors. Baseline - high income factor, larger positive change (bright yellow) and larger negative change (dark blue).}
    \label{fig:covariates_2016_2020_florida}
\end{figure}

Figure \ref{fig:covariates_2016_2020_florida} displays the changes in mid and low factor contrasts between 2016 and 2020 for Florida. The top row displays the educational attainment effects, the middle row the race effects and the bottom row the gender effects. The heat scales provide a guide for identifying the largest negative (dark blue) and positive contrast (bright yellow). Recall that positive contrasts indicate that large covariate effects are associated with larger weights on the mid and low income factors, whereas negative contrasts indicate a larger weight on the high income factor. 

\begin{figure}[htbp]
    \centering
    \hspace*{0.05\textwidth}\includegraphics[width=0.7\textwidth]{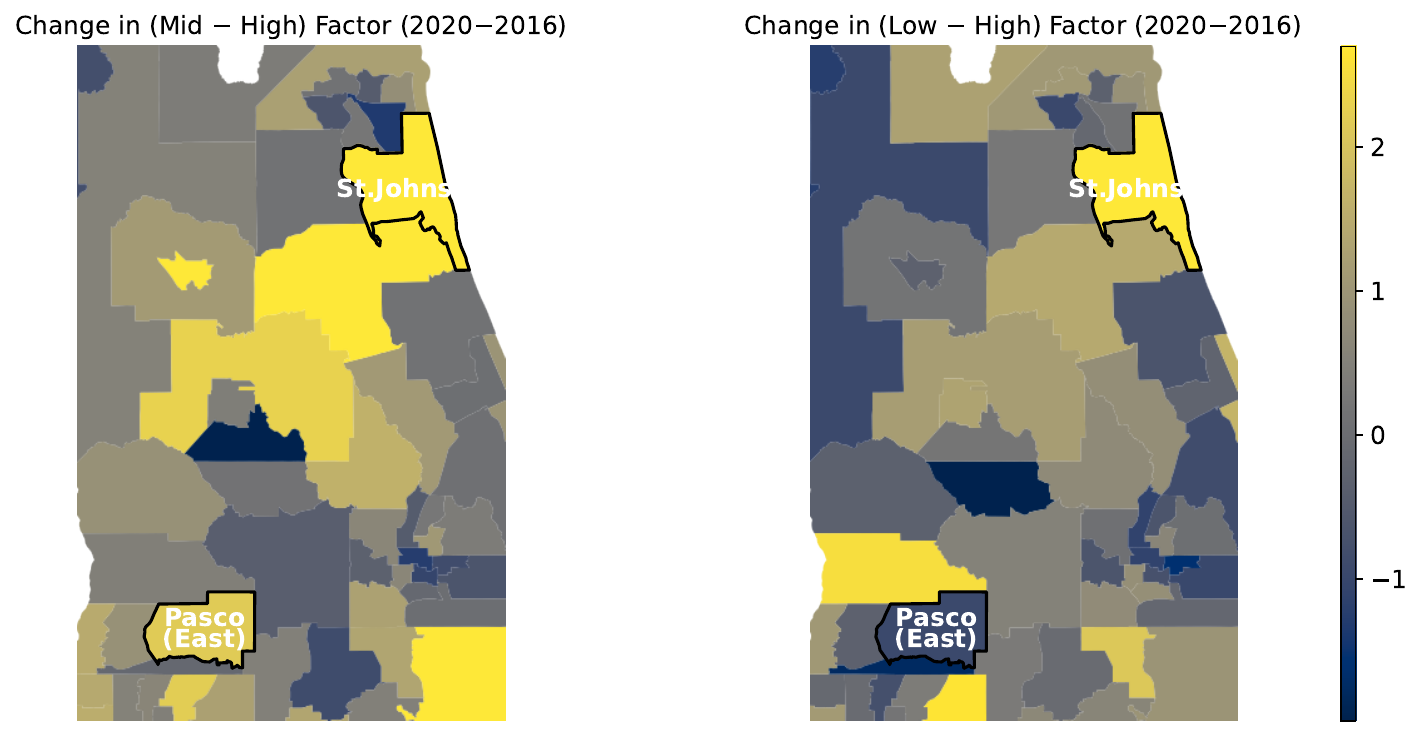}
    \vspace{0.25cm}
    
    \begin{tabular}{cc}
    Pasco East & St.\ Johns \\
    \includegraphics[width=0.35\textwidth]{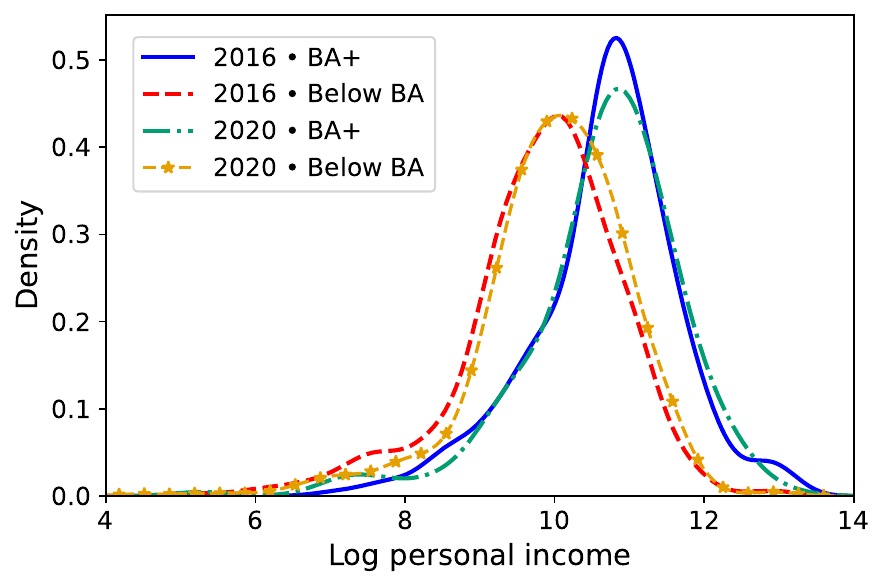}
    &
    \includegraphics[width=0.35\textwidth]{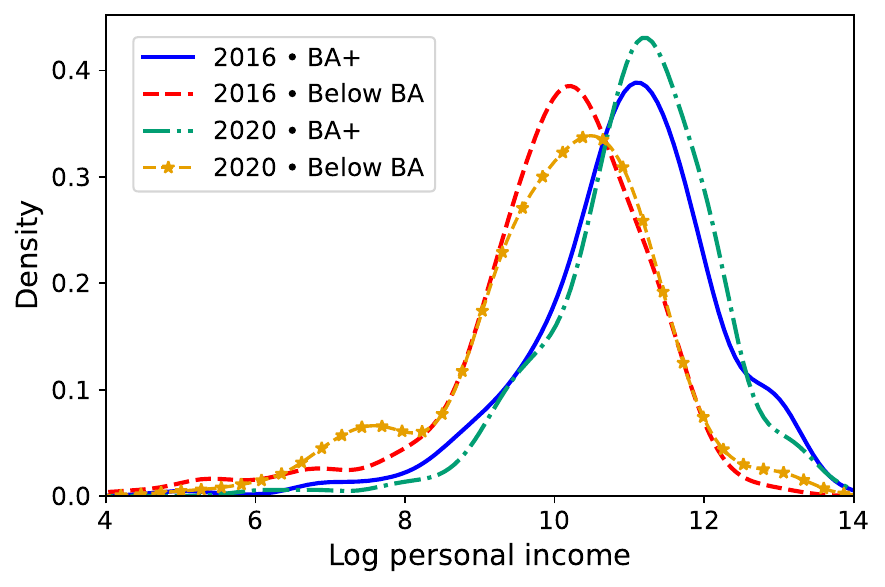}
    \end{tabular}
    
    \caption{Florida (zoom-in on Pasco and St. Johns): Changes in PUMA-specific effects of Education (high income factor as baseline) from 2016 to 2020 (top row). Posterior mean densities of log personal income by Education (BA+: bachelor's degree or higher) (bottom row).}
    \label{fig:florida_schl_combined}
\end{figure}

We begin our in-depth analysis with educational attainment and Figure \ref{fig:florida_schl_combined} presents a zoomed-in view of the biggest contrast changes between 2016 and 2020 for Pasco (East) and St.\ Johns. The corresponding posterior mean densities of log personal income by educational attainment for 2016 and 2020 are displayed in the second row. Pasco (East) shows a positive change in the mid  factor contrast and a negative change in the low factor contrast. The former indicates that educational attainment contributed more to the factor weight of the mid-income factor and less to the factor weight of the low-income factor. This shift towards the mid-income factor is supported by the corresponding density plots. This is reasonable given that a third of Pasco's residents have a high-school diploma or equivalent with only around 5\%  having a Bachelors degree. In addition the key labor market sectors are construction, hospitality services and health and education services paying close to the median wages of around \$ 40000. St.\ Johns exhibits positive changes in both the mid and low income contrasts over time. Educational attainment contributes more to the factor weight of the mid and low income factors. The density estimates provide additional evidence of this as there is a shift of the mode towards the mid and lower log-income values. 

\begin{figure}[htbp]
    \centering
    \hspace*{0.05\textwidth}\includegraphics[width=0.8\textwidth]{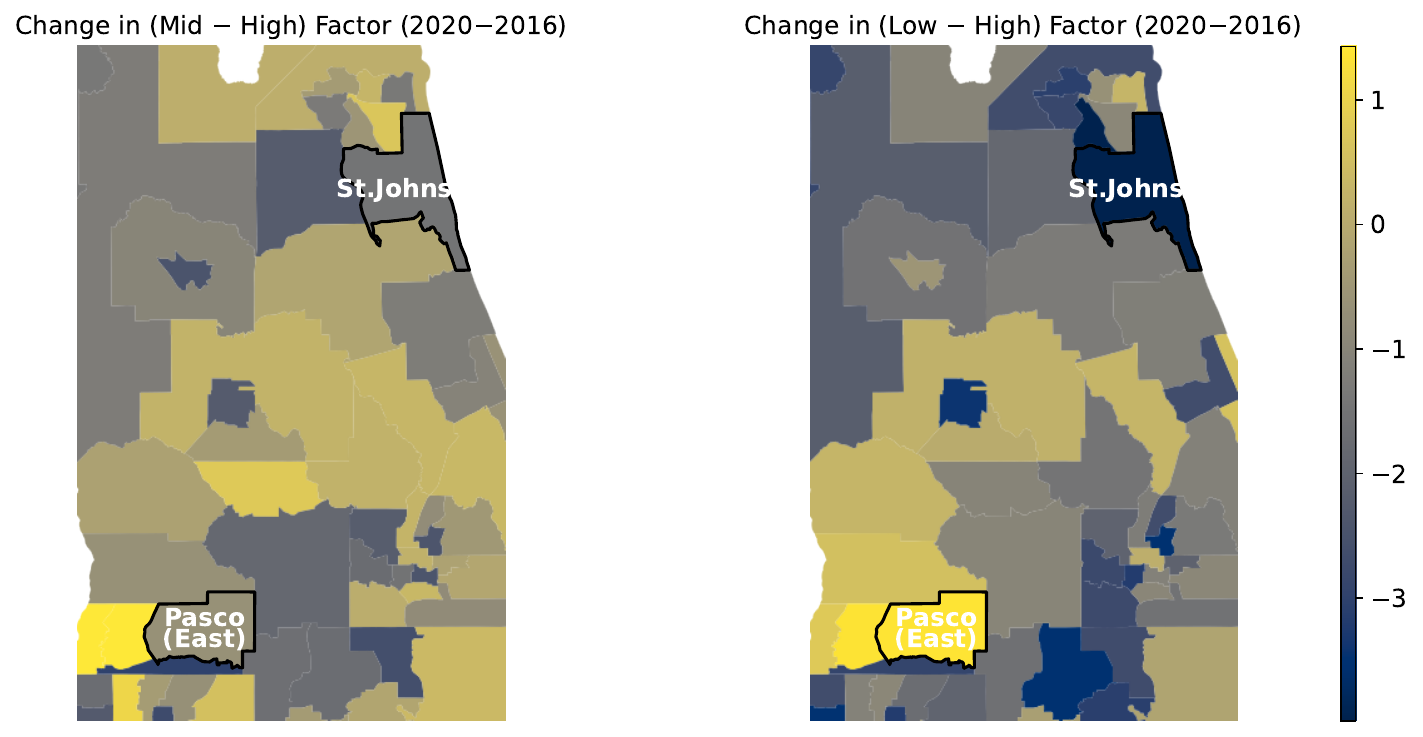}
    \vspace{0.25cm}
    \begin{tabular}{cc}
    \\[-0.1cm]
    Pasco East & St.\ Johns \\
    \includegraphics[width=0.35\textwidth]{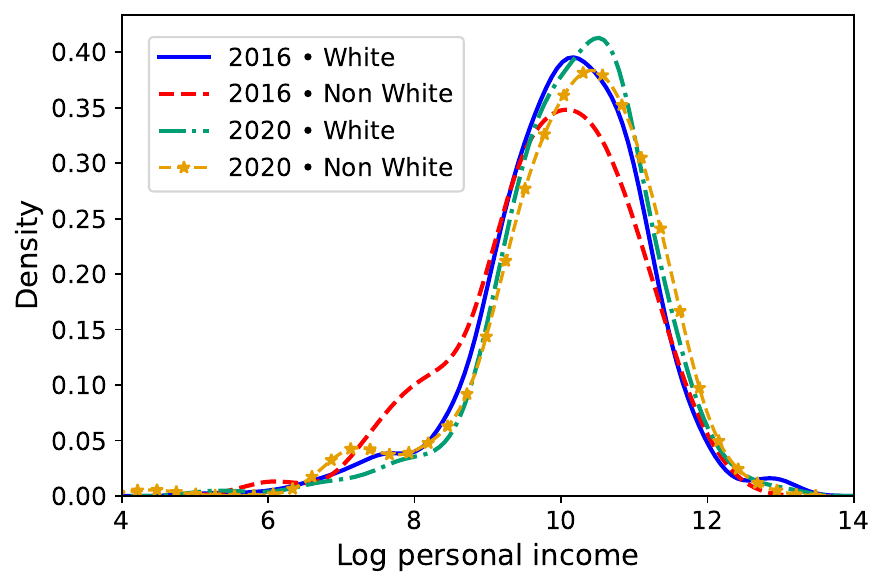}
    &
    \includegraphics[width=0.35\textwidth]{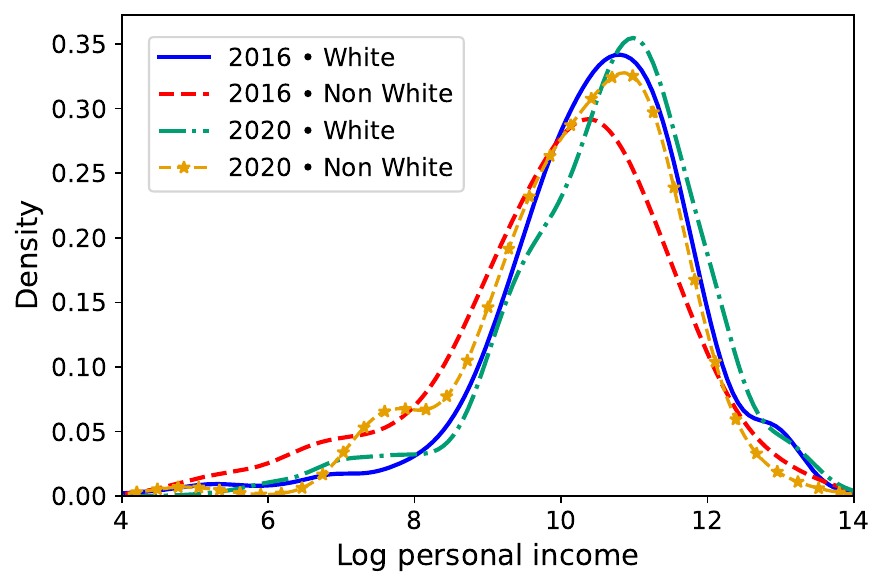}
    \end{tabular}
    \caption{Florida (zoom-in on Pasco and St. Johns): Changes in PUMA-specific effects of Race (high income factor as baseline) from 2016 to 2020 (top row). Posterior mean densities of log personal income by Race (bottom row).}

    \label{fig:florida_race_combined}
\end{figure}

Moving to the race-related effect, Figure \ref{fig:florida_race_combined} displays a zoomed-in view 
the biggest contrast changes between 2016 and 2020 for Pasco (East) and St.\ Johns. The corresponding posterior mean densities of log personal income for White and Non-White populations for 2016 and 2020 are displayed in the second row.
In St.\ Johns, both the mid and low contrasts show negative changes from 2016 to 2020, with the decline appearing especially strong for the low contrast. This suggests that the race effect contributes more to the weight of the high-income factor rather than the mid and low income factors. This is in accordance with the labor market make up of the country as non-white residents' income tends to be either at or below the median income. Pasco (East) exhibits a positive change in the low factor contrast, suggesting that the race effect contributed more to the weight of the low income factor. The posterior mean density estimates in Figure \ref{fig:florida_race_combined} provide complementary evidence on how the income distributions for White and Non-White populations evolved in these two PUMAs between 2016 and 2020.

\begin{figure}[htbp]
    \centering
    \hspace*{0.05\textwidth}\includegraphics[width=0.8\textwidth]{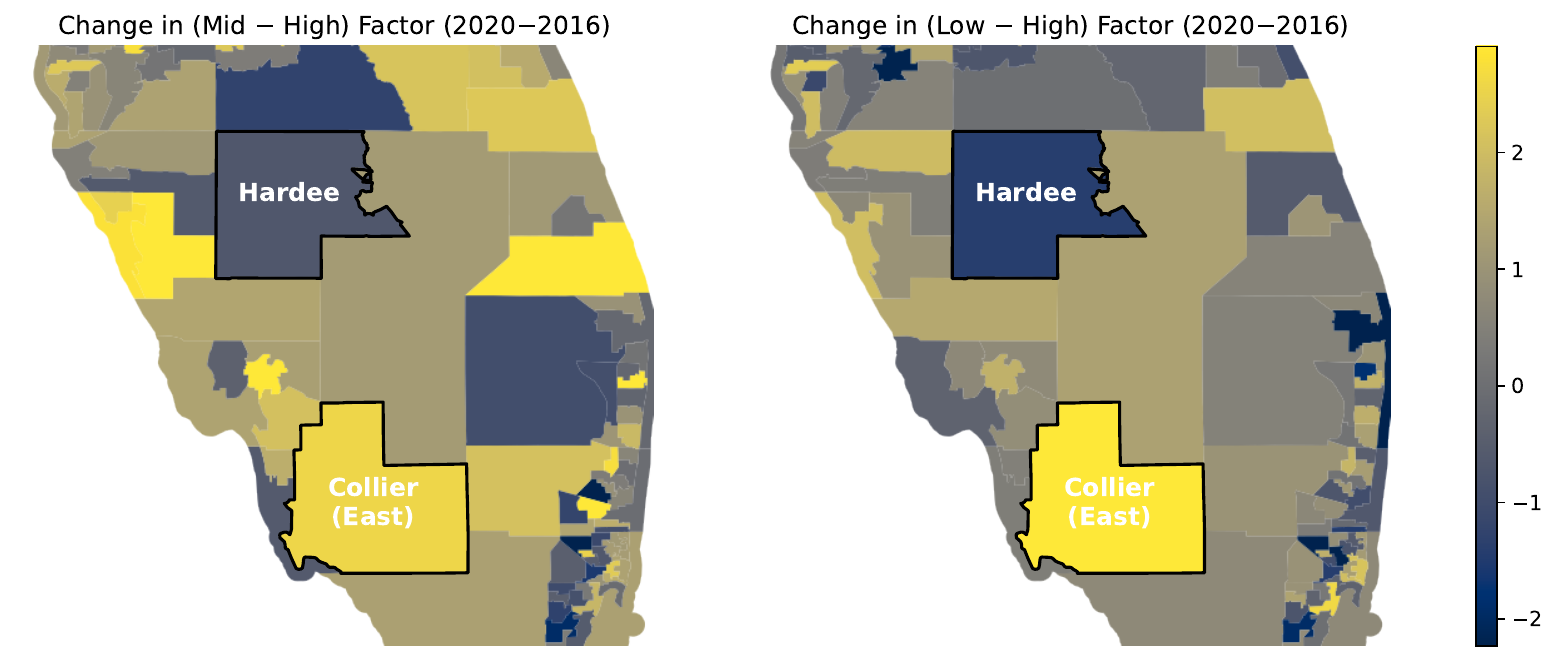}
    \vspace{0.25cm}
    \begin{tabular}{cc}
    \\[-0.1cm]
    Collier East & Hardee \\
    \includegraphics[width=0.35\textwidth]{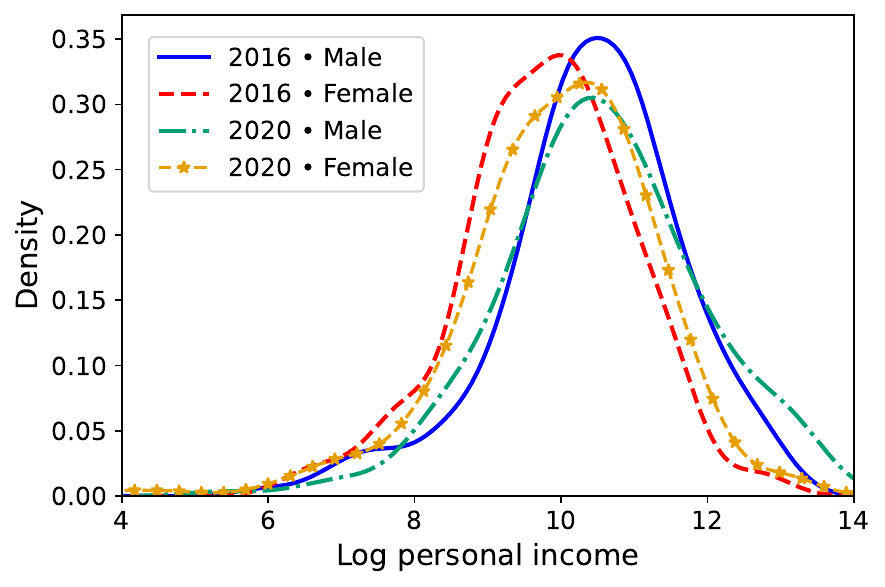}
    &
    \includegraphics[width=0.35\textwidth]{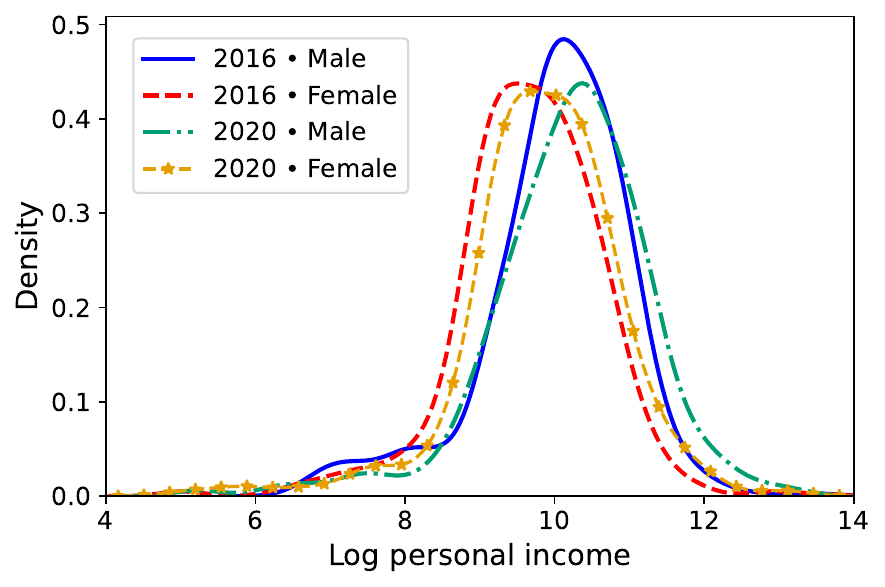}
    \end{tabular}
    \caption{Florida (zoom-in on Collier (East) and Hardee): Changes in PUMA-specific effects of Gender (high income factor as baseline) from 2016 to 2020 (top row). Posterior mean densities of log personal income by Gender (bottom row).}

    \label{fig:florida_gender_combined}
\end{figure}

\begin{figure}[htbp]
    \centering
    \begin{subfigure}[b]{0.7\linewidth}
        \includegraphics[width=\linewidth]{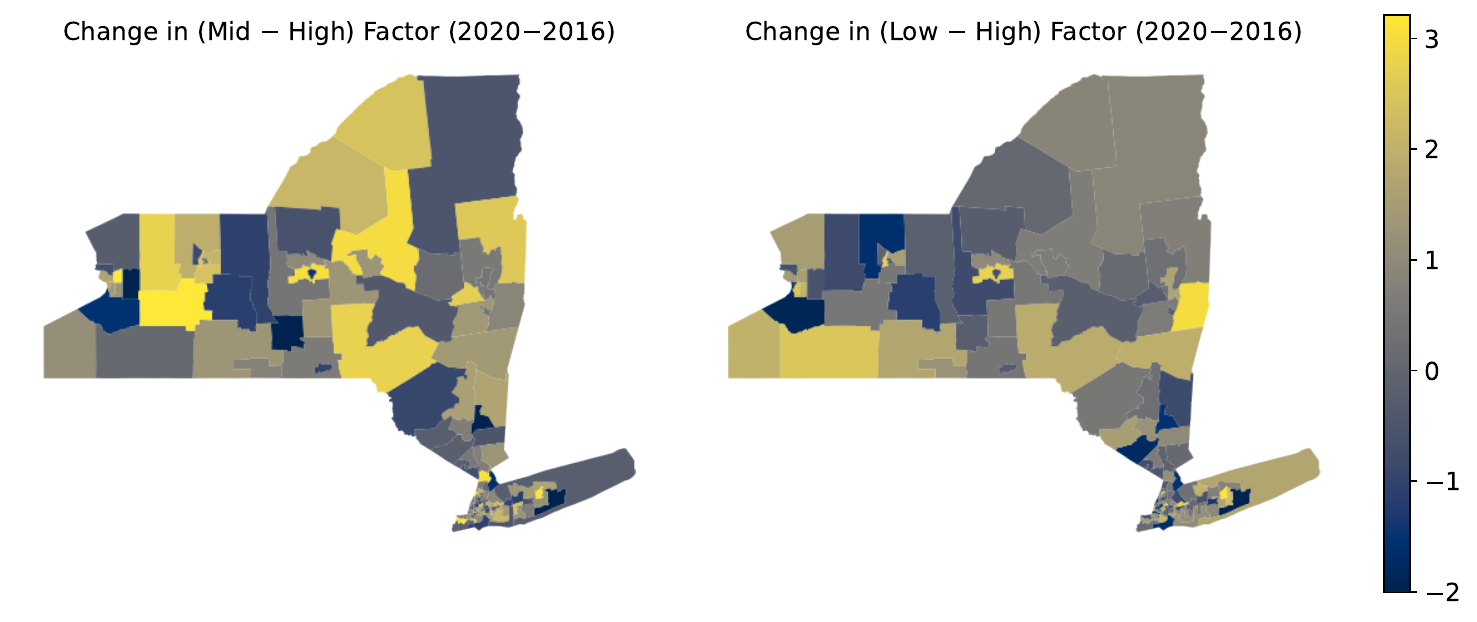}
        \caption{Education-related effects}
        \label{fig:schl_2016_2020_ny}
    \end{subfigure}
    \hfill
    \begin{subfigure}[b]{0.7\linewidth}
        \includegraphics[width=\linewidth]{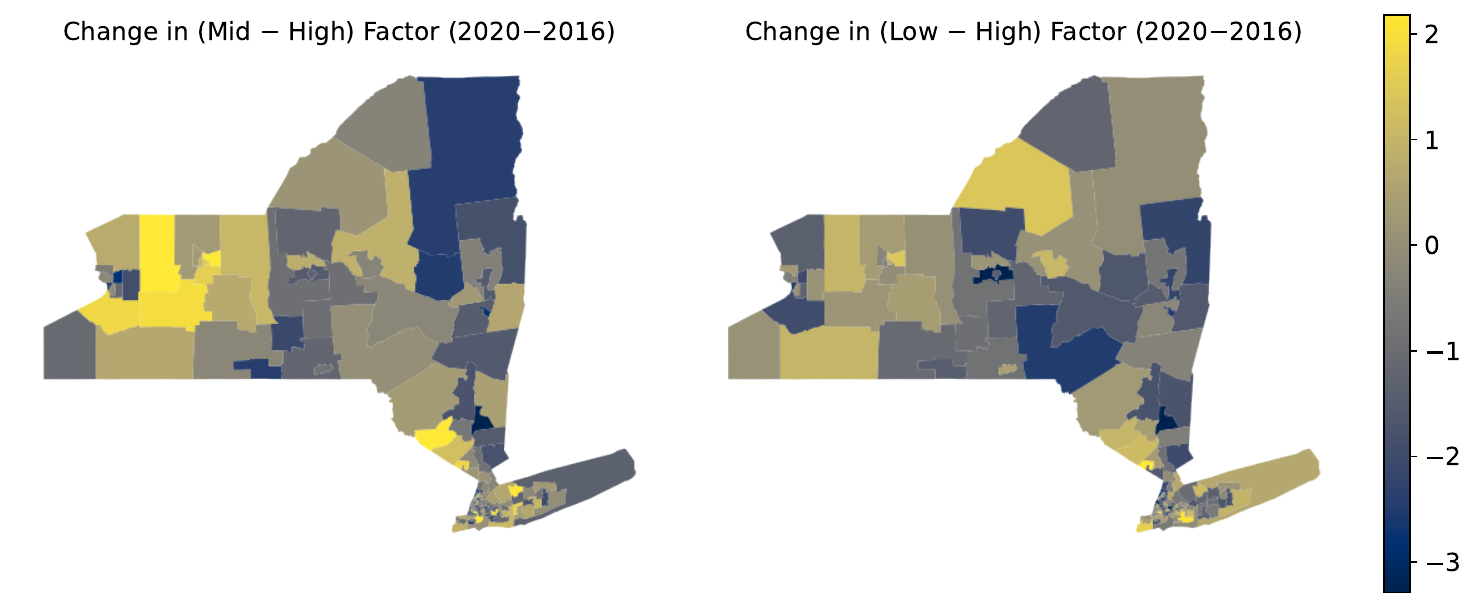}
        \caption{Race-related effects}
        \label{fig:race_2016_2020_ny}
    \end{subfigure}
    \hfill
    \begin{subfigure}[b]{0.7\linewidth}
        \includegraphics[width=\linewidth]{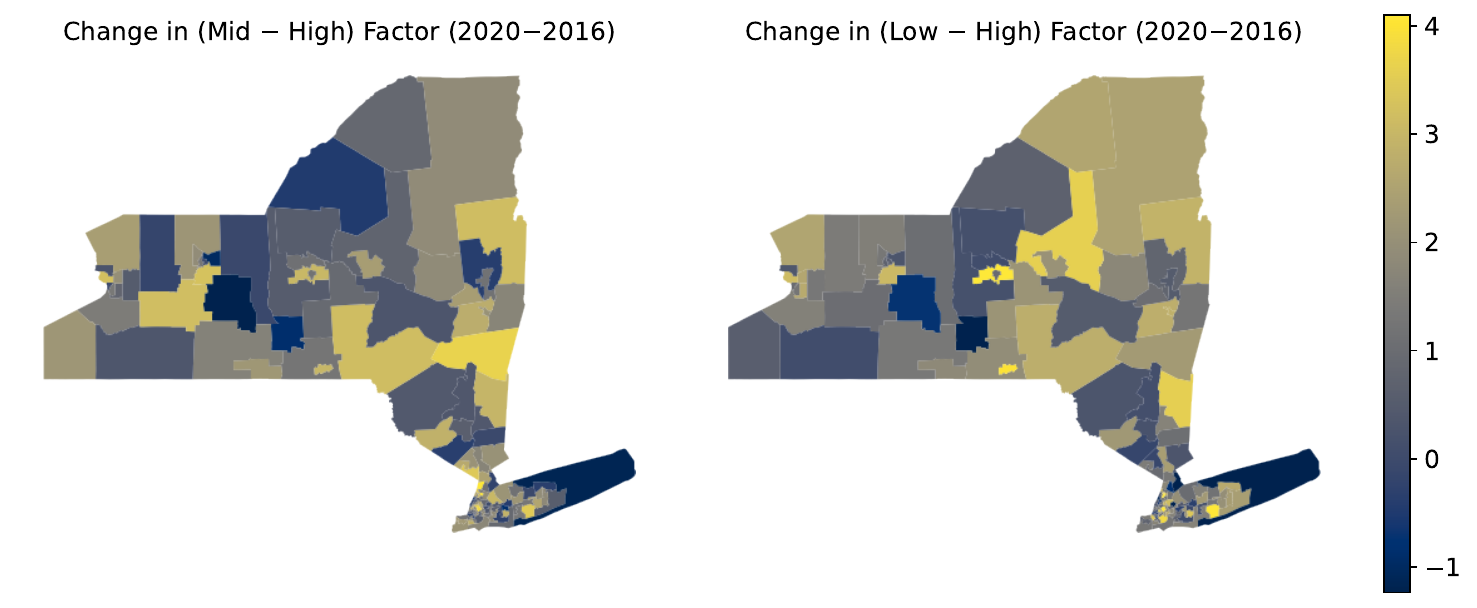}
        \caption{Gender-related effects}
        \label{fig:gender_2016_2020_ny}
    \end{subfigure}
    \caption{
    New York: Heat maps of contrast changes of PUMA specific covariate effects from 2016 to 2020 for mid and low income factors. Baseline - high income factor, larger positive change (bright yellow) and larger negative change (dark blue).}

    \label{fig:covariates_2016_2020_ny}
\end{figure}

In terms of the gender related effects Figure \ref{fig:florida_gender_combined} presents a zoomed-in view of the contracts' maps for Collier (East) and Hardee between 2016 and 2020 in the top row, whereas the bottom row provides the posterior mean densities of log personal income for males and females in 2016 and 2020. The change maps indicate contrasting temporal patterns across the two areas. In Hardee, both mid and low contrasts are negative from 2016 to 2020, suggesting that the gender effect contributes more to the weight of the high-income factor over time. Collier (East) shows positive changes in both contrasts, indicating that the gender effect contributes more to the weight of mid and low income factors. The posterior mean density estimates provide complementary evidence on how the male and female income distributions evolved in these two PUMAs between 2016 and 2020. It is worth noting that about 60\% of the workforce in Collier and Hardee is male.
\newpage

\subsubsection{New York} 
\begin{figure}[htbp]
    \centering
    \includegraphics[width=0.8\textwidth]{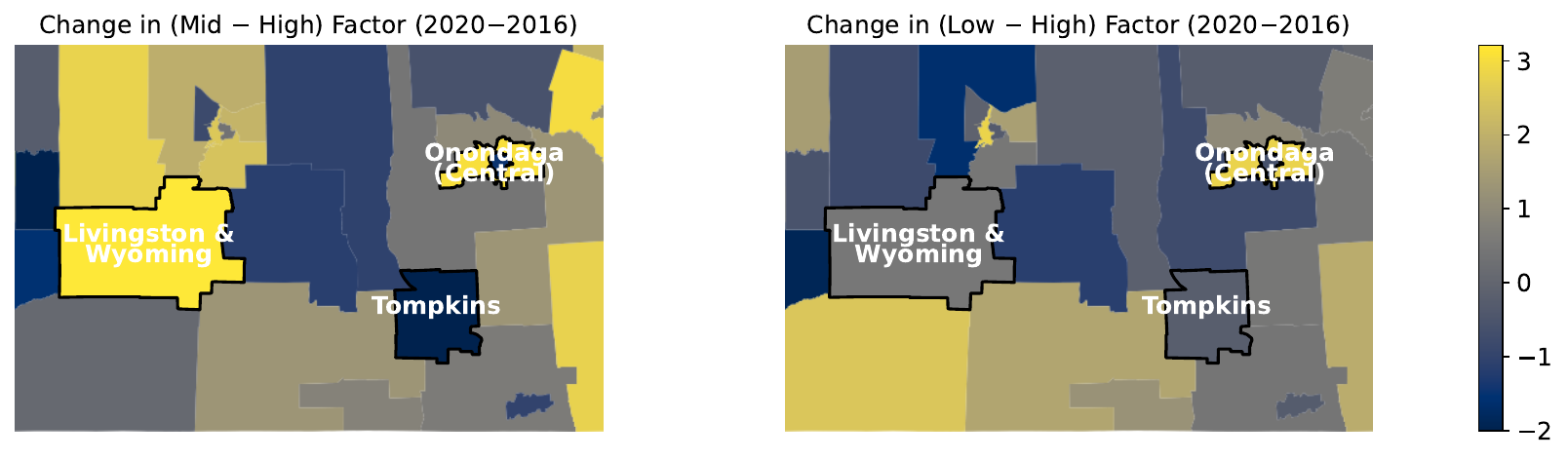}
      \begin{tabular}{ccc}
      \\[-0.1cm]
  Livingston \& Wyoming & Onondaga (Central) & Tompkins 
  \\
  \includegraphics[width=0.31\textwidth]{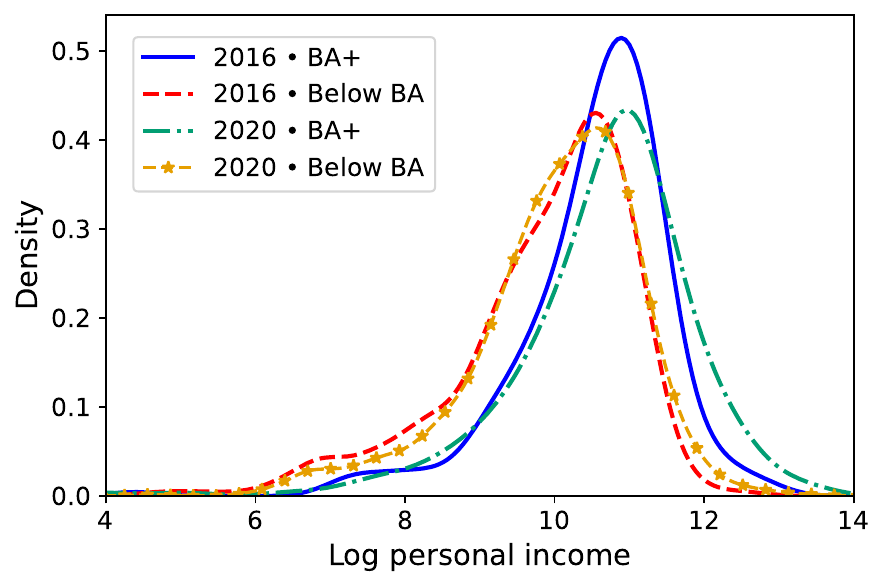}
  &
  \includegraphics[width=0.31\textwidth]{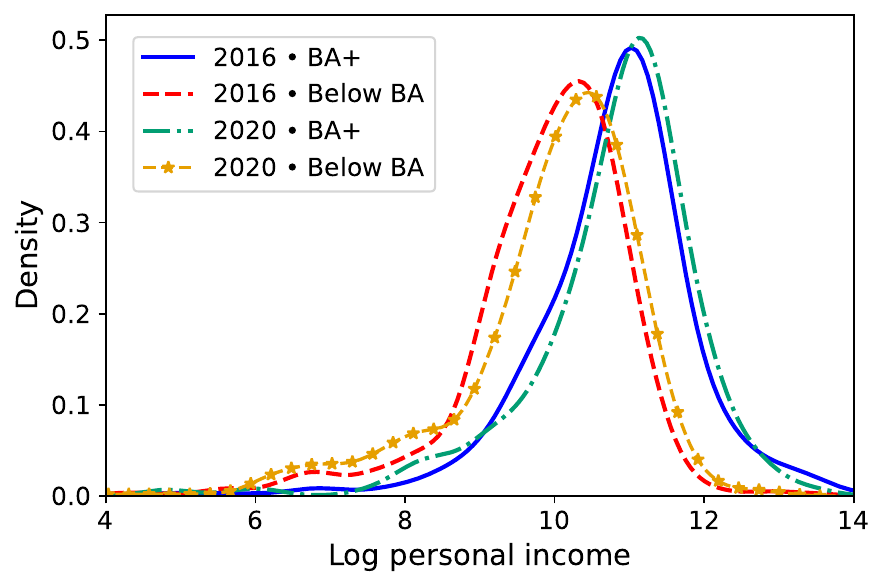}
  &
   \includegraphics[width=0.31\textwidth]{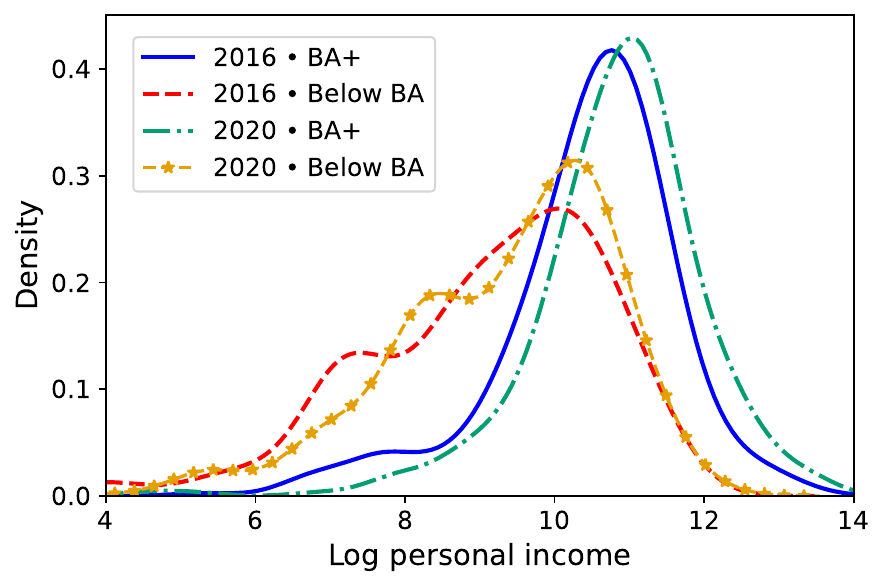}
   \end{tabular}
    \caption{New York (zoom-in on Livingston \& Wyoming, Onondaga Central and Tompkins): Changes in PUMA-specific effects of Education (high income factor as baseline) from 2016 to 2020 (top row). Posterior mean densities of log personal income by Education (BA+: bachelor's degree or higher) (bottom row).}

    \label{fig:ny_schl_combined}
\end{figure}

\begin{figure}[htbp]
    \centering
    \includegraphics[width=0.8\textwidth]{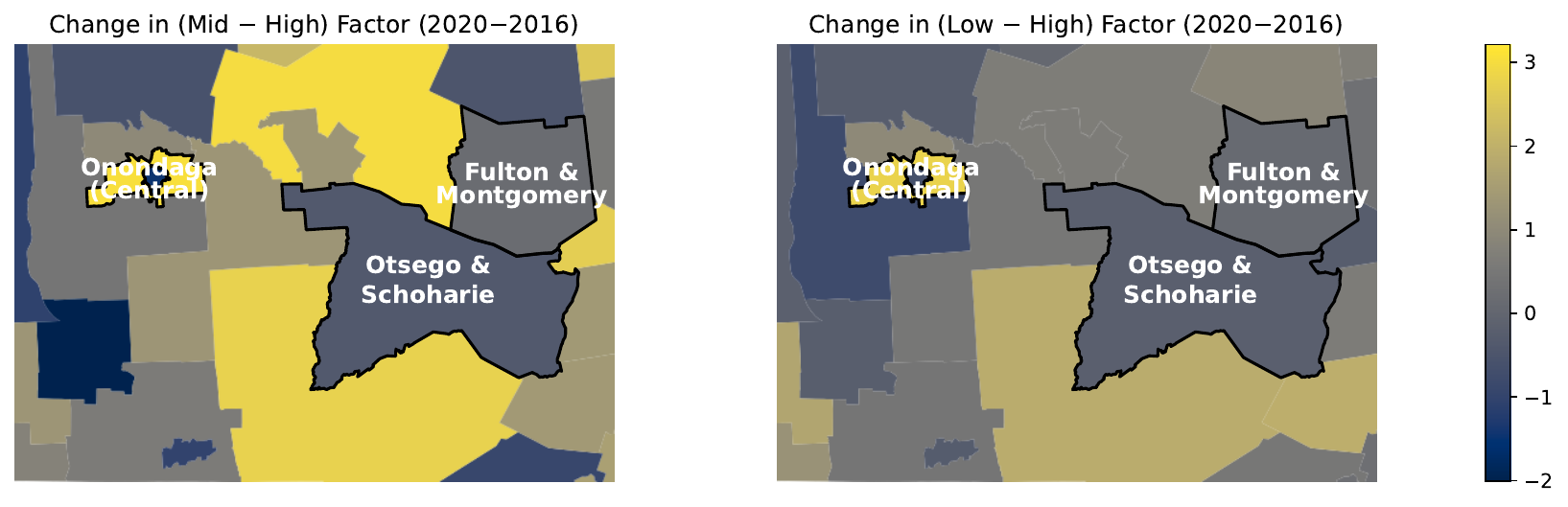}
   \begin{tabular}{ccc}
   \\[-0.1cm]
  Onondago (Central) & Otesgo  \& Schoharie & Fulton \& Montegomery 
  \\
    \includegraphics[width=0.31\textwidth]{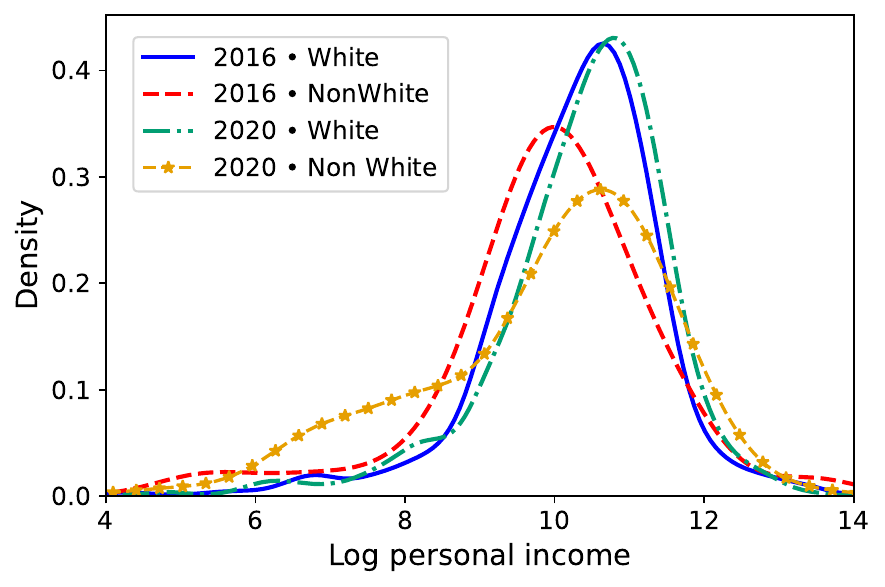}
  &
  \includegraphics[width=0.31\textwidth]{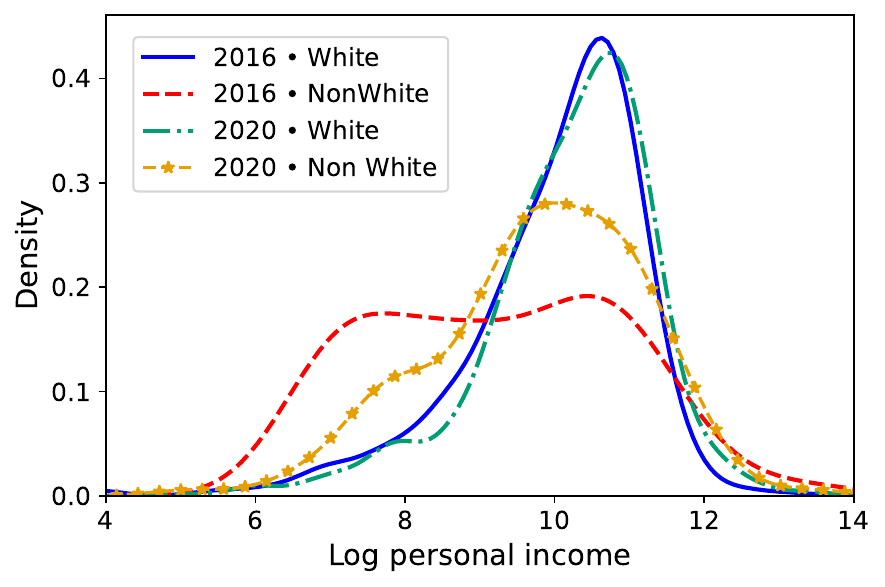}
  &
  \includegraphics[width=0.31\textwidth]{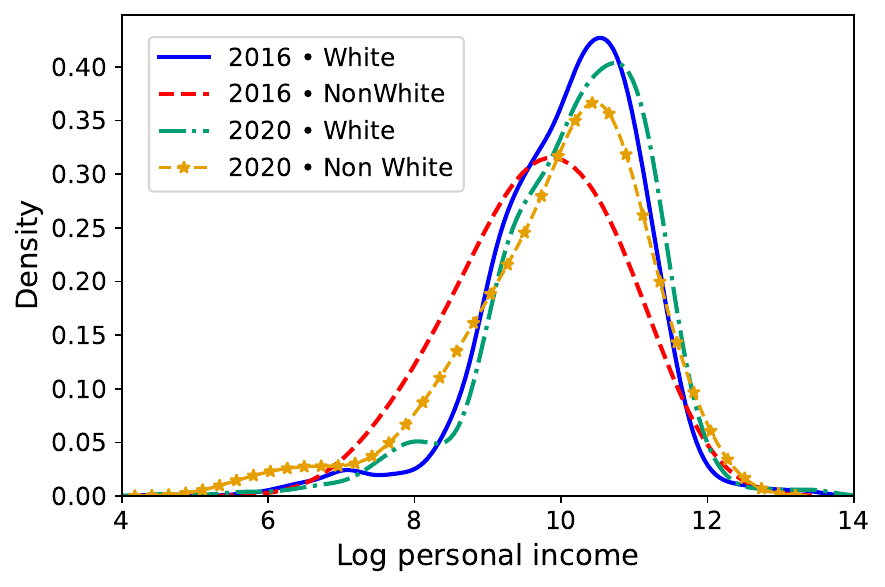}
   \end{tabular}
    \caption{New York (zoom-in on Onondago Central, Otesgo \& Schoharie and Fulton \& Montgomery): Changes in PUMA-specific effects of Race (high income factor as baseline) from 2016 to 2020 (top row). Posterior mean densities of log personal income by Race  (bottom row).}

    \label{fig:ny_race_combined}
\end{figure}

\begin{figure}[htbp]
    \centering
    \hspace*{0.1\textwidth}\includegraphics[width=0.8\textwidth]{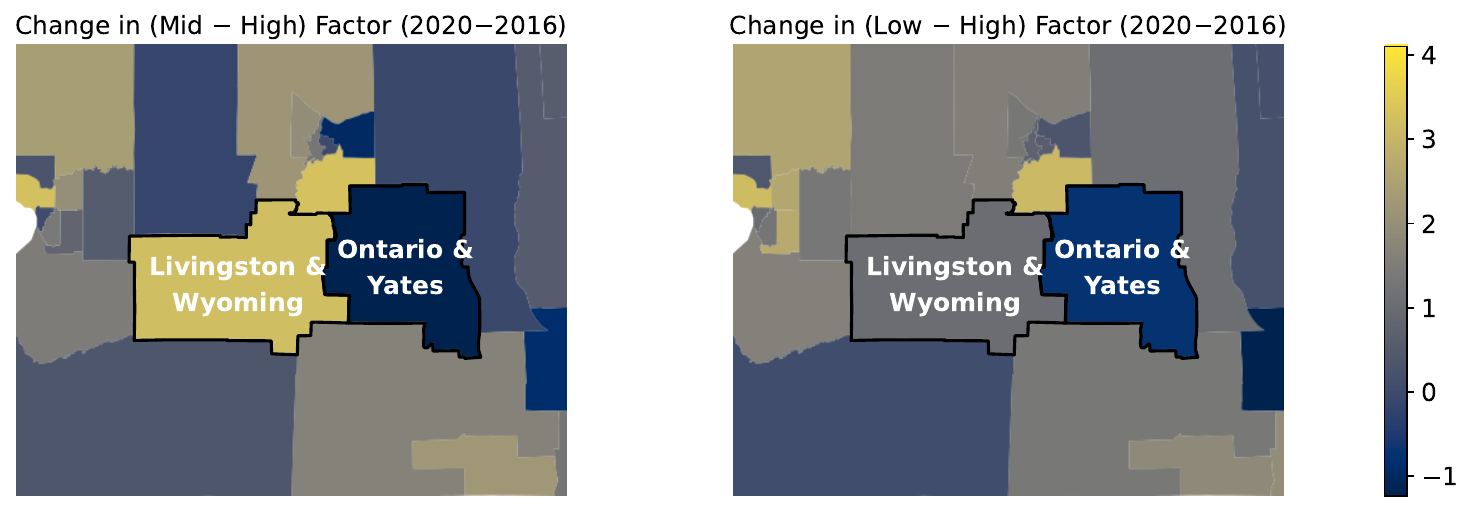}
    \begin{tabular}{cc}
    \\[-0.1cm]
    Livingston \& Wyoming & Ontario \& Yates \\
    \includegraphics[width=0.35\textwidth]{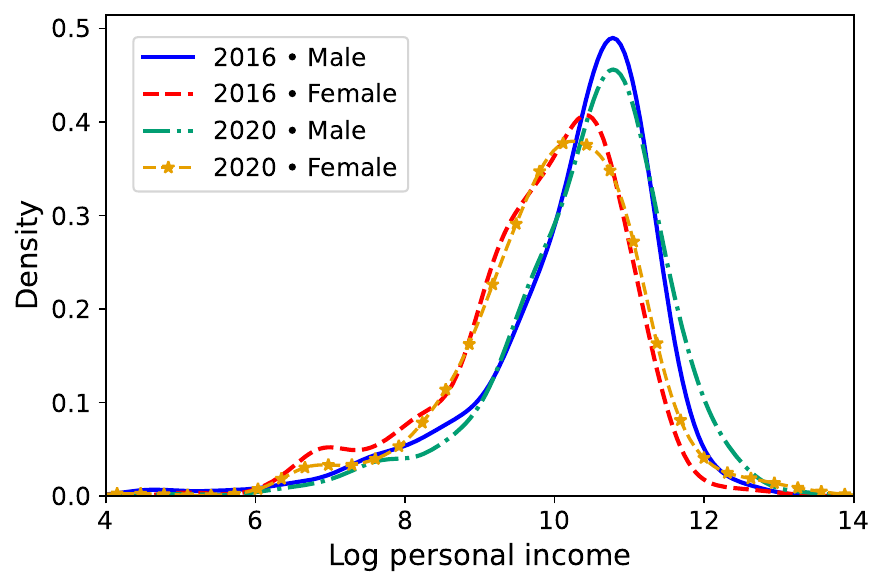}
    &
    \includegraphics[width=0.35\textwidth]{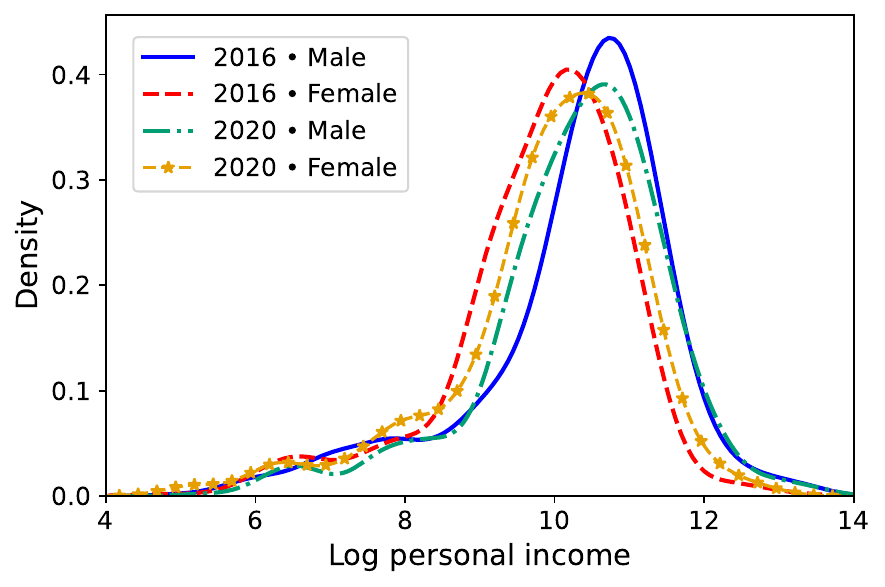}
    \end{tabular}
    \caption{New York (zoom-in on Livingston \& Wyoming, and Ontario \& Yates): Changes in PUMA-specific effects of Gender (high income factor as baseline) from 2016 to 2020 (top row). Posterior mean densities of log personal income by Gender  (bottom row).}

    \label{fig:ny_gender_combined}
\end{figure}

\begin{figure}[htbp]
    \centering
    \includegraphics[width=\textwidth]{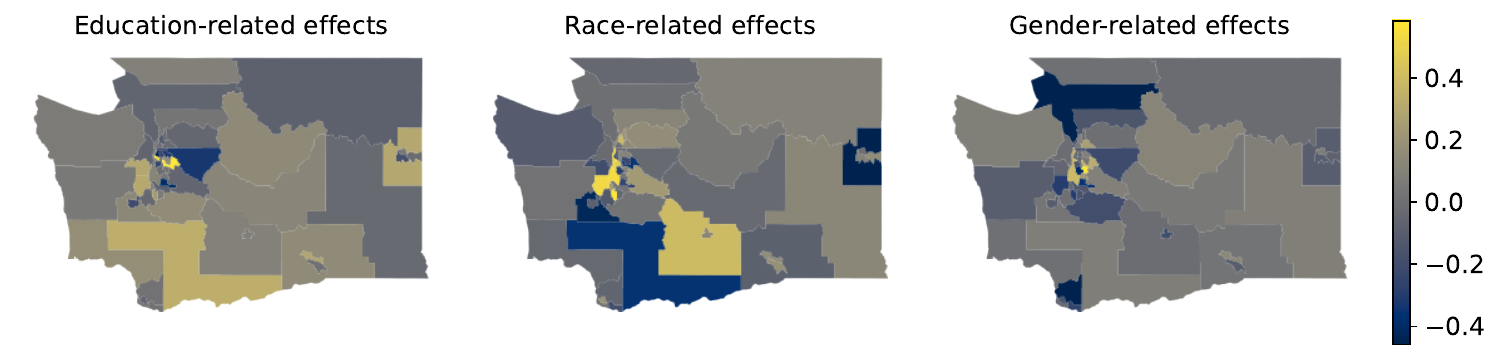}
    \caption{Washington: Heat maps of contrast changes of PUMA specific covariate effects from 2016 to 2020 for low income factors. Baseline - high income factor, larger positive change (bright yellow) and larger negative change (dark blue).}
    
    \label{fig:covariates_2016_2020_washington}
\end{figure}

Figures \ref{fig:ny_schl_combined}-\ref{fig:ny_gender_combined} zoom in on PUMAs exhibiting notable contrasts in PUMA specific covariate effects and provide related estimated density plots between 2016 and 2020.

The PUMAs with the largest educational attainment contrast changes are displayed in Figure \ref{fig:ny_schl_combined}. The zoomed-in views for Livingston \& Wyoming, Onondaga Central, and Tompkins are displayed in the top row whereas the corresponding posterior mean densities of log personal income for 2016 and 2020 in the bottom row. Livingston \& Wyoming and Onondaga Central exhibit positive changes in both the mid and low factor contrasts between 2016 to 2020. Education attainment contributed more to the factor weight of both the mid and low income factors over time. In contrast, Tompkins shows a negative change in the mid income contrast suggesting that the temporal shift in the effect of educational attainment has placed more weight on the mid-income factor. Looking at the posterior mean density estimates we can see a mode shift towards mid and lower incomes in 2020, which complements the evidence from the the heat maps. It is worth noting that in all PUMAs the biggest sectors are hospitality, health and educational services and construction where wages are at or below the median income of around \$50000 and the percentage of employees with a high school or equivalent diploma is close to 80\%. 

The PUMAs with the largest changes in race related contrast between 2016 and 2020 are displayed in Figure \ref{fig:ny_race_combined}. The zoom-ins for Onondaga (Central), Otsego \& Schoharie, and Fulton \& Montgomery are diaplyed in the tope row, with the corresponding posterior mean densities of log personal income for White and Non-White populations in 2016 and 2020 displayed in the bottom row. Onondaga (Central) and Fulton \& Montgomery show negative changes in both the mid and low income contrasts, indicating that more weight is placed on the high income factor in 2020 compared to 2016. Otsego \& Schoharie exhibits a pronounced positive change in both low and mid contrasts, suggesting that in 2020 the race effect contributed more to the weight of the low and mid income factors. The shift in the modes of density estimates provide additional evidence on the evolution of income distributions across the three PUMAs between 2016 and 2020.

The PUMAs with the largest contrast changes in gender effects between 2016 and 2020 are displayed in Figure \ref{fig:ny_gender_combined}. These PUMAs are Livingston \& Wyoming and Ontario \& Yates the heatmaps fo which are displayed in the top row with the corresponding posterior mean densities of log personal income for males and females displayed in bottom row. Ontario \& Yates exhibits negative changes in both the mid and low income contrasts from 2016 to 2020, indicating that over time the gender effect placed more weight on the high-income factor. In contrast, Livingston \& Wyoming shows a positive change in the mid income contrast. The posterior mean density estimates in the bottom row of in Figure \ref{fig:ny_gender_combined} providen evidence of the gender pay gap and how these has changed over time with a clear shift of mode towards the mid to lower income levels for females in 2020.

\subsubsection{Washington} The PUMAs with the biggest contrast changes in area specific covariate effects are zoomed in for detailed interpretation, following the same approach as in Florida and New York. 

\begin{figure}[htbp]
    \centering
    \begin{tabular}{cc}
  \includegraphics[width=0.4\textwidth]{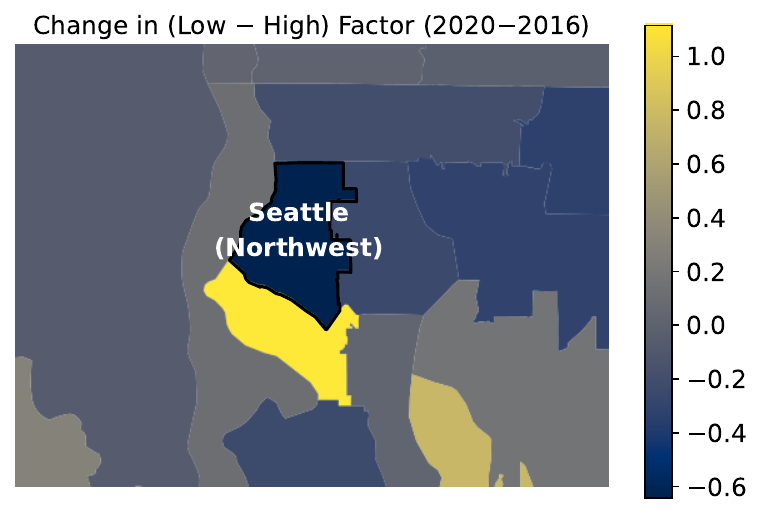}
  &
  \includegraphics[width=0.35\textwidth]{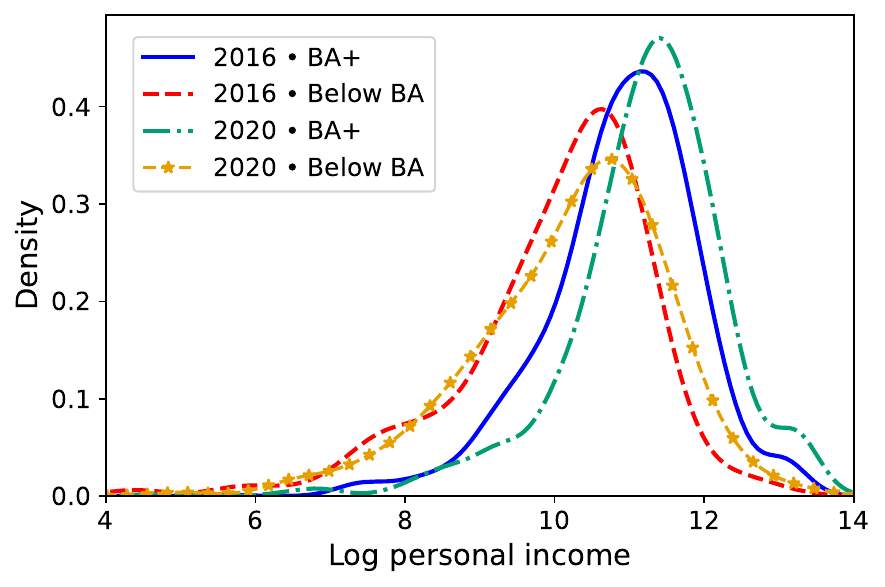}
   \end{tabular}
    \caption{Washington (zoom-in on Seattle Northwest): Changes in PUMA-specific effects of Education (high income factor as baseline) from 2016 to 2020 (top row). Posterior mean densities of log personal income by Education (BA+: bachelor's degree or higher) (bottom row).}
        \label{fig:wa_schl_combined}
\end{figure}

\begin{figure}[htbp]
    \centering
    \begin{tabular}{cc}
  \includegraphics[width=0.4\textwidth]{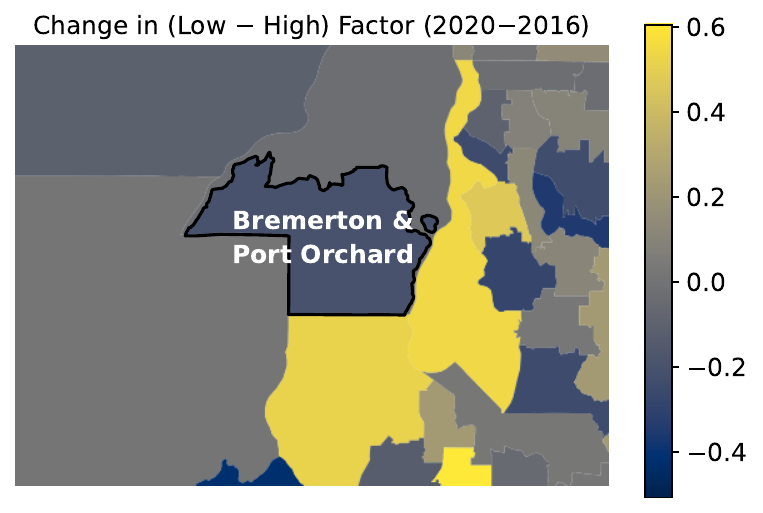}
  &
  \includegraphics[width=0.35\textwidth]{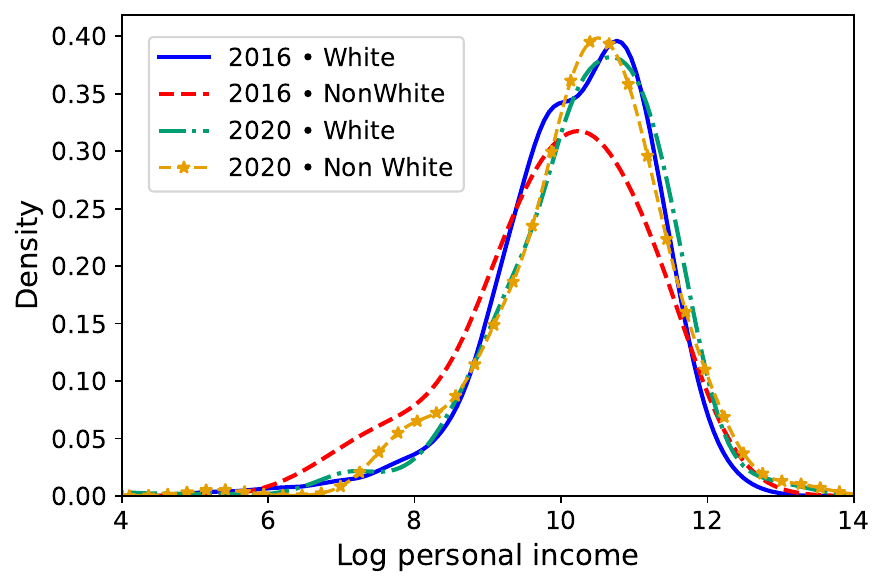}
   \end{tabular}
    \caption{
    Washington (zoom-in on Bremerton \& Port Orchard): Changes in PUMA-specific effects of Race (high income factor as baseline) from 2016 to 2020 (top row). Posterior mean densities of log personal income by Education (bottom row).}
    \label{fig:wa_race_combined}
\end{figure}
Figure \ref{fig:wa_schl_combined} presents a zoomed-in view of the education-related change in the low factor contrast for Seattle Northwest, together with the corresponding posterior mean densities of log personal income for individuals with and without a bachelor’s degree in 2016 and 2020. The highlighted PUMA exhibits a negative change in the low income factor contrast from 2016 to 2020, indicating that the education effect became more strongly associated with the high-income factor relative to the low-income factor over time. The posterior mean density estimates further illustrate this pattern, showing a widening separation between the income distributions of the two educational groups, particularly in the lower tail, consistent with the strengthened association with the low-income factor.


Figure \ref{fig:wa_race_combined} presents a zoomed-in view of the race-related change in the (Low–High) contrast for the Bremerton \& Port Orchard area, together with the corresponding posterior mean densities of log personal income for White and Non-White populations in 2016 and 2020. The highlighted PUMA appears to show a negative change in the low factor contrast from 2016 to 2020, indicating that the race effect became less associated with the low-income factor over time. The posterior mean density estimates provide complementary evidence on how the income distributions of the White and Non-White populations evolved between 2016 and 2020.

The gender-related effects in Washington are generally not pronounced and exhibit limited spatial heterogeneity, and for this reason we do not pursue a separate zoomed-in analysis.

\end{document}